\documentclass[a4paper,11pt]{article}
\usepackage{jheppub}

\pdfoutput=1
\usepackage[T1]{fontenc}
\usepackage{graphicx}
\usepackage{amsmath}
\usepackage{xspace}
\usepackage{subcaption}
\usepackage{xcolor}
\usepackage{mathtools}
\usepackage{braket}
\usepackage{multirow}
\usepackage{colortbl}
\usepackage{setspace}
\usepackage{slashed}
\usepackage{appendix}
\usepackage{booktabs}
\usepackage{cancel}
\usepackage{array}
\usepackage{tikz}
\usepackage{wasysym}
\usepackage{amssymb}
\usepackage{hyperref}
\usepackage{cleveref}
\usepackage{makecell}
\usetikzlibrary{tikzmark}
\usetikzlibrary{positioning}
\usepackage{url}

\usepackage{ragged2e}

\usepackage{float}

\newcommand{\ubarL}{\overline{u_L}}

\newcommand{\ubarR}{\overline{u_R}}

\newcommand{\dbar}{\overline{d}}

\newcommand{\dbarR}{\overline{d_R}}

\newcommand{\ebar}{\overline{e}}
\newcommand{\ebarL}{\overline{e_L}}

\newcommand{\Lbar}{\overline{L}}

\newcommand{\Qbar}{\overline{Q}}

\newcommand{\eps}{\epsilon}

\newcommand{\befune}[1]{ \mu \frac{{\rm d}#1}{{\rm d}\mu}}

\allowdisplaybreaks

\usepackage[detect-all]{siunitx}
\definecolor{newgray}{gray}{0.85}
\definecolor{newgray2}{gray}{0.95}

\definecolor{darkblue}{rgb}{0.0, 0.0, 0.75}
\definecolor{darkred}{rgb}{0.55, 0.0, 0.0}
\definecolor{palatinatepurple}{rgb}{0.41, 0.16, 0.38}
\definecolor{burntorange}{rgb}{0.8, 0.33, 0.0}
\definecolor{darkcyan}{rgb}{0.0, 0.666666, 0.666666}
\usepackage{hyperref}
\hypersetup{
    colorlinks=true,
    linkcolor=darkcyan,
    citecolor=darkred,
    filecolor=darkcyan,
    urlcolor=darkcyan
}

\newcommand{\0}[0]{$0\nu\beta\beta$}

\Crefname{table}{}{}
\Crefname{figure}{}{}
\definecolor{newgray}{gray}{0.85}
\definecolor{newgray2}{gray}{0.95}
\newcolumntype{P}[1]{>{\centering\arraybackslash}p{#1}}
\newcolumntype{M}[1]{>{\centering\arraybackslash}m{#1}}
\newcolumntype{L}[1]{>{\raggedright\arraybackslash}m{#1}}
\newcolumntype{R}[1]{>{\raggedleft\arraybackslash}m{#1}}
\newcolumntype{?}{!{\vrule width 0.8pt}}
\newcolumntype{w}{!{\vrule width 1.6pt}}

\title{
Flavorful Lepton-Number-Violating SMEFT:\\ Loop Mixing, $0\nu\beta\beta$ and Rare Meson Decays
}

\author[a,b]{Luk\'a\v{s} Gr\'af}
\author[c]{Chandan Hati}
\author[c]{Ana Mart\'in-Gal\'an}
\author[d]{Oliver Scholer}

\emailAdd{lukas.graf@matfyz.cuni.cz}
\emailAdd{chandan@ific.uv.es}
\emailAdd{ana.martin@ific.uv.es}
\emailAdd{scholer@berkeley.edu}

\affiliation[a]{Institute of Particle and Nuclear Physics (IPNP), Faculty of Mathematics and Physics, Charles University in Prague, V Hole\v{s}ovi\v{c}k\'{a}ch 2, 180 00 Praha 8, Czech Republic}
\affiliation[b]{Institute of Physics, Silesian University in Opava, Bezru{\v{c}}ovo n{\'a}m{\v{e}}st{\'i} 1150/13, 746 01 Opava, Czech Republic}
\affiliation[c]{Instituto de F\'isica Corpuscular (IFIC), Universitat de Val\`encia–CSIC, E-46980 Val\`encia, Spain}
\affiliation[d]{Department of Physics, University of California, Berkeley, CA 94720, USA}

\abstract{
We investigate how one-loop operator mixing reshapes the flavor sensitivity of lepton-number-violating dimension-7 SMEFT interactions. Using both bottom-up and top-down EFT approaches, we study the interplay of neutrinoless double beta decay ($0\nu\beta\beta$) and neutral current rare meson decays $K\to\pi\nu\nu$, and $B\to K^{(*)}\nu\nu$, which at tree level probe apparently disconnected quark- and lepton-flavor sectors. We show that loop mixing qualitatively changes this picture: flavor structures radiatively feed into one another, substantially extending the sensitivity to higher new-physics scales across operators and flavors. In the bottom-up EFT approach, this can make $0\nu\beta\beta$ highly sensitive even to operators involving heavier-generation quarks, while rare kaon and $B$ decays retain crucial complementary sensitivity to other flavor combinations. A top-down EFT approach example using a ultraviolet model involving scalar-leptoquark illustrates how the same effects can reorder the relative importance of different observables.
}

\begin{document}	
\maketitle
\flushbottom

\section{Introduction}

Observing lepton-number violation (LNV) would have profound implications for particle physics. In particular, violation of lepton number by two units ($\Delta L=2$) can be intimately connected to the potential Majorana nature of neutrino mass and offers a key experimental avenue for probing physics beyond the Standard Model (SM)~\cite{Schechter:1981bd, Takasugi:1984xr, Duerr:2011zd, Deppisch:2012nb, Graf:2020cbf, Graf:2022lhj, Fridell:2023rtr}. The canonical low-energy observable that currently provides some of the most stringent constraints on lepton-number-violating interactions involving first-generation leptons and quarks is neutrinoless double beta decay ($0\nu\beta\beta$)~\cite{Graf:2018ozy, Deppisch:2020ztt, KamLAND-Zen:2024eml}, including its other modes~\cite{Doi:1987rx,Brune:2018sab,Cepedello:2018zvr,Graf:2023dzf,deVries:2025hqa,Bolton:2026gku}. At the same time, neutral current rare meson decays with missing energy, such as $K\to\pi\nu\nu$ and $B\to K^{(*)}\nu\nu$, probe different combinations of quark and lepton flavors and can therefore provide complementary information on the underlying lepton-number-violating new physics~\cite{Li:2019fhz,Deppisch:2020oyx,Belle-II:2023esi,NA62:2024pjp,KOTO:2024zbl}.

If the new heavy degrees of freedom responsible for LNV reside well above the electroweak scale, their effects can be described systematically within the Standard Model Effective Field Theory (SMEFT). Gauge invariance implies that $\Delta L=2$ interactions arise at odd operator dimensions~\cite{Kobach:2016ami}. The leading contribution is the dimension-five Weinberg operator~\cite{Weinberg:1979sa}, while dimension-7 operators provide the first set of interactions that can generate a much broader range of low-energy lepton-number-violating processes~\cite{Babu:2001ex,deGouvea:2007qla,Fridell:2023rtr}. A complete non-redundant basis of such operators is now well established~\cite {Lehman:2014jma,Henning:2015alf,Liao:2016hru}, and their matching onto the Low-energy Effective Field Theory (LEFT)~\cite{Jenkins:2017jig,Liao:2020zyx}, as well as their renormalization group evolution, have been developed in recent years~\cite{Liao:2019tep,Zhang:2023kvw,Zhang:2023ndw}. These developments made it possible to consistently connect new physics at a high energy scale to observables ranging from nuclear processes to rare meson decays~\cite{Fridell:2023rtr, Graf:2025cfk, Liao:2025lxg, Esser:2026ehc, deVries:2026ujz}.

A key feature of the SMEFT framework is that the flavor structure at the new-physics scale is generally not preserved under renormalization group evolution. Operators that are absent at the matching scale can be generated at loop level, and Wilson coefficients involving different quark and lepton flavors can feed into one another through Standard-Model interactions~\cite{Zhang:2023kvw,Zhang:2023ndw}. This observation is particularly relevant for the phenomenology of lepton-number-violating interactions. A high-scale interaction involving second- or third-generation quarks, for example, can induce operators that contribute to $0\nu\beta\beta$ through SMEFT operator mixing. Conversely, operators constrained at tree level only by nuclear observables may mix into structures accessible in meson decays. The resulting bounds can therefore differ substantially from those inferred from tree-level or single-flavor analyses. The importance of such loop effects for $0\nu\beta\beta$ has recently been emphasized for dimension-seven SMEFT operators in~\cite{Graf:2025cfk}; see also~\cite{Liao:2025lxg,deVries:2026ujz} for some follow-up analyses. 

In this work, we present a systematic and comprehensive exploration of loop-level operator matching and mixing effects for lepton-number-violating dimension-7 SMEFT operators across quark and lepton flavors, showing how loop-level results severely alter the landscape of constraints obtained at tree level from nuclear and flavor observables. We show that the loop effects can not only remarkably improve the tree-level sensitivity of the concerned observables, but also can propagate across other flavors of quarks and leptons via SM Yukawa couplings, making the interplay of observables like $0\nu\beta\beta$, $K\to\pi\nu\nu$, and $B\to K^{(*)}\nu\nu$ decays extremely interesting and complementary in distinguishing the flavor structure of heavy new physics interactions. This is so because, at tree level, \0 is directly sensitive to electron-number violation and first-generation hadronic currents, whereas rare kaon and $B$-meson decays probe flavor-changing $s\to d$ and $b\to s$ transitions, respectively (for different neutrino flavors). However, once we account for loop-level operator mixing effects, this simple picture blurs, and a new picture of flavored constraints emerges across operators, depending on the underlying flavor structure of the new physics interactions. 

We present the relevant detailed constraints from the lepton-number-violating observables in the bottom-up EFT picture for single operator dominance and for two very distinct underlying physical flavor structures of the new physics interactions: (i) a lepton flavor universal and diagonal scenario and (ii) a scenario where all combinations of lepton and quark flavors are (one at a time and) independent. We also present the implications in a top-down EFT picture, with a concrete ultraviolet (UV) model example, to show how relevant loop matching and mixing effects change the picture of tree-level constraints from different lepton-number-violating observables and propagate across flavors.

Our findings show that when loop-induced contributions are included, \0 can provide the leading constraints on a wide range of dimension-seven SMEFT coefficients across different quark and lepton flavor combinations, including second- and third-generation quarks. For some operators, this effect is especially pronounced for third-generation quark flavors, whose evolution is enhanced by large third-generation SM Yukawa couplings. At the same time, rare kaon decays remain the most sensitive probe for selected flavor combinations, while $B$-meson decays provide complementary constraints, particularly when the underlying lepton flavor combination of new physics interactions is lepton flavor non-universal. To illustrate how this interplay arises in a top-down EFT picture, we also discuss a concrete UV model example involving two scalar leptoquarks that can generate lepton-number-violating dimension-7 SMEFT operators at tree level. By exploring different representative flavor textures
that naturally suppress tree-level lepton-number-conserving meson-decay contributions, we show how the picture of relative sensitivities among different lepton-number-violating observables changes drastically when we
consistently account for one-loop matching and mixing effects.

Our results highlight the importance of including one-loop SMEFT operator mixing effects in a consistent bottom-up or top-down EFT approach to interpret constraints from lepton-number-violating observables across different energy scales and flavors. While rare meson decays and \0 appear to probe mutually isolated flavor sectors of the new-physics interactions underlying lepton-number-violating dimension-7 SMEFT operators at tree level, at one loop the sensitivities of these observables propagate non-trivially across flavors and energy scales due to operator mixing. Therefore, the synergy and complementarity of these lepton-number-violating observables can play a crucial role in identifying the scale and flavor structure of possible lepton-number-violating new physics if a positive signal is seen in one or more of these experiments. 

The remainder of this paper is organized as follows. In Sec.~\ref{sec:eft} we introduce and define the EFT frameworks relevant for our analysis at various energy scales. We also discuss the SMEFT operator matching and RGE-governed operator mixing frameworks. Sections~\ref{sec:kaons} and~\ref{sec:bmeson} discuss the formalisms for lepton-number-violating contributions to rare kaon and $B$-meson decays, respectively, while Sec.~\ref{sec:ovbb} reviews the formalism for $0\nu\beta\beta$. In Sec.~\ref{sec:eftlimits}, we present our main results for the bottom-up EFT approach under single-operator dominance in two distinct physical limits of the underlying flavor structure.  Section~\ref{sec:uvmodel} illustrates the results for the top-down EFT approach for a concrete UV model example. Finally, we conclude in Sec.~\ref{sec:conclusions}.

\section{Effective Field Theory Framework for Operator Matching and Mixing}
\label{sec:eft}

In the presence of New Physics (NP) at a heavy energy scale above electroweak symmetry breaking, the light SM degrees of freedom can be used to construct a systematic, general effective description that captures any potential effect of heavy NP in a model-independent manner. Such a framework is known as the Standard Model Effective Field Theory (SMEFT), where the SM Lagrangian can be extended by SM gauge group invariant higher-dimensional operators as
\begin{equation}
	\label{eq:wilson}
	\mathcal{L}=\mathcal{L}_{\text{SM}} + \mathcal{L}_\text{eff}\, ,
\end{equation}
where
\begin{equation}
\label{eq:EFTlag}
    \mathcal{L}_\text{eff} = \sum_{i}\sum_a C_a^{(d)}\mathcal{O}_a^{(d)}\, ,
\end{equation}
with $C_a^{(d)}$ being the dimensionful Wilson Coefficient associated with the operator labeled as $\mathcal{O}_a$ of mass dimension $d$. The Wilson coefficients of higher-dimensional operators are suppressed by the NP scale as $\Lambda^{d-4}$ to preserve the correct dimensionality of the Lagrangian. 
On the other hand, at scales below electroweak symmetry breaking, $\Lambda_{EW} \approx v$, one can similarly construct the Low-Energy Effective Field Theory (LEFT), where the heavy SM degrees of freedom like $W$ and $Z$ gauge bosons, the $t$ quark, and the Higgs boson $h$ are also integrated out, giving a suppression of inverse powers of $v$, so the Lagrangian can be described as
\begin{equation*}
    \mathcal{L}=\frac{1}{v^{D-4}}\sum_{i,D\geqslant 5}c_i^{(D)}O_i^{(D)}.
\end{equation*}
where the relevant Wilson coefficients are dimensionless.

From a top-down perspective, to study the effect of a typical UV-complete model containing heavy BSM states (with mass $M\sim \Lambda$) on an observable at the electroweak scale, one can simply use the full UV model itself. However, computations involving the full model quickly become involved as the number of heavy degrees of freedom and the loop order increase, and they must be performed case by case for each model. On the other hand, the EFT approach allows estimating the same observable up to corrections of order $\mathcal{O}\left(\frac{E_O^n}{\Lambda^n}\right)$, where  $E_O$ is the energy scale of the observable with $n$ fixed by the loop order of the EFT calculation (and operator dimension) and can be chosen to meet the desired experimental precision for the observable. The major advantage is that the EFT results can be derived generally up to a given loop order and applied to any number of models. To map a given UV model to SMEFT, the first step is to perform a ``matching'' up to a given order in a loop expansion such that at each loop order the $S$-matrix elements in the UV model and the EFT are the same at the scale $\mu=\Lambda$. Having obtained the Wilson coefficients at the matching scale $\mu=\Lambda$, to connect them with a measurement at the electroweak scale $\mu=m_W$, it is then necessary to evolve them with the scale using the renormalization group equations (RGEs). We will sometimes also refer to such effects as ``mixing effects'', since the RGEs are, physically, nothing but a systematic way to capture operator mixing effects by summing log series to different leading orders. At the leading order, the solution to these RGEs is governed by the anomalous dimension matrix $\gamma_{ij}$, such that the RGEs are of the form
\begin{equation}
\frac{d C_i(\mu)}{d \log \mu} = \sum_j \frac{1}{16\pi^2} \gamma_{ij} C_j, \label{eqn:RGE}
\end{equation}
gives at leading order
\begin{equation}
C_i(m_W) = C_i(\Lambda) - \sum_j \frac{1}{16\pi^2} \gamma_{ij} C_j(\Lambda) \log \left(\frac{\Lambda}{m_W}\right) \, . \label{eqn:RGrelation}
\end{equation}
For an observable at a scale much lower than the electroweak symmetry breaking scale, the SMEFT can be matched onto the LEFT and then evolved down using the corresponding RGEs to the observable scale or to the matching scale of the next EFT, e.g., chiral EFT relevant for studying kaon decays and $0\nu\beta\beta$ decay~\cite{Cirigliano:2017djv}.

lepton-number-violating SMEFT operators that violate lepton number by two units ($\Delta L = 2$) can be written in terms of odd-dimensional lepton-number-violating operators \cite{Kobach:2016ami}
\begin{equation}
	\label{eq:wilsonLNV}
	\mathcal{L}=\mathcal{L}_{\text{SM}} + C^{(5)}\mathcal{O}^{(5)} + \sum_a C_a^{(7)}\mathcal{O}_a^{(7)} + \sum_a C_a^{(9)}\mathcal{O}_a^{(9)} + \dots \, .
\end{equation}
We will particularly focus on $\Delta L=2$ dimension-7 SMEFT operators in this work. Historically, the first complete set of $\Delta L=2$ dimension-7 operators was derived in~\cite{Lehman:2014jma}, although it included some redundant operators. This was later corrected in \cite{Liao:2016hru}, removing these redundant operators. The operator basis adopted in this work is shown explicitly in Table~\ref{tab:SMEFT_Dim_7}, and it follows the convention of \cite{Cirigliano:2017djv, Lehman:2014jma, Liao:2016hru}. On the other hand, Table~\ref{tab:LEFT_matching} lists the lepton-number-violating $\Delta L=2$ operators in the LEFT at dimensions 3, 5, and 6, together with the tree-level matching to them from the SMEFT operators as well as those dimension-7 and 9 operators that directly contribute to \0 induced at SMEFT dimension-7~\cite{Cirigliano:2017djv, Scholer:2023bnn}\footnote{Note that the full LEFT RGEs for dimension-7 and 9 have not been calculated in the literature. We therefore only implement the QCD running of the 1st generation in the dimension-7 and 9 LEFT operators relevant for \0~\cite{Cirigliano:2018yza} while the remaining dimension-6 and lower LEFT operators are evolved under the full QCD + QED RGEs~\cite{Jenkins:2017dyc}.}.

While many existing studies have accounted for LEFT RG evolution, they have largely ignored RG running of SMEFT operators and loop-matching effects. Recently, the anomalous dimensions and the RGEs for the $\Delta L=2$ SMEFT dimension-7 operators have been derived in Refs.~\cite{Liao:2019tep,Zhang:2023kvw,Zhang:2023ndw}. In Ref.~\cite{Graf:2025cfk}, the impact of loop-matching and RG running contributions for the $\Delta L=2$ SMEFT dimension-7 operators was pointed out. In Refs.~\cite{Liao:2025lxg,deVries:2026ujz}, similar results were also reported for RG running contributions. While Ref.~\cite{Graf:2025cfk} mainly focused on $0\nu\beta\beta$ decays, here we are interested in the systematic exploration of the operator matching and mixing effects for the flavored lepton-number-violating meson decays in complement to the $0\nu\beta\beta$ decays.

To quantify the RG running effects of $\Delta L=2$ SMEFT dimension-7 operators from the matching scale ($\Lambda$) to the electroweak scale, we will use the Renormalization Group Equations (RGEs) up to order $\mathcal{O}(\Lambda^{-3})$. The RGEs for the relevant $\Delta L=2$ lepton-number-violating operators follow the simplified structure~\cite{Zhang:2023kvw}
\begin{eqnarray}\label{eq:rge-structure}
	\befune{C^{}_5} &=& \gamma^{(5,5)} C^{}_5 + \hat{\gamma}^{(5,5)} C^{}_5 C^{}_5 C^{}_5+ \gamma^{(5,6)}_i C^{}_5 C^{i}_6 + \gamma^{(5,7)}_i C^{i}_7 \;,
	\nonumber
	\\
	\befune{C^{i}_7} &=& \gamma^{(7,7)}_{ij} C^{j}_7 + \gamma^{(7,5)}_i C^{}_5 C^{}_5 C^{}_5 + \gamma^{(7,6)}_{ij} C^{}_5 C^{j}_6 \;.
\end{eqnarray}
where, $\gamma^{ij}$ stands for the anomalous dimension matrix for WCs of dimension-$i$ operators resulting from dimension-$j$ ones. Equation \ref{eq:rge-structure} is the most general form for the RGEs of lepton-number-violating operators at $\mathcal{O}(\Lambda^{-3})$. As Eq.~\eqref{eq:rge-structure} shows, the RGE for the Wilson Coefficient of the Weinberg operator receives contributions from itself, triple insertions of itself, mixed insertions of dimension-5 and 6, and single insertions of dimension-7 operators. Similarly, one can interpret the RGE for the dimension-7 operators. Since dimension-6 operators do not directly mix with the $\Delta L=2$ operators at one loop, their running is decoupled and can be found, for instance, in Ref.~\cite{Jenkins:2013zja, Jenkins:2013wua, Alonso:2013hga}.

In what follows, we investigate the effects of operator mixing for $\Delta L=2$ SMEFT dimension-7 operators across flavors and their implications for meson-decay constraints and their complementarity with constraints from $0\nu\beta\beta$ decay. To this end, we will use the bottom-up EFT approach with the single-operator dominance limit, where we apply the relevant low-energy constraints bottom-up, assuming that a single SMEFT operator is responsible at a time for generating the low-energy effect.  We will also discuss one top-down example with an explicit UV model to show how to implement such effects in realistic model case studies. We will use different flavor-structure assumptions to derive generally applicable constraints for various physically interesting scenarios. 

To perform the numerical computations of SMEFT RGE-governed mixing contributions for all the observables discussed in this work, we implement the EFT framework, including the relevant one-loop SMEFT and LEFT evolution, using an updated, in-development version of the open-source code $\nu$DoBe~\cite{Scholer:2023bnn}. We use $\nu$DoBe to evolve operators from the high-energy new-physics scale $\Lambda$ to the electroweak scale, match onto the corresponding LEFT interactions, and then evolve the LEFT operators to the low-energy scales relevant for the next matching to another low-energy EFT or simply the energy scale of the processes concerned.
{\renewcommand{\arraystretch}{1.35}
\begin{table}[htbp]
    \centering
    \small
    \begin{tabular}{@{}llll@{}}
        \toprule
        \multicolumn{2}{c}{Class 1: $\psi^2 H^4$}
        &
        \multicolumn{2}{c}{Class 5: $\psi^4 D$}
        \\
        \cmidrule(lr){1-2}\cmidrule(lr){3-4}
        $\mathcal{O}^{(7),\ell_1\ell_2}_{LH}$
        &
        $\eps_{ij}\eps_{mn}(L_i^{T,\ell_1} C L^{\ell_2}_m)H_jH_n(H^\dagger H)$
        &
        $\mathcal{O}^{(7),q_1q_2\ell_1\ell_2}_{LL d u D1}$
        &
        $\eps_{ij}(\dbar^{q_1}\gamma_\mu u^{q_2})
        (L_i^{T,\ell_1} C(D^\mu L^{\ell_2})_j)$
        \\
        \addlinespace

        \multicolumn{2}{c}{Class 2: $\psi^2 H^2 D^2$}
        &
        \multicolumn{2}{c}{Class 6: $\psi^4 H$}
        \\
        \cmidrule(lr){1-2}\cmidrule(lr){3-4}
        $\mathcal{O}^{(7),\ell_1\ell_2}_{LHD1}$
        &
        $\eps_{ij}\eps_{mn}
        (L_i^{T,\ell_1} C(D_\mu L^{\ell_2})_j)H_m(D^\mu H)_n$
        &
        $\mathcal{O}^{(7),\ell_1\ell_2\ell_3\ell_4}_{LL e H}$
        &
        $\eps_{ij}\eps_{mn}
        (\ebar^{\ell_1} L_i^{\ell_2})(L_j^{T,\ell_3} C L_m^{\ell_4})H_n$
        \\
        $\mathcal{O}^{(7),\ell_1\ell_2}_{LHD2}$
        &
        $\eps_{im}\eps_{jn}
        (L_i^{T,\ell_1} C(D_\mu L^{\ell_2})_j)H_m(D^\mu H)_n$
        &
        $\mathcal{O}^{(7),q_1\ell_1q_2\ell_2}_{LLQ d H1}$
        &
        $\eps_{ij}\eps_{mn}
        (\dbar^{q_1} L_i^{\ell_1})(Q_j^{T,q_2} C L_m^{\ell_2})H_n$
        \\
        &&
        $\mathcal{O}^{(7),q_1\ell_1q_2\ell_2}_{LLQ d H2}$
        &
        $\eps_{im}\eps_{jn}
        (\dbar^{q_1} L_i^{\ell_1})(Q_j^{T,q_2} C L_m^{\ell_2})H_n$
        \\
        \addlinespace

        \multicolumn{2}{c}{Class 3: $\psi^2 H^3 D$}
        &
        $\mathcal{O}^{(7),q_1q_2\ell_1\ell_2}_{LL Q uH}$
        &
        $\eps_{ij}(\Qbar^{q_1}_m u^{q_2})(L_m^{T,\ell_1} C L_i^{\ell_2})H_j$
        \\
        \cmidrule(lr){1-2}
        $\mathcal{O}^{(7),\ell_1\ell_2}_{LHDe}$
        &
        $\eps_{ij}\eps_{mn}
        (L_i^{T,\ell_1} C\gamma_\mu e^{\ell_2})H_jH_m(D^\mu H)_n$
        &
        $\mathcal{O}^{(7),\ell_1\ell_2q_1q_2}_{Leu d H}$
        &
        $\eps_{ij}(L_i^{T,\ell_1} C\gamma_\mu e^{\ell_2})
        (\dbar^{q_1}\gamma^\mu u^{q_2})H_j$
        \\
        &&
        &
        \\
        \addlinespace

        \multicolumn{2}{c}{Class 4: $\psi^2 H^2 X$}
        &
        &
        \\
        \cmidrule(lr){1-2}
        $\mathcal{O}^{(7),\ell_1\ell_2}_{LHB}$
        &
        $\eps_{ij}\eps_{mn}g'
        (L_i^{T,\ell_1} C\sigma^{\mu\nu}L_m^{\ell_2})H_jH_nB_{\mu\nu}$
        &
        &
        \\
        $\mathcal{O}^{(7),\ell_1\ell_2}_{LHW}$
        &
        $\eps_{ij}(\eps\tau^I)_{mn}g_2
        (L_i^{T,\ell_1} C\sigma^{\mu\nu}L_m^{\ell_2})H_jH_nW^I_{\mu\nu}$
        &
        &
        \\
        \bottomrule
    \end{tabular}
    \caption{Lepton-number-violating $\Delta L=2$ operators at SMEFT
    dimension-7.}
    \label{tab:SMEFT_Dim_7}
\end{table}
}

{\renewcommand{\arraystretch}{1.25}
\begin{table}[htbp]
    \centering
    \small
    \resizebox{\textwidth}{!}{ %
    \begin{tabular}{@{}lll@{}}
        \toprule
        \textbf{Label}
        &
        \textbf{Operator}
        &
        \textbf{Matching relation}
        \\
        \toprule

        \multicolumn{3}{l}{\textit{Dimension 3}}
        \\
        \midrule
        $O_{\nu}^{(3),pr}$
        &
        $\overline{\nu_L^{C,p}}\nu_L^r$
        &
        $m_\nu^{pr}
        =
        \dfrac{v^2}{2}C_{LH}^{(5),pr}
        +
        \dfrac{v^4}{4}C_{LH}^{(7),pr}$
        \\
        \addlinespace

        \multicolumn{3}{l}{\textit{Dimension 5}}
        \\
        \midrule
        $O_{\nu\gamma}^{(5),pr}$
        &
        $\overline{\nu_L^{C,p}}\sigma^{\mu\nu}\nu_L^rF_{\mu\nu}$
        &
        $c_{\nu\gamma}^{(5),pr}
        =
        \dfrac{ev^3}{4}
        \left(
        2C_{LHB}^{(7),pr}
        +C_{LHW}^{(7),rp}
        -C_{LHW}^{(7),pr}
        \right)$
        \\
        \addlinespace

        \multicolumn{3}{l}{\textit{Dimension 6}}
        \\
        \midrule
        $O_{\nu d,prst}^{SLL(6)}$
        &
        $\big[\overline{\nu_L^{C,p}}\nu_L^r\big]
         \big[\overline{d_R^s}d_L^t\big]$
        &
        $c_{\nu d,prst}^{SLL(6)}
        =
        \dfrac{v^3}{4\sqrt{2}}
        \left(
        -C_{LLQdH1}^{(7),srtp}
        -C_{LLQdH1}^{(7),sprt}
        \right)$
        \\

        $O_{\nu u,prst}^{SLL(6)}$
        &
        $\big[\overline{\nu_L^{C,p}}\nu_L^r\big]
         \big[\overline{u_R^s}u_L^t\big]$
        &
        \\

        $O_{\nu e,prst}^{SLL(6)}$
        &
        $\big[\overline{\nu_L^{C,p}}\nu_L^r\big]
         \big[\overline{e_R^s}e_L^t\big]$
        &
        $c_{\nu e,prst}^{SLL(6)}
        =
        \dfrac{\sqrt{2}v^3}{8}
        \left(
        2C_{LLeH}^{(7),stpr}
        +C_{LLeH}^{(7),sptr}
        +2C_{LLeH}^{(7),strp}
        +C_{LLeH}^{(7),srtp}
        \right)$
        \\

        $O_{\nu d,prst}^{SLR(6)}$
        &
        $\big[\overline{\nu_L^{C,p}}\nu_L^r\big]
         \big[\overline{d_L^s}d_R^t\big]$
        &
        \\

        $O_{\nu u,prst}^{SLR(6)}$
        &
        $\big[\overline{\nu_L^{C,p}}\nu_L^r\big]
         \big[\overline{u_L^s}u_R^t\big]$
        &
        $c_{\nu u,prst}^{SLR(6)}
        =
        \dfrac{\sqrt{2}v^3}{4}
        \left(
        C_{LLQuH}^{(7),stpr}
        +C_{LLQuH}^{(7),strp}
        \right)$
        \\

        $O_{\nu e,prst}^{SLR(6)}$
        &
        $\big[\overline{\nu_L^{C,p}}\nu_L^r\big]
         \big[\overline{e_L^s}e_R^t\big]$
        &
        $c_{\nu e,prst}^{SLR(6)}
        =
        i\dfrac{v^3}{\sqrt{2}}
        \left(
        C_{LHDe}^{(7),pt}\delta^{sr}
        +C_{LHDe}^{(7),rt}\delta^{ps}
        \right)$
        \\

        $O_{\nu d,prst}^{TLL(6)}$
        &
        $\big[\overline{\nu_L^{C,p}}\sigma^{\mu\nu}\nu_L^r\big]
         \big[\overline{d_L^s}\sigma_{\mu\nu}d_L^t\big]$
        &
        $c_{\nu d,prst}^{TLL(6)}
        =
        \dfrac{v^3}{16\sqrt{2}}
        \left(
        C_{LLQdH1}^{(7),sptr}
        -C_{LLQdH1}^{(7),srtp}
        \right)$
        \\

        $O_{\nu u,prst}^{TLL(6)}$
        &
        $\big[\overline{\nu_L^{C,p}}\sigma^{\mu\nu}\nu_L^r\big]
         \big[\overline{u_L^s}\sigma_{\mu\nu}u_L^t\big]$
        &
        \\

        $O_{\nu e,prst}^{TLL(6)}$
        &
        $\big[\overline{\nu_L^{C,p}}\sigma^{\mu\nu}\nu_L^r\big]
         \big[\overline{e_L^s}\sigma_{\mu\nu}e_L^t\big]$
        &
        $c_{\nu e,prst}^{TLL(6)}
        =
        \dfrac{\sqrt{2}v^3}{32}
        \left(
        C_{LLeH}^{(7),sptr}
        -C_{LLeH}^{(7),srtp}
        \right)$
        \\

        $O_{\nu edu,prst}^{SLL(6)}$
        &
        $\big[\overline{\nu_L^{C,p}}e_L^r\big]
         \big[\overline{d_R^s}u_L^t\big]$
        &
        $c_{\nu edu,prst}^{SLL(6)}
        =
        \dfrac{v^3}{2\sqrt{2}}
        \left(
        -C_{LLQdH1}^{(7),srtp}
        +C_{LLQdH2}^{(7),srtp}
        -C_{LLQdH2}^{(7),sptr}
        \right)$
        \\

        $O_{\nu edu,prst}^{SLR(6)}$
        &
        $\big[\overline{\nu_L^{C,p}}e_L^r\big]
         \big[\overline{d_L^s}u_R^t\big]$
        &
        $c_{\nu edu,prst}^{SLR(6)}
        =
        \dfrac{v^3}{\sqrt{2}}C_{LLQuH}^{(7),strp}$
        \\

        $O_{\nu edu,prst}^{VRL(6)}$
        &
        $\big[\overline{\nu_L^{C,p}}\gamma^\mu e_R^r\big]
         \big[\overline{d_L^s}\gamma_\mu u_L^t\big]$
        &
        $c_{\nu edu,prst}^{VRL(6)}
        =
        -i\dfrac{v^3}{\sqrt{2}}
        C_{LHDe}^{(7),pr}\delta^{st}$
        \\

        $O_{\nu edu,prst}^{VRR(6)}$
        &
        $\big[\overline{\nu_L^{C,p}}\gamma^\mu e_R^r\big]
         \big[\overline{d_L^s}\gamma_\mu u_L^t\big]$
        &
        $c_{\nu edu,prst}^{VRR(6)}
        =
        \dfrac{v^3}{\sqrt{2}}
        C_{LeudH}^{(7),prst}$
        \\

        $O_{\nu edu,prst}^{TLL(6)}$
        &
        $\big[\overline{\nu_L^{C,p}}\sigma^{\mu\nu}e_L^r\big]
         \big[\overline{d_R^s}\sigma_{\mu\nu}u_L^t\big]$
        &
        $c_{\nu edu,prst}^{TLL(6)}
        =
        \dfrac{v^3}{8\sqrt{2}}
        \left(
        -C_{LLQdH1}^{(7),srtp}
        +C_{LLQdH2}^{(7),srtp}
        +C_{LLQdH2}^{(7),sptr}
        \right)$
        \\
        \addlinespace

        \multicolumn{3}{l}{\textit{Dimension 7}}
        \\
        \midrule
        $O_{ude\nu}^{VLR(7)}$
        &
        $\big[\ubarL \gamma^\mu d_L\big]\big[\ebarL i\overset{\leftrightarrow}{\partial}_\mu \nu_L^C\big]$
        &
        $c_{ude\nu}^{VLR(7)} = \frac{v^3}{2}V_{ud} \big(2 C_{LHD1}^{*(7),ee}
                                          -     C_{LHD2}^{*(7),ee}
                                          + 8 C_{LHW}^{*(7),ee}\big)$
        \\
        $O_{ude\nu}^{VRR(7)}$
        &
        $\big[\ubarR \gamma^\mu d_R\big]\big[\ebarL i\overset{\leftrightarrow}{\partial}_\mu \nu_L^C\big]$
        &
        $c_{ude\nu}^{VRR(7)} = -iv^3 C_{LeudH}^{*(7), eeud} $
        \\
        \addlinespace
        \multicolumn{3}{l}{\textit{Dimension 9}}
        \\
        \midrule
        $O_{1L}^{(9)}$
        &
        $\big[\ebarL e_L^C\big]\big[\ubarL \gamma^\mu d_L\big]\big[\ubarL \gamma_\mu d_L\big]$
        & 
        $c_{1L}^{(9)} = v^3 V_{ud}^2 \big(2C_{LHD1}^{*(7),ee} + 8C_{LHW}^{*(7),ee}\big)$
        \\
        $O_{4L}^{(9)}$
        &
        $\big[\ebarL e_L^C\big]\big[\ubarL \gamma^\mu d_L\big]\big[\ubarR \gamma_\mu d_R\big]$
        & 
        $c_{4L}^{(9)} = -2i v V_{ud}C_{LLduD1}^{*(7),eedu}$
        
        \\
        \bottomrule
    \end{tabular}%
    }
    \caption{$\Delta L=2$ LEFT operators and their matching onto
    dimension-five and dimension-seven SMEFT operators. The matching
    relations are given in the basis of diagonal down-type Yukawa interactions; the appropriate CKM factors must be included when transforming to the quark mass basis. Terms proportional to $m_f/v$ arising from equations of motion are omitted here, but can be relevant for neutrinoless double beta decay~\cite{Scholer:2023bnn}.}
    \label{tab:LEFT_matching}
\end{table}
}

\section{Lepton-Number-Violating Kaon Decays}
\label{sec:kaons}

In the SM, the kaon decays into a pion and a neutrino-antineutrino pair are mediated by flavor-changing neutral currents suppressed by the GIM mechanism, with the branching ratio predictions for charged and neutral meson modes~\cite{Buras:2006gb,Brod:2010hi,Buras:2015qea,Buras:2026hbe} $\mathrm{Br}(K^+ \to \pi^+ \nu \overline{\nu})_{\mathrm{SM}} = (8.65 \pm 0.42) \times 10^{-11}$ and $ \mathrm{Br}(K_L \to \pi^0 \nu \overline{\nu})_{\mathrm{SM}} = (3.05 \pm 0.17) \times 10^{-11}$, respectively. While in the SM the final-state neutrinos are necessarily a lepton-number-conserving pair of neutrino-antineutrino, the experiments searching for these modes, namely NA62 and KOTO, cannot distinguish between the neutrinos and antineutrinos in the final states. These experiments have reported the latest result for the $K^+ \to \pi^+ \nu \overline{\nu}$ branching ratio $  \mathrm{Br}(K^+ \to \pi^+ \nu \hat{\nu})_{\mathrm{exp}} = 13.0^{+3.3}_{-3.0} \times 10^{-11}$~\cite{NA62:2024pjp} and $K_L \to \pi^0 \nu \hat{\nu}$ branching ratio $  \mathrm{Br}(K_L \to \pi^0 \nu \hat{\nu})_{\mathrm{exp}} < 2.2 \times 10^{-9}$, respectively. We have used $\hat{\nu}$ in the previous line to denote the fact that these measurements can be interpreted as a combination of the lepton-number-violating and -conserving modes. Therefore, the lepton-number-violating new physics can contribute on top of the SM-allowed lepton-number-conserving mode, with a total branching fraction consistent with these current best measurements. This makes observables like $\mathrm{Br}(K^+ \to \pi^+ \nu \hat{\nu})$ and $\mathrm{Br}(K_L \to \pi^0 \nu \hat{\nu})$ excellent probes for constraining lepton-number-violating new physics. Interestingly, these modes are sensitive to complementary quark flavors for lepton-number-violating SMEFT operators compared with the first-generation sensitivity at tree level in $0\nu\beta\beta$ experiments.

Recently, there has been considerable interest in lepton-number-violating kaon decay modes as a probe of lepton-number-violating SMEFT operators; see, e.g., ~\cite{Li:2019fhz, Deppisch:2020oyx, Fridell:2023rtr, Gonzalez:2023him, Buras:2024ewl, Endo:2026qof, deVries:2026ujz}. Among these studies, only some very recent ones~\cite{Endo:2026qof, deVries:2026ujz} touch upon the SMEFT operator flavor mixing effects to some extent in the context of the lepton-number-violating kaon decays following the study in Ref.~\cite{Graf:2025cfk}, which pointed out the striking impact of such effects for $0\nu\beta\beta$ decay. In this work, we will present a dedicated and detailed exploration of the SMEFT operator mixing effects for various flavor ans\"{a}tze in the model-independent bottom-up EFT approach, and in the top-down EFT approach using an explicit model example. 

To take into account consistently the SMEFT operator mixing effects, we will use the full system of RG evolution equations for lepton-number-violating dimension-7 and dimension-5 SMEFT operators, c.f. Eq.~\eqref{eq:rge-structure}, supplemented by the set of equations for dimension-6 SMEFT and evolution of relevant couplings. Following this, we will match the relevant lepton-number-violating dimension-7 SMEFT operators into the relevant LEFT operators. The lepton-number-violating dimension-6 LEFT operators that contribute to the short-range kaon decay modes of our interest are 
\begin{equation}\label{Eq:LEFT-d6-LNV}
\begin{aligned}
    O^{SLL(6)}_{\nu u,prst}&= (\overline{\nu^c_{L,p}}\nu_{L,r})(\overline{u_{R,s}}u_{L,t}) + \text{h.c.}
    ,\quad\qquad\;\;
    O^{SLR(6)}_{\nu u,prst} = (\overline{\nu^c_{L,p}}\nu_{L,r})(\overline{u_{L,s}}u_{R,t}) + \text{h.c.}, 
    \\
    O^{SLL(6)}_{\nu d,prst}&= (\overline{\nu^c_{L,p}}\nu_{L,r})(\overline{d_{R,s}}d_{L,t}) + \text{h.c.}
    ,\quad\qquad\,\;\;
    O^{SLR(6)}_{\nu d,prst} = (\overline{\nu^c_{L,p}}\nu_{L,r})(\overline{d_{L,s}}d_{R,t}) + \text{h.c.},
    \\
    O^{TLL(6)}_{\nu u,prst}&= (\overline{\nu^c_{L,p}}\sigma^{\mu\nu}\nu_{L,r})(\overline{u_{R,s}}\sigma_{\mu\nu}u_{L,t}) + \text{h.c.}
    ,\;\;
    O^{TLL(6)}_{\nu d,prst} = (\overline{\nu^c_{L,p}}\sigma^{\mu\nu}\nu_{L,r})(\overline{d_{R,s}}\sigma_{\mu\nu}d_{L,t}) + \text{h.c.}.
\end{aligned}
\end{equation}
There are also lepton number conserving operators at dimension-6 in LEFT
\begin{equation}\label{Eq:LEFT-d6-LNC}
\begin{aligned}
    O^{VLL(6)}_{\nu u,prst}&= (\overline{\nu_{L,p}} \Gamma^\mu \nu_{L,r})(\overline{u_{L,s}} \Gamma_\mu u_{L,t}) + \text{h.c.}
    ,\,\;\;
    O^{VLR(6)}_{\nu u,prst} = (\overline{\nu_{L,p}} \Gamma^\mu \nu_{L,r})(\overline{u_{R,s}} \Gamma_\mu u_{R,t}) + \text{h.c.}, 
    \\
    O^{VLL(6)}_{\nu d,prst}&= (\overline{\nu_{L,p}} \Gamma^\mu \nu_{L,r})(\overline{d_{L,s}} \Gamma_\mu d_{L,t}) + \text{h.c.}
    ,\,\;\;
    O^{VLR(6)}_{\nu d,prst} = (\overline{\nu_{L,p}} \Gamma^\mu \nu_{L,r})(\overline{d_{R,s}} \Gamma_\mu d_{R,t}) + \text{h.c.},
\end{aligned}
\end{equation}
which we will keep for completeness in our general formulae. Following the matching to LEFT operators, we evolve the LEFT operator down to the chiral symmetry breaking scale using the relevant RGEs, c.f.~\cite{Jenkins:2013wua, Jenkins:2013zja, Alonso:2013hga}, where the QCD running provides the dominant effects. 
Then, for the calculation of the kaon decay rate, we match these LEFT operators to chiral perturbation theory ($\chi$PT) at the chiral symmetry breaking scale $\Lambda_\chi\sim2\,\mathrm{GeV}$ to take into account non-perturbative QCD effects. Following the prescription of~\cite{Li:2019fhz}, the chiral matching yields the relevant effective Lagrangians for $K^+ \to \pi^+ \nu \hat{\nu}$ and $K_L \to \pi^0 \nu \hat{\nu}$ modes. At leading order, the effective Lagrangians only involve the lepton-number-conserving vector and lepton-number-violating scalar LEFT Wilson coefficients as
\begin{eqnarray}
\mathcal{L}_{K^+\to \pi^+\nu\hat{\nu}}&=&
-\frac{B}{2}\Big[
\left(c_{\nu d, \alpha\beta sd}^{SLL}+c_{\nu d, \alpha\beta sd}^{SLR}\right)(\overline{\nu^c_\alpha}\nu_\beta)
+\left(c_{\nu d, \alpha\beta ds}^{SLL*}+c_{\nu d, \alpha\beta ds}^{SLR*}\right)(\overline{\nu_\alpha}\nu_\beta^c)\Big]K^+\pi^-
\nonumber\\
&&
+\frac{i}{2}\left(c_{\nu d,\alpha\beta sd}^{VLL}+c_{\nu d,\alpha\beta sd}^{VLR}\right)(\overline{\nu_\alpha}\gamma^\mu\nu_\beta)(K^+ \partial_\mu \pi^--\partial_\mu K^+ \pi^-)\; ,
\end{eqnarray}
and
\begin{eqnarray}
\mathcal{L}_{K_L\to \pi^0\nu\hat{\nu}}&=&
\frac{B}{4}\Big[
\left(c_{\nu d, \alpha\beta sd}^{SLL}+c_{\nu d, \alpha\beta sd}^{SLR}
+c_{\nu d,\alpha\beta ds}^{SLL}+c_{\nu d, \alpha\beta ds}^{SLR}\right)(\overline{\nu^c_\alpha}\nu_\beta) \nonumber \\
&&+\left(c_{\nu d, \alpha\beta ds}^{SLL*}+c_{\nu d,\alpha\beta ds}^{SLR*}
+c_{\nu d,\alpha\beta sd}^{SLL*}+c_{\nu d,\alpha\beta sd}^{SLR*}\right)(\overline{\nu_\alpha}\nu_\beta^c)\Big]K_L\pi^0 \\
&&
-\frac{i}{4}\left(c_{\nu d,\alpha\beta sd}^{VLL}+c_{\nu d,\alpha\beta sd}^{VLR}-c_{\nu d,\alpha\beta ds}^{VLL}-c_{\nu d,\alpha\beta ds}^{VLR}\right)(\overline{\nu_\alpha}\gamma^\mu\nu_\beta)(K_L \partial_\mu \pi^0-\partial_\mu K_L \pi^0) \; , \nonumber
\end{eqnarray}
where $B$ is a dimensionful parameter relating quark masses to the pion mass, with $B\equiv m_pi^2/(m_u+m_d)\simeq 2.8$ GeV~\cite{Cirigliano:2017djv}. The final branching fractions can be expressed in the form~\cite{Li:2019fhz}
\begin{align}
    \mathrm{Br}\left(K_L\rightarrow\pi^0 \nu\nu\right) &= 40.4\,{G_F}^{-2} \sum_{\alpha\leq\beta}\bigg(1-\frac{1}{2}\delta_{\alpha\beta}\bigg)\left|c_{\nu d,\alpha\beta sd}^{SLL(6)} + c_{\nu d,\alpha\beta sd}^{SLR(6)} + s\leftrightarrow d\right|^2\nonumber\\
    &+0.247\,{G_F}^{-2} \sum_{\alpha\leq\beta}\left|c_{\nu d,\alpha\beta sd}^{VLL(6)} + c_{\nu d,\alpha\beta sd}^{VLR(6)} - s\leftrightarrow d\right|^2 \; ,
    \\
    \mathrm{Br}\left(K^+\rightarrow\pi^+ \nu\nu\right) &= 17.9\,{G_F}^{-2}\sum_{\alpha\leq\beta}\bigg(1-\frac{1}{2}\delta_{\alpha\beta}\bigg)\left(\left|c_{\nu d,\alpha\beta sd}^{SLL(6)} + c_{\nu d,\alpha\beta sd}^{SLR(6)}\right|^2 + s\leftrightarrow d\right)\nonumber\\
   & +0.22\,{G_F}^{-2}\sum_{\alpha\leq\beta}\left|c_{\nu d,\alpha\beta sd}^{VLL(6)} + c_{\nu d,\alpha\beta sd}^{VLR(6)}\right|^2 \; ,
\end{align}
with the prefactors normalized to the inverse square of the Fermi constant $G_F$ arising from the kinematics of the three-body decays. Since we are mainly interested in lepton-number-violating new physics, the dimension-6 LEFT scalar Wilson coefficients are of primary interest. Note that the lepton-number-violating dimension-6 LEFT tensor operators match to the chiral Lagrangian at next-to-leading order $\mathcal{O}(p^4)$ and are suppressed by an additional factor of $p^2/\Lambda_\chi^2$, leading only to sub-leading contributions. There are also contributions to $K\rightarrow \pi \nu\nu$ transitions from a lepton number conserving tensor dimension-7 LEFT operator $O^{TLL(7)}_{\nu d,prst}= (\overline{\nu_{L,p}} \gamma^{\left[\mu \right.} i \overleftrightarrow{\partial}^{\left.\nu\right]}\nu_{L,r})(\overline{d_{L,s}}\sigma_{\mu\nu}d_{R,t})$ and two lepton-number-violating vector dimension-7 LEFT operators $c_{\nu d,\alpha\beta sd}^{VLX(7)}= (\overline{\nu^c_{L,p}}  i \overleftrightarrow{\partial^\mu}\nu_{L,r})(\overline{d_{L,s}}\gamma_\mu d_{L,t})$, with $X=L,R$. However, they also face additional suppressions and are sub-leading compared to the dimension-6 scalar operators. 

Finally, $K\rightarrow(\pi)\nu\hat{\nu}$ can also receive contributions from lepton-number-violating dimension-7 SMEFT operators. Such contributions involve the dimension-5 LEFT operator corresponding to the neutrino magnetic moment given by  ${O}^{5\gamma}_{\nu\nu F}=(\overline{\nu^C}i\sigma_{\mu\nu}\nu)F^{\mu\nu}$ such that $c^{5\gamma}_{\nu\nu F} {O}^{5\gamma}_{\nu\nu F}\equiv \mu \,e \,{O}^{5\gamma}_{\nu\nu F}$, with $F_{\mu\nu}$ denoting the electromagnetic field strength tensor, $\mu$ denoting the magnetic dipole moment of neutrino and $e$ denotes the electric charge.
Here $F_{\mu\nu}$ is the electromagnetic field strength tensor.  The Wilson coefficient for the lepton-number-violating dimension-5 LEFT operator matches the lepton-number-violating dimension-7 SMEFT operators as~\cite{Cirigliano:2017djv}
\begin{eqnarray}\label{numm2}
c^{5\gamma}_{\nu\nu F}/e \equiv\mu_{i j} =  \frac{1}{2 v} \, \left(v^3  C^{ij}_{LHB} -v^3 \frac{ C^{ij}_{LHW}- C^{ji}_{LHW}}{2}\right)\,.
\end{eqnarray}
Note that $\mu$ and $\mathcal C_{LHB}$ are anti-symmetric in flavor indices. However, $c_{\nu\nu F}$ is subject to stringent constraints from solar and reactor neutrino-electron scattering  experiments~\cite{Bell:2006wi,Giunti:2014ixa,Canas:2015yoa}), astrophysical limits from globular clusters~\cite{Raffelt:1998xu}, Coherent Elastic Neutrino Nucleus Scattering (CE$\nu$NS)~\cite{Bolton:2021pey,Miranda:2021kre}, etc., allowing for rather small values for the relevant Wilson coefficient $|c_{\nu\nu F}|\leq 4\times10^{-9}~\text{GeV}^{-1}$~\cite{Canas:2015yoa}. On the other hand, the relevant quark part of the long-distance $s\to d\gamma$ transition operator $O_{s\to d\gamma} \equiv \bar s\sigma_{\mu\nu}P_{L/R}dF^{\mu\nu}$ is also very small in the SM, with $c_{sdF}\sim 10^{-9}~\text{GeV}^{-1}$~\cite{Tandean:1999mg}. Consequently, in the absence of additional ingredients that could enhance these and avoid the abovementioned constraints, the potential long-distance contributions are also subleading and will not be pursued further in this work.

\section{Lepton-Number-Violating $B$-meson Decays}
\label{sec:bmeson}
Like the lepton-number-violating kaon decays, the $B$ meson decay into a kaon and neutrinos involving the $b\to s \nu \hat{\nu}$ transition provides yet another interesting probe for lepton-number-violating new physics. These modes are also mediated by flavor-changing neutral currents and are GIM-suppressed within the SM. The $b\to s \nu \hat{\nu}$ transitions cleanly factorize into hadronic and leptonic parts, giving generally excellent theoretical control over the predictions. On the other hand, SM predictions depend on $| V _ {cb} |$, which shows tensions between inclusive and exclusive determinations~\cite{FlavourLatticeAveragingGroupFLAG:2021npn, Finauri:2023kte}. While the experimental limits from $B$ meson decays are known to be weaker than those from kaon decays, the clean angular distributions provide better potential for distinguishing various underlying NP cases. The current best experimental bounds on $b\to s \nu \hat{\nu}$ modes are due to Belle and its successor Belle-II
\begin{eqnarray}
\mathrm{Br}\left(B^+\,\rightarrow\,K^+\nu \hat{\nu}\right) &=& \left(2.3 \pm 0.5^{+0.5}_{-0.4}\right)\times10^{-5}\; ~\text{[Belle-II]}\text{~\cite{Belle-II:2023esi}}
        \\
\mathrm{Br}\left(B^0\,\rightarrow\,K^0\nu \hat{\nu}\right)  &<& 2.6\times10^{-5}\; ~\text{[Belle 90\% C.L.]}\text{~\cite{Belle:2017oht}}
        \\
\mathrm{Br}\left(B^{+}\,\rightarrow\,K^{*+}\nu\hat{\nu}\right) &<& 4.0\times10^{-5}\; ~\text{[Belle 90\% C.L.]}\text{~\cite{Belle:2013tnz}}
        \\
\mathrm{Br}\left(B^0\,\rightarrow\,K^{*0}\nu\hat{\nu}\right) &<& 1.8\times10^{-5}\; ~\text{[Belle 90\% C.L.]}\text{~\cite{Belle:2017oht}}
\end{eqnarray}
with the corresponding SM predictions~\cite{Bharucha:2015bzk, Gubernari:2018wyi, Felkl:2021uxi}
\begin{eqnarray}
\mathrm{Br}^{\text{SM}}\left(B^+\,\rightarrow\,K^+\nu\nu\right) &=& 
        (4.4 \pm 0.7) \times 10^{-6}
        \\
\mathrm{Br}^{\text{SM}}\left(B^0\,\rightarrow\,K^0\nu\nu\right) &=&(4.1 \pm 0.5)\times10^{-6}
        \\
\mathrm{Br}^{\text{SM}}\left(B^{+}\,\rightarrow\,K^{*+}\nu\nu\right) &=&
        (12.4 \pm 1.2) \times 10^{-6}
        \\
\mathrm{Br}^{\text{SM}}\left(B^0\,\rightarrow\,K^{*0}\nu\nu\right) &=&
        (11.6 \pm 1.1) \times 10^{-6}
\end{eqnarray}
As in the kaon decay case, we start with the lepton-number-violating SMEFT operators and consistently include SMEFT operator mixing effects before matching the SMEFT operators to the lepton-number-violating LEFT operator at the electroweak symmetry breaking scale. The relevant LEFT operators in this case are the ones given by Eqs.~\eqref{Eq:LEFT-d6-LNV} and~\eqref{Eq:LEFT-d6-LNC}, with the relevant down quark flavors updated to reflect the $b\rightarrow s$ transitions. Since we are mainly interested in lepton-number-violating contributions and to simplify the analytic formulae in what follows, we will drop the lepton-number-conserving vector contributions. We will follow the prescription of~\cite{Felkl:2021uxi}, which employs the helicity formalism~\cite{Jacob:1959at, Gratrex:2015hna} and includes appropriate corrections due to finite width effects~\cite{Descotes-Genon:2019bud, Das:2017ebx}. The relevant differential decay rate distributions due to lepton-number-violating LEFT interactions in the massless neutrino limit can then be expressed as
\begin{align}
    \frac{\mathrm{d}\Gamma\left(B\rightarrow K\nu_\alpha\nu_\beta\right)}{\mathrm{d}q^2} &= \frac{\sqrt{\lambda_{BK}}q^2}{(4\pi)^3 m_B^3(1+\delta_{\alpha\beta})v^4}\nonumber
    \\
    &\times\Bigg[\frac{(m_B^2-m_K^2)^2}{8(m_b-m_s)^2}|f_0|^2\left(\left|c^{SLL(6)}_{\nu d,\alpha\beta sb} + c^{SLR(6)}_{\nu d,\alpha\beta sb} \right|^2 + s\leftrightarrow b\right) \nonumber
    \\
    &\quad\;\;+\frac{2\lambda_{BK}}{3(m_B+m_K)^2}|f_T|^2\left(\left|c_{\nu d,\alpha\beta sb}^{TLL(6)}\right|^2 + s\leftrightarrow b\right) + \alpha\leftrightarrow\beta\Bigg] ,
    \\
    \frac{\mathrm{d}\Gamma\left(B\rightarrow K^*\nu_\alpha\nu_\beta\right)}{\mathrm{d}q^2} &= \frac{\sqrt{\lambda_{BK^*}}q^2}{(4\pi)^3 m_B^3(1+\delta_{\alpha\beta})v^4}\nonumber
    \\
    &\times\Bigg[\frac{\lambda_{BK^*}}{8(m_b+m_s)^2}|A_0|^2\left(\left|c^{SLR(6)}_{\nu d,\alpha\beta sb} - c^{SLL(6)}_{\nu d,\alpha\beta sb}\right|^2 + s \leftrightarrow b\right) \nonumber
    \\
    &\quad\;\;+\left(\frac{32m_B^2m_{K^*}^2|T_{23}|^2}{3(m_B+m_{K^*})^2} + \frac{4\lambda_{BK^*}|T_1|^2 + 4(m_B^2 - m_{K^*}^2)^2|T_{2}|^2}{3q^2}\right)\nonumber
    \\
    &\quad\;\;\times\left(\left|c^{TLL(6)}_{\nu d,\alpha\beta sb}\right|^2+s\leftrightarrow b\right) +\alpha\leftrightarrow\beta\Bigg],
    \\
    \frac{\mathrm{d}\Gamma\left(B\rightarrow X_s\nu_\alpha\nu_\beta\right)}{\mathrm{d}q^2} &= \frac{\sqrt{\lambda(m_b^2,m_s^2,q^2)}}{768\pi^3 m_b^3(1+\delta_{\alpha\beta})v^4} \nonumber
    \\
    &\times\Bigg[2\bigg(3\frac{q^2}{m_b^2}(m_b^2 + m_s^2 - q^2)\left[\left|c^{SLL(6)}_{\nu d,\alpha\beta sb}\right|^2 + \left|c^{SLR(6)}_{\nu d,\alpha\beta sb}\right|^2 + s \leftrightarrow b\right]\nonumber
    \\
    &\quad\;\;+12q^2\frac{m_s}{m_b}\mathrm{Re}\left[c^{SLL(6)}_{\nu d,\alpha\beta sb}c^{SLR(6)*}_{\nu d,\alpha\beta sb} + s\leftrightarrow b\right]\bigg) \nonumber
    \\
    &\quad\;\;+32\left(3\frac{q^2}{m_b^2}(m_b^2+m_s^2-q^2) + \frac{2}{m_b^2}\lambda(m_b^2,m_s^2,q^2)\right)\nonumber
    \\
    &\quad\;\;\times\left(\left|c^{TLL(6)}_{\nu d,\alpha\beta sb}\right|^2 + s \leftrightarrow b\right)\Bigg],
\end{align}
with the Kaellen function
\begin{align}
    \lambda_{BK} = \lambda(m_B^2,m_K^2,q^2),\qquad \lambda(x,y,z)=x^2 + y^2 + z^2 - 2(xy + yz + xz),
\end{align}
and the relevant form factors given by~\cite{Bharucha:2015bzk, Gubernari:2018wyi}
\begin{align}
    f_0(0) &=0.25, & f_T(0) &=0.27, & A_0(0) &=0.356,\nonumber\\
    T_1(0) &= 0.282, & T_2(0) &=T_1(0), & T_{23}(0) &=0.668 \, .
\end{align}
In general, there is no interference between the vector, scalar, and tensor contributions in the massless neutrino case. We also note that the longitudinal polarization fraction can provide excellent complementarity to the $B\rightarrow K^*$ branching fraction, with a potentially valuable handle on distinguishing different types of operators; however, a detailed discussion of this aspect is beyond the scope of the current work and can be found, for instance, in~\cite{Felkl:2021uxi}.
We can calculate the branching ratios $\mathrm{Br}(B\rightarrow X
\nu\nu)$ by numerically performing the integral over $q^2$
\begin{align}
    \mathrm{Br}(B\rightarrow X\nu\nu) = \sum_{\alpha\leq \beta}\int d(q^2) \frac{\mathrm{d}\Gamma\left(B\rightarrow X\nu_\alpha\nu_\beta\right)}{\mathrm{d}q^2}.
\end{align}
%

\section{Neutrinoless Double Beta Decay}
\label{sec:ovbb}

Neutrinoless double beta decay~\cite{Cirigliano:2022oqy, Agostini:2022zub, GERDA:2020xhi, Augier:2022znx, CUORE:2024ikf, KamLAND-Zen:2024eml, LEGEND:2021bnm, nEXO:2021ujk, CUPID:2022wpt, SNO:2021xpa, NuDoubt:2024jax},
\begin{equation}
    0\nu\beta^\pm\beta^\pm:\qquad(A,Z)\to (A,Z\mp2)+2e^\pm ,
\end{equation}
is a particularly sensitive probe of $\Delta L=2$ interactions involving first-generation fermions. Within the EFT framework, contributions to $0\nu\beta\beta$ can arise from operators of different dimensions and through qualitatively different mechanisms~\cite{Deppisch:2012nb, Graf:2022lhj, Graf:2023dzf, Graf:2026rie}, including light-neutrino exchange, long-range semileptonic interactions, and short-range hadronic operators. A systematic connection between lepton-number-violating interactions above the electroweak scale and the nuclear decay rate has been developed using SMEFT, LEFT, and $\chi$EFT~\cite{Cirigliano:2017djv,Cirigliano:2018yza}.

Schematically, the inverse half-life can be written as
\begin{equation}
    \left[T^{0\nu}_{1/2}\right]^{-1}
    =
    g_A^4\sum_k G_{0k}\,
    {\cal F}_k\!\left(\{{\cal A}_i\}\right),
    \label{eq:0nbb-master}
\end{equation}
where $G_{0k}$ denote the phase-space factors and the functions ${\cal F}_k$ contain the different subamplitudes ${\cal A}_i$ and their interference terms. The latter depend on the low-energy Wilson coefficients, hadronic low-energy constants (LECs)\footnote{A large set of the short-range contact LECs is currently unknown and only NDA estimates exist. In order to keep our analysis on the conservative side, we follow $\nu$DoBe's default setting~\cite{Scholer:2023bnn} which turns all off unknown LEC, except for $g_{6,7}^{NN}, g_V^{\pi N}, \tilde{g}_V^{\pi N}$ which are set to their NDA value of $1$ in order to keep the contributions of short-range vector operators in the LEFT framework proportional to these unknown LECs. Because no SMEFT dim-7 operator matches onto any relevant dim-9 short-range vector operator in LEFT, this effectively turns all relevant unknown LECs for our analysis to 0.}, and nuclear matrix elements (NMEs). The complete expressions relevant for lepton-number-violating SMEFT operators up to dimension nine were derived in Ref.~\cite{Cirigliano:2018yza}.

Below the electroweak scale, the LEFT coefficients are evolved to the hadronic scale, $\mu\simeq 2~\mathrm{GeV}$, where they are matched onto $\chi$EFT. The resulting lepton-number-violating pion--pion, pion--nucleon, and nucleon--nucleon interactions, together with the long-range neutrino-exchange potentials, determine the nuclear transition operators~\cite{Cirigliano:2017djv,Cirigliano:2018yza}. Their nuclear matrix elements have been calculated using a variety of many-body approaches; representative results used in $\nu$DoBe are given, for example, in Refs.~\cite{Hyvarinen:2015bda,Menendez:2017fdf,Deppisch:2020ztt}.

The strongest present limit for the isotope considered in our analysis is provided by KamLAND-Zen,
\begin{equation}
    T^{0\nu}_{1/2}\left(^{136}\mathrm{Xe}\right)
    >
    3.8\times10^{26}\ {\rm yr}
    \qquad (90\%~{\rm C.L.}),
    \label{eq:kamland-limit}
\end{equation}
obtained from the complete KamLAND-Zen dataset~\cite{KamLAND-Zen:2024eml}. We use this bound to constrain the dimension-seven SMEFT coefficients at $\Lambda_{\rm NP}$, consistently including the operator mixing between $\Lambda_{\rm NP}$ and the electroweak scale.

As we will see in Sec.~\ref{sec:eftlimits}, this running has an important impact on the comparison with rare meson decays. In particular, Wilson coefficients containing strange, charm, bottom, or top quarks can radiatively generate first-generation lepton-number-violating interactions. Consequently, $0\nu\beta\beta$ can constrain flavor structures that do not induce the decay at tree level, and in several cases these loop-induced limits are stronger than the direct constraints from $K\to\pi\nu\nu$ or $B\to K^{(*)}\nu\nu$.

{\renewcommand{\arraystretch}{1.2}
\begin{table}[htbp]
    \centering
    \small
    \begin{tabular}{@{}lll@{}}
        \toprule
        \textbf{Observable}
        &
        \textbf{Experimental limit}
        &
        \textbf{Standard Model expectation}
        \\
        \toprule

        \multicolumn{3}{l}{$\mathit{0\nu\beta\beta}$ \textit{Decay}}
        \\
        \midrule
        $T_{1/2}^{0\nu}
        \left(^{136}\mathrm{Xe}\to{}^{136}\mathrm{Ba}+2e^-\right)$
        &
        $>3.8\times10^{26}\,\mathrm{yr}$~
        \cite{KamLAND-Zen:2024eml}
        &
        --- 
        \\
        \addlinespace

        \multicolumn{3}{l}{
        \textit{Kaon decays:} $\mathit{K\to\pi\nu\nu}$}
        \\
        \midrule
        $\mathrm{Br}(K_L\to\pi^0\nu\nu)$
        &
        $2.1\times10^{-9}$~\cite{Kitahara:2019lws}
        &
        $(3.4\pm0.6)\times10^{-11}$~
        \cite{Buras:2006gb,Brod:2010hi,Buras:2015qea}
        \\
        $\mathrm{Br}(K^+\to\pi^+\nu\nu)$
        &
        $<1.7\times10^{-10}$~\cite{NA62:2020fhy}
        &
        $(8.4\pm1.0)\times10^{-11}$~
        \cite{Buras:2006gb,Brod:2010hi,Buras:2015qea}
        \\
        \addlinespace

        \multicolumn{3}{l}{
        \textit{$B$-meson decays:} $\mathit{B\to K^{(*)}\nu\nu}$}
        \\
        \midrule
        $\mathrm{Br}(B^0\to K^0\nu\nu)$
        &
        $<2.6\times10^{-5}$~\cite{Belle:2017oht}
        &
        $(4.1\pm0.5)\times10^{-6}$~
        \cite{Bharucha:2015bzk,Gubernari:2018wyi,Felkl:2021uxi}
        \\
        $\mathrm{Br}(B^0\to K^{*0}\nu\nu)$
        &
        $<1.8\times10^{-5}$~\cite{Belle:2017oht}
        &
        $(11.6\pm1.1)\times10^{-6}$~
        \cite{Bharucha:2015bzk,Gubernari:2018wyi,Felkl:2021uxi}
        \\
        $\mathrm{Br}(B^+\to K^+\nu\nu)$
        &
        $=(2.3\pm0.5^{+0.5}_{-0.4})\times10^{-5}$~\cite{Belle-II:2023esi}
        &
        $(4.4\pm0.7)\times10^{-6}$~
        \cite{Bharucha:2015bzk,Gubernari:2018wyi,Felkl:2021uxi}
        \\
        $\mathrm{Br}(B^+\to K^{*+}\nu\nu)$
        &
        $<4.0\times10^{-5}$~\cite{Belle:2013tnz}
        &
        $(12.4\pm1.2)\times10^{-6}$~
        \cite{Bharucha:2015bzk,Gubernari:2018wyi,Felkl:2021uxi}
        \\
        \bottomrule
    \end{tabular}
    \caption{Summary table of experimental constraints and Standard Model predictions for the observables considered in this work.}
    \label{tab:observables}
\end{table}
}

\section{Results in bottom-up EFT}
\label{sec:eftlimits}
To explore the impact of mixing of the lepton-number-violating SMEFT operators above the electroweak symmetry breaking scale, this section focuses on the bottom-up EFT approach under the single operator dominance scenario. The single operator dominance assumes that only a given lepton-number-violating dimension-7 SMEFT operator (with a specified flavor structure) is generated after integrating out the heavy new physics at the scale $\Lambda$. We then study its evolution with energy scale, accounting for the mixing with all relevant operators up to $\mathcal{O}(\Lambda^3)$ down to the electroweak symmetry breaking scale. At the electroweak symmetry breaking scale, we match onto LEFT and evolve it further down (if necessary through a subsequent chain of applicable EFTs) to the relevant energy scale of the concerned observable, as described in the previous sections. For the convenience of the readers, we summarize all relevant experimental constraints on \0, $B\rightarrow K\nu\nu$ and $K\rightarrow\pi\nu\nu$ in Table~\ref{tab:observables}.

We will explore two physically interesting limits in the flavor structure under the single-operator-dominance scenario. One flavor scenario, FS-I, is the lepton flavor universal (LFU) limit, a commonly observed feature in many observables in many precision experimental measurements so far. For simplicity, we also further impose a flavor-diagonal structure in the lepton flavor sector in this case. For this choice, the Wilson coefficients for a given lepton-number-violating dimension-7 SMEFT operator are diagonal, with each entry equal in lepton-flavor space while the active quark flavors are free to choose. The second flavor scenario, FS-II, is the limit where all possible flavor combinations are allowed for both quarks and leptons, and are completely independent of each other. That is, in FS-II we assume only a single SMEFT operator with fixed flavor indices to be non-zero, with flavor symmetries enforced subsequently; see, e.g., the discussion about the flavor symmetry of the operators in~\cite{Liao:2019tep, Zhang:2023kvw}. In what follows, we discuss our findings for these two flavor scenarios above for the 12 lepton-number-violating dimension-7 SMEFT operators listed in Table~\ref{tab:SMEFT_Dim_7}. Since each Wilson coefficient allows several quark- and lepton-flavor combinations, we use a color-coded matrix-style plot with blocks corresponding to different flavor combinations to organize and maximize information. We include only new-physics sensitivity scale results greater than 0.5 TeV ($\Lambda\geq 0.5\,\mathrm{TeV}$) to ensure the SMEFT framework remains valid. 

\begin{figure}[t]
    \centering
    \includegraphics[width=0.99\linewidth]{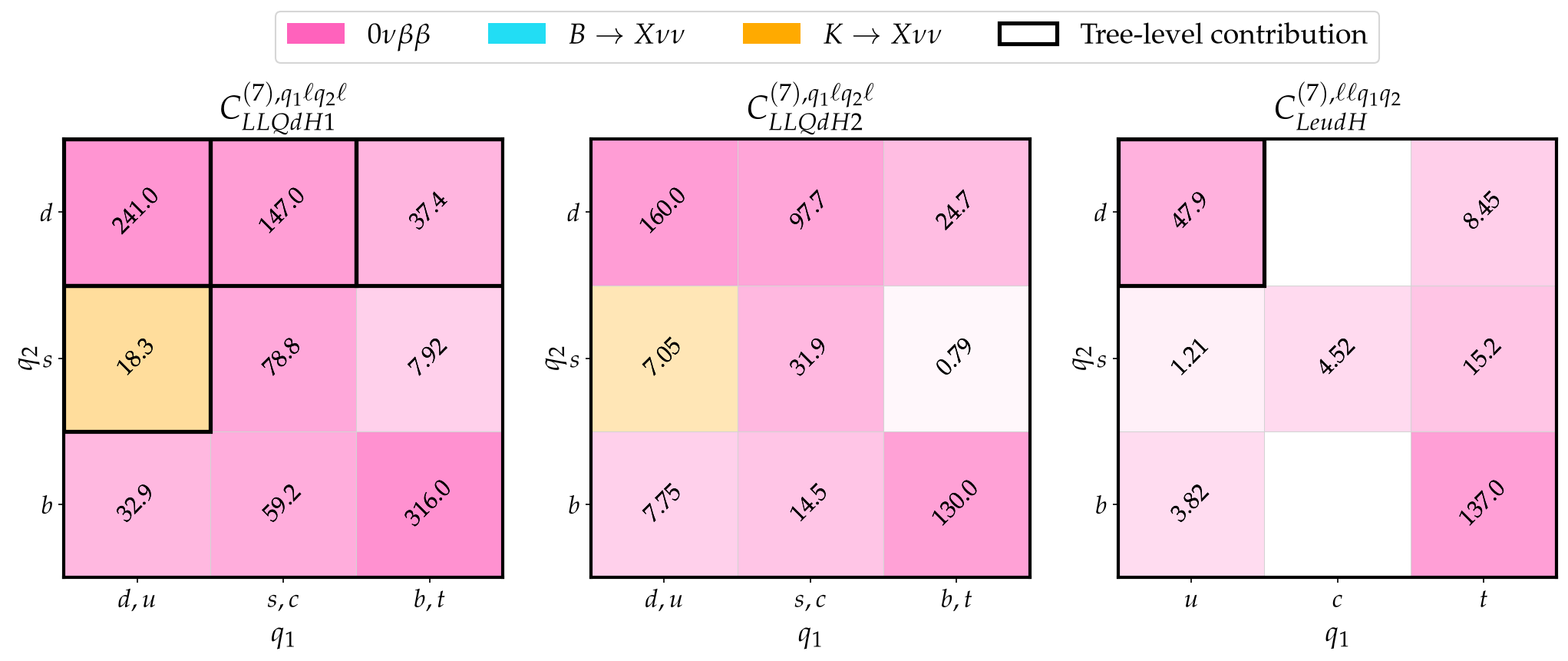}\\
    \includegraphics[width=0.66\linewidth]{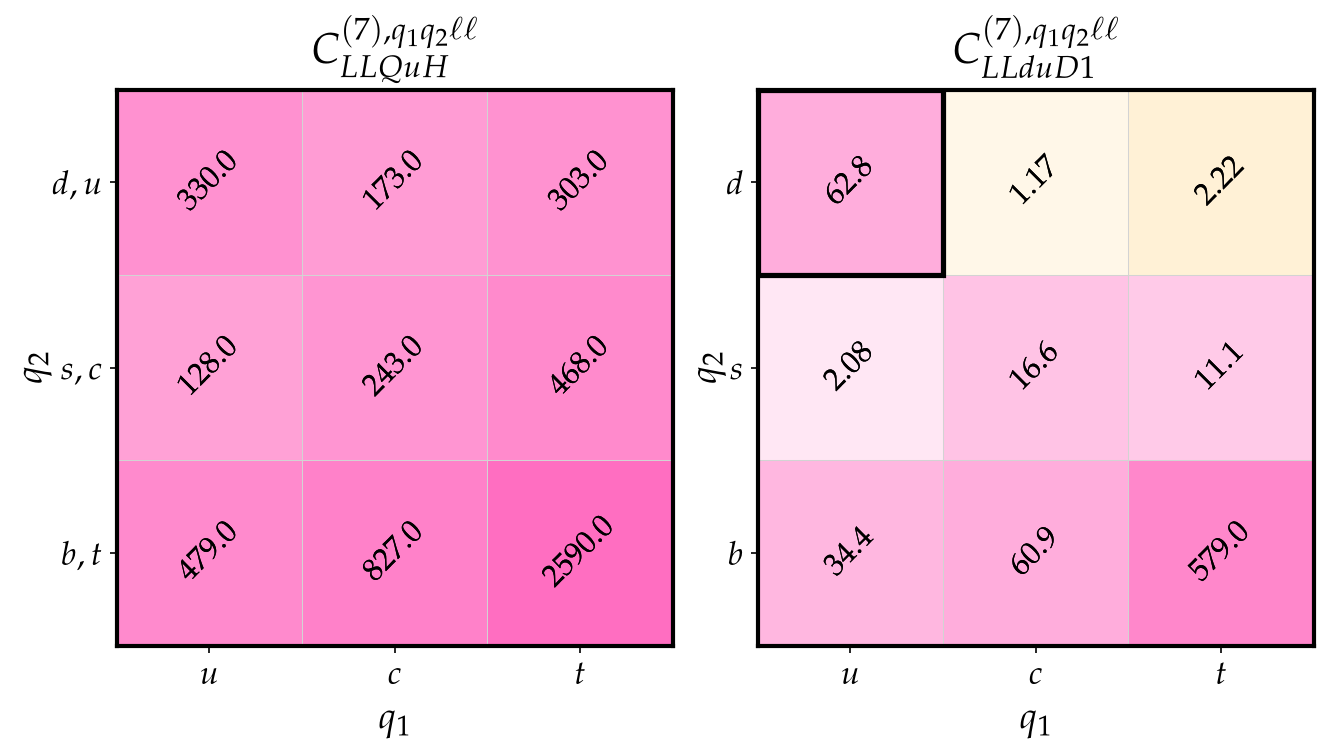}\\
    \includegraphics[width=0.80\linewidth]{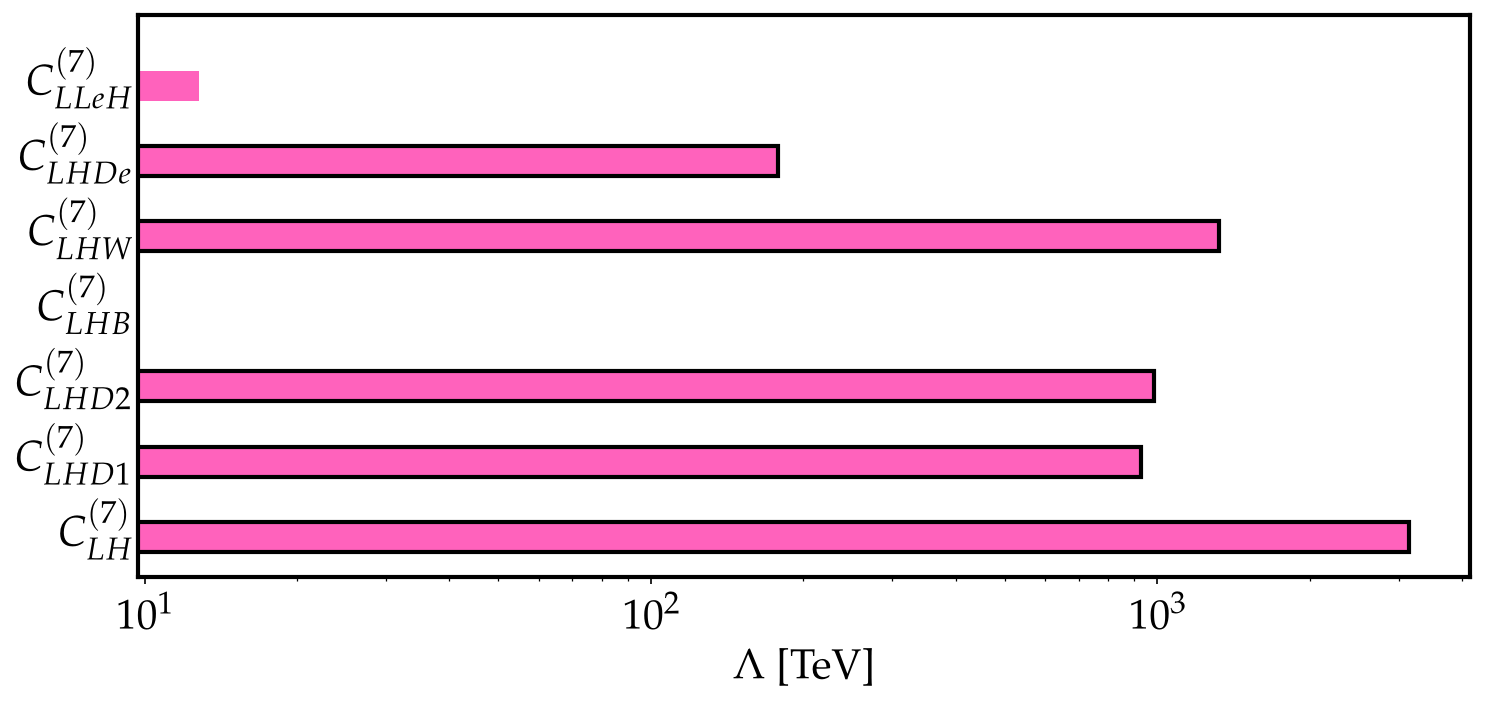}
    \caption{Bottom-up EFT constraints on the new-physics scale $\Lambda\,[\mathrm{TeV}]$ in flavor scenario FS-I (Wilson coefficients are lepton flavor universal and diagonal) for $\Delta L=2$ SMEFT dimension-7 operators, assuming single operator dominance. Only limits with $\Lambda>0.5\,\mathrm{TeV}$ are included.}
    \label{fig:LFUscan}
\end{figure}

Fig.~\ref {fig:LFUscan} shows the constraints on the new-physics scales corresponding to the Wilson coefficients for lepton-number-violating dimension-7 SMEFT operators in the lepton-flavor-universal and diagonal case of FS-I. In these plots, the x- and y-axes correspond to different choices for the two quark generations involved. The color code for each block indicates the observable that provides the strongest limit for that quark-flavor combination. The number inside a given colored block denotes the constraint on the new physics scale $\Lambda$ in units of TeV from the color-coded observable. A thick boundary around a block implies that the relevant color-coded process is induced at tree level (the relevant constraint always includes any one-loop mixing effects). To keep the results compact, we show here only the most strongly constraining observable for each flavor block and provide the full observable-wise results in the Appendix \ref{Ap:EFT_Limits} for interested readers.

Our results for the lepton-flavor-universal case FS-I show that \0 generally provides the most stringent constraints for most quark-flavor combinations of the relevant lepton-number-violating dimension-7 SMEFT operators, except for three operators and in a total of four cases where kaon decays provide the strongest constraints. One of these three operators is the one corresponding to $C_{LLQdH1}^{(7)d\ell s\ell}$ which contributes to $K\rightarrow \pi\nu\hat{\nu}$ at tree level. The other two contribute to $K\rightarrow \pi\nu\hat{\nu}$ at one-loop level through loop mixing, which provides the strongest constraints in a total of three quark flavor settings $C_{LLQdH2}^{(7)d\ell s\ell}$, $C_{LLduD1}^{(7)su\ell\ell}$, and $C_{LLduD1}^{(7)bu\ell\ell}$. Finally, the $B$ decays always provide much weaker limits than $K$ and/or \0 in all cases; see Appendix \ref{Ap:EFT_Limits} for more details on the relevant constraints. We note that for the cases $C_{LeudH}^{(7)\ell\ell su}$ and $C_{LeudH}^{(7)\ell\ell st}$ the best constraints fall below our cut-off of 0.5 TeV and hence are not included here. The dominance of \0 over meson decays in constraining most quark-flavor settings in the lepton-flavor-universal scenario FS-I can be attributed to one-loop SMEFT operator mixing, as first pointed out in~\cite{Graf:2025cfk}. For instance, the operator corresponding to $C_{LLQdH1}^{(7)s\ell d\ell}$ provides a significantly stronger limit from the loop-level contribution (involving heavier-generation SM Yukawas) to \0 as compared to the tree-level contribution to $K\rightarrow\pi\nu\nu$. The situation is similar for the operators corresponding to $C_{LLQdH1}^{(7)b\ell s\ell}$ and $C_{LLQdH1}^{(7)s\ell b\ell}$ which induces $B\rightarrow K\nu\nu$ at the tree level. Another interesting feature in these plots is that constraints from \0 are sometimes significantly stronger for the third quark generation compared to the first quark generation, as can be seen in the cases of $C_{LLQdH1}^{(7)}$, $C_{LeudH}^{(7)}$, $C_{LLQuH}^{(7)}$, $C_{LLduD1}^{(7)}$. This feature can again be attributed to loop mixing with other operators, which \0 constrains much more severely and which are also constrained by large third-generation SM Yukawa couplings. Finally, for all other $\Delta L=2$ SMEFT dimension-7 operators involving no quarks, \0 again gives the strongest constraints; see also~\cite{Graf:2025cfk}. 

\begin{figure}[h!]
    \centering
    \includegraphics[width=\linewidth]{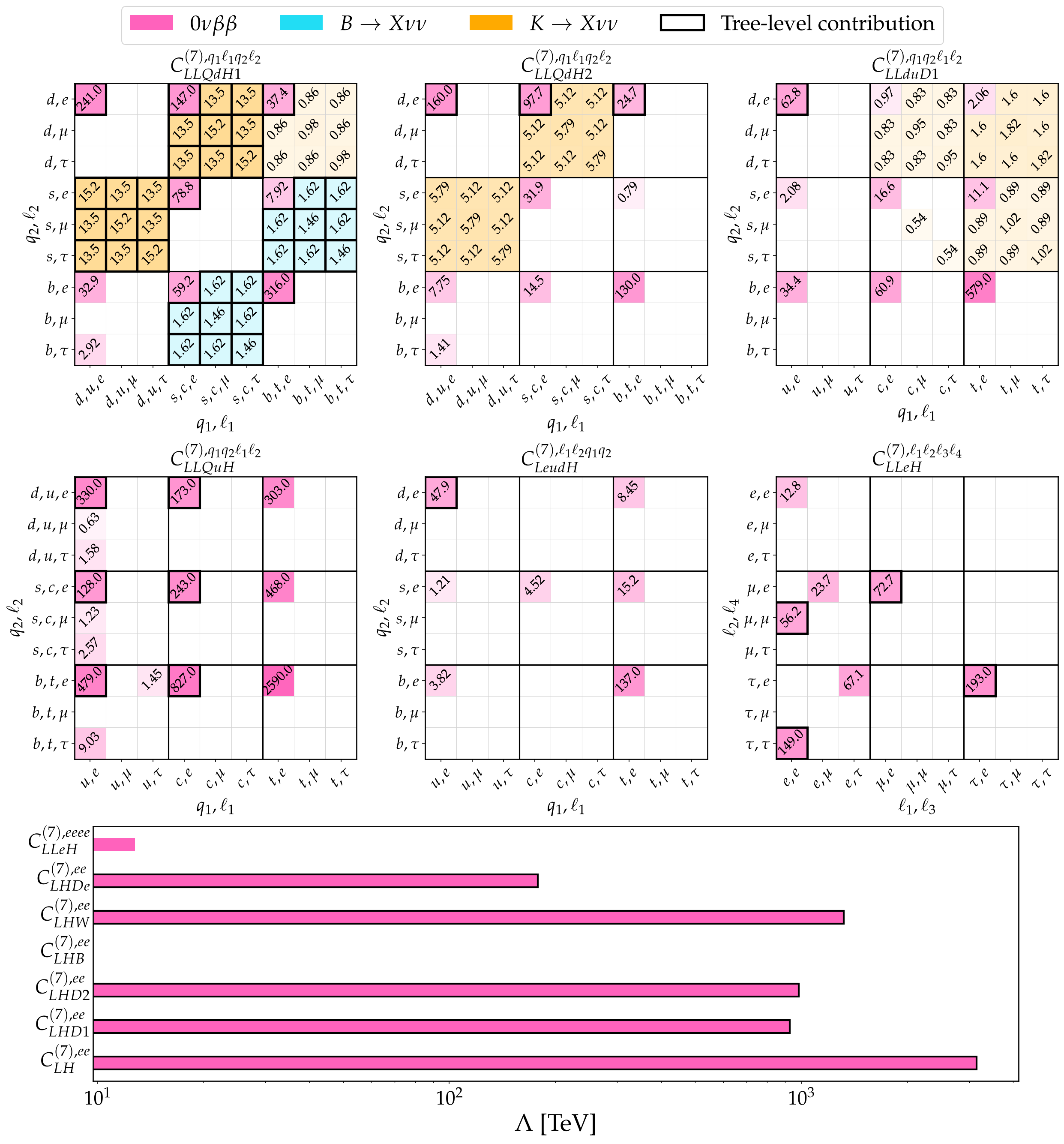}
    \caption{Bottom-up EFT constraints on the new-physics scale $\Lambda\,[\mathrm{TeV}]$ in flavor scenario FS-II (all possible quark and lepton flavor combinations for the Wilson coefficients are independent) for $\Delta L=2$ SMEFT dimension-7 operators, assuming single operator dominance. Only limits with $\Lambda>0.5\,\mathrm{TeV}$ are included.}
    \label{fig:UniversalScan}
\end{figure}
\begin{figure}
    \centering
    \includegraphics[width=\linewidth]{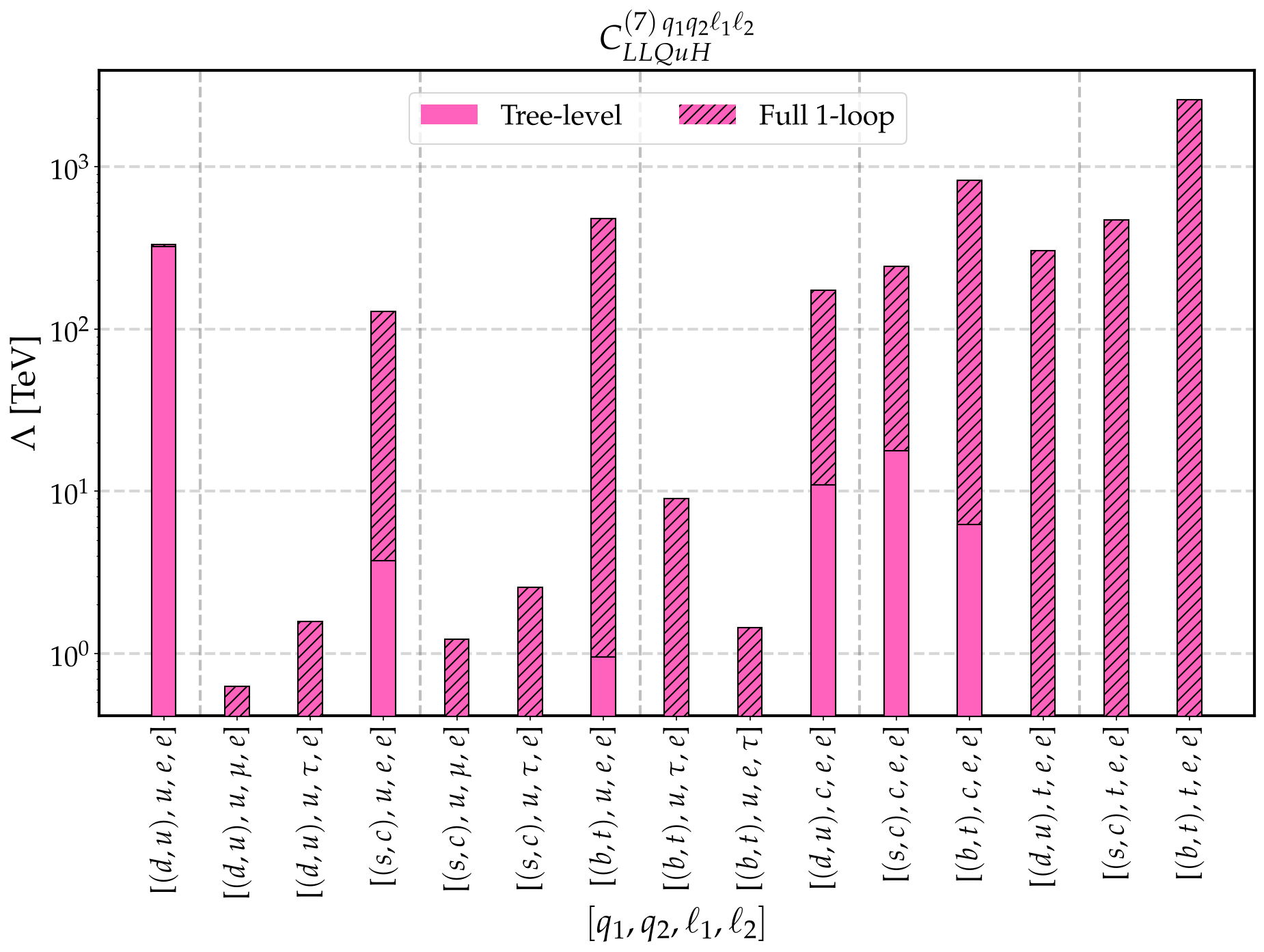}
    \caption{Constraints on the new physics scale $\Lambda$ for $C^{(7)}_{LLQuH}$ from \0 at tree-level (if present) and including one-loop mixing effects.}
    \label{fig:barplot_LLQuH}
\end{figure}

The results for the FS-II case, where all possible quark and lepton flavor combinations are independent, are shown in Fig.~\ref{fig:UniversalScan} assuming single-operator dominance (with a chosen quark-lepton flavor combination). We follow the same style and conventions as already described above for the FS-I case to present the results. In contrast to the FS-I scenario, where the lepton flavor universal and diagonal structure makes \0 the most stringent in most cases, FS-II constraints from rare neutral-current $K$ and $B$ meson decays provide much more competitive bounds for many flavor combinations. Notably, the operator corresponding to $C_{LLQdH1}^{(7)}$ induces $K\rightarrow \pi\nu\nu$ and $B\rightarrow K\nu\nu$ at tree level, and consequently they provide the stringent constraints for quark flavor combinations corresponding to $s\rightarrow d$ and $b\rightarrow d$ transitions, respectively, with the exception of some cases when both leptons are of electron flavor. For the operators corresponding to $C_{LLQdH2}^{(7)}$ and $C_{LLduD1}^{(7)}$ also, $K\rightarrow \pi\nu\nu$ gives the dominant constraints for most of the cases, with again some exceptions when both leptons are of electron flavor. In contrast, for first-generation leptons and operators without quarks, \0 still gives the most stringent constraints. As in the FS-I case, some \0 constraints are stronger for the third quark generation than for the first quark generation due to loop mixing with other operators subject to much more severe \0 constraints and large third-generation SM Yukawa couplings. Fig.~\ref{fig:barplot_LLQuH} shows a comparison of constraints from \0 on the new physics scale $\Lambda$ for the case of $C^{(7)}_{LLQuH}$ at tree-level (if present) and including one-loop mixing effects, demonstrating that the loop mixing effects can totally dominate over the tree-level constraints. Notably, due to CKM mixing effects in the SMEFT-LEFT matching process, \0 can provide considerable constraints on second- and third-generation quark coefficients even at tree level. The same is true for other SMEFT operators, as can be seen in Appendix~\ref{Ap:EFT_Limits}.

Again, for all other $\Delta L=2$ SMEFT dimension-7 operators without any quarks, \0 gives the strongest constraints in the first lepton generation, while 2nd and 3rd lepton generation coefficients are not constrained beyond our threshold of $\Lambda_\mathrm{NP}>0.5\,\mathrm{TeV}$. We note that while the main text shows only the most stringent constraints and the corresponding observables for each Wilson-coefficient flavor-combination block, we provide all constraints for each observable separately in the Appendix \ref{Ap:EFT_Limits}.

\section{Results in top-down EFT for a UV complete model example}
\label{sec:uvmodel}
To show how lepton-number-violating SMEFT operator loop-mixing effects affect constraint extraction from the low-energy observables discussed in a top-down EFT approach, we study one UV-complete model example in detail. The model example we will use is the extension of the SM by two scalar leptoquarks with the following quantum numbers under the SM gauge group $SU(3)_c\times SU(2)_L \times U(1)_Y$
\begin{align}
    \tilde R_2 &\in \left(3,2,1/6\right)\nonumber,\\
    S_1 &\in \left(\bar 3,1,1/3\right).
\end{align}
The relevant Lagrangian for the new fields is given by
\begin{align}
\mathcal{L} &= \mathcal{L}_{SM} - \tilde R_2^{\dagger a} (\square + M^2_{\tilde R_2})\tilde R_{2a} - S_1^\star (\square + M_{S_1}^2)S_1 + \mu S_1 H^{\dagger a} \tilde R_{2 a}
\nonumber \\ 
&\quad - g_1^{\alpha\beta} \bar L_{\alpha a} i \sigma_2^{ab} \tilde R^\star_{2 b} \overline{d^c}_{\beta} - g_2^{\alpha \beta} Q_{\beta}^{a} L_{\alpha}^b \epsilon_{ab} S_1 - g_3^{\alpha \beta} \overline{u^c}_{\beta} e_\alpha S_1 + \text{h.c.},
\end{align}
where greek letters denote flavor indices, roman letters denote $SU(2)_L$ indices, and $\sigma_2$ denotes the second Pauli matrix. Note that according to our conventions, $\tilde R_2$ carries lepton number $L=-1$ and baryon number $B=\frac{1}{3}$, while $S_1$ carries $L=-1$ and $B=-\frac{1}{3}$. The coupling between both leptoquarks and the Higgs, $\mu$, is a dimensionful coupling of order $\mathcal{O}(\Lambda_{NP})$.

The first task in the top-down EFT approach is to compute all matching contributions to the relevant SMEFT operators for our model example up to a given order, in our case $\mathcal{O}(\Lambda^ {-3})$. For the lepton-number-violating dimension-7 and lepton-number-conserving dimension-6 SMEFT operators, we compute the matching at tree level. On the other hand, for this example model, the tree-level contribution to the lepton-number-violating dimension-5 Weinberg operator vanishes, and the lowest-order contribution appears at one loop. Therefore, for the lepton-number-violating dimension-5 Weinberg operator, we compute the one-loop matching, and we make some important remarks shortly. We use the packages \texttt{Matchete} \cite{Fuentes-Martin:2022jrf} and \texttt{MatchMakerEFT} \cite{Carmona:2021xtq} for verifying the matching results. As discussed in section~\ref{sec:eft}, for one-loop operator mixing up to $\mathcal{O}(\Lambda^ {-3})$, dimension-7 lepton-number-violating operators can mix among themselves and with combinations of dimension-5 lepton-number-violating and dimension-6 lepton-number-conserving operators.  

The tree-level matching for relevant lepton-number-violating dimension-7 SMEFT Wilson coefficients is given by
\begin{align}
C_{LLQdH1}^{(7),\alpha\beta\gamma\delta}
&=
i\frac{\mu^\dagger}{M_{\tilde R_2}^2 M_{S_1}^2}
\left(g_1\right)_{\alpha\beta}\left(g_2\right)_{\gamma\delta},
\\
C_{LeudH}^{(7),\alpha\beta\gamma\delta}
&=
\frac{\mu^\dagger}{2 M_{\tilde R_2}^2 M_{S_1}^2}
\left(g_1\right)_{\alpha\beta}\left(g_3\right)_{\gamma\delta}.
\end{align}
The NP couplings $g_1$ and $g_2$ induce tree-level contributions to lepton-number-conserving dimension-6 SMEFT Wilson coefficients
\begin{align}
    C_{Ld}^{(6),\alpha \beta\gamma\delta} = \frac{g_{1,\alpha\delta}g_{1,\beta\gamma}^*}{2M_{\tilde R_2}^2},\qquad
    C_{LQ1}^{(6), \alpha\beta\gamma\delta} = \frac{g_{2,\alpha\delta}g^*_{2,\beta\gamma}}{4M_{ S_1}^2},\qquad
    C_{LQ3}^{(6), \alpha\beta\gamma\delta} = - \frac{g_{2,\alpha\delta}g^*_{2,\beta\gamma}}{4M_{S_1}^2},
\end{align}
which can induce lepton-number-conserving contributions for meson decays if $g_{1,2}$ couple to multiple quark generations simultaneously (e.g., $d$ and $s$ or $s$ and $b$). Here we follow the standard Warsaw basis conventions for SMEFT dimension-6 operators~\cite{Grzadkowski:2010es, Jenkins:2013zja}
\begin{align}
    \mathcal{O}_{Ld}^{(6)} = \Big[\Lbar \gamma^\mu L\Big]\Big[\dbarR \gamma_\mu d_R\Big],
    \quad
    \mathcal{O}_{LQ1}^{(6)} = \Big[\Lbar \gamma^\mu L\Big]\Big[\Qbar \gamma_\mu Q\Big],
    \quad
    \mathcal{O}_{LQ2}^{(6)} = \Big[\Lbar \gamma^\mu \tau^I L\Big]\Big[\Qbar \gamma_\mu \tau^I Q\Big].
\end{align}
On the other hand, if each coupling $g_{1,2}$ only couples to one quark generation, no lepton-number-conserving meson decays are induced at tree level. The lepton-number-violating meson-decay-inducing dimension-7 Wilson coefficients always depend on a combination of two couplings ($g_1$ and $g_2$ or $g_1$ and $g_3$). We focus on the latter scenario, where the lepton-number-conserving dimension-6 SMEFT contributions to meson decays are absent, and on lepton-number-violating modes mediated by the lepton-number-violating dimension-7 SMEFT.  

At tree-level, this model does not generate a contribution to the lepton-number-violating dimension-5 Weinberg operator. The relevant contribution appears at one-loop order, leading to the matching relation
\begin{align}
    C^{(5) \alpha \beta}_{W} = \frac{6\mu y_d^{\gamma \delta}}{32(M^2_{\tilde R_2} - M^2_{S_1})\pi^2} &\left[ \left(g^{\beta \gamma}_3 g^{\delta\alpha}_2 - g^{\beta \gamma}_2 g^{\delta\alpha}_1\right) \log\left( \frac{M_{\tilde R_2}}{M_{S_1}}\right) \right. \nonumber\\
    +& \left.\left(g^{\alpha \gamma}_3 g^{\delta\beta}_2 - g^{\alpha \gamma}_2 g^{\beta}_1\right) \log\left( \frac{M_{\tilde R_2}}{M_{S_1}}\right) \right].
\end{align}
 Naively, one might question the consistency of the one-loop treatment of dimension-5 operators, compared with tree-level dimension-6 and -7 operators. However, this can be understood from different power counting for how the different dimensional contributions enter the evolution of the Wilson coefficients. Let us explain in a bit more detail why and where it is important to include the above one-loop matching contribution.  It is well known that in the EFT formalism depending on whether the low-energy degrees of freedom are weakly or strongly coupled to the new heavy degrees of freedom, a systematic expansion or truncation depends not only on the inverse power of the heavy mass scale, but also on the number of loops~\cite{Weinberg:1978kz,Manohar:1983md,Jenkins:2013sda,Buchalla:2013eza,Gavela:2016bzc,Buchalla:2022vjp}. Considering a general SMEFT interaction involving scalars $\phi$, gauge fields $A$, fermions $\psi$, derivatives and generic weak coupling $\kappa$ given by
\begin{align}\label{generic}
\partial^{N_p}\phi^{N_\phi}A^{N_A}\psi^{N_\psi}\, \kappa^{N_\kappa}\, ,
\end{align}
power counting can estimate and compare the coefficients of different tree- and loop-level operator structures in SMEFT for contributions to a given observable. For the generic operator given above, the order of the relevant coefficient is given by
\begin{align}\label{cl16pi}
C(d_c, d_\chi) = \frac{1}{\Lambda^{d_c-4}}
\left(\frac{1}{16\pi^2}\right)^{\frac{(d_\chi -d_c)}{2}+1},
\end{align}
where the canonical dimension $d_c$ determines the usual $1/\Lambda$ power dependence and the chiral dimension $d_\chi$ together with $d_c$ determines the loop order correctly. The relevant expressions for them are given by
\begin{align}
d_c &= N_p + N_\varphi + N_A + \frac{3}{2} N_\psi \label{dimc},\\
d_\chi &= N_p + \frac{1}{2} N_\psi + N_\kappa\label{dimchi}.
\end{align}
Under this power counting, the one-loop matching contribution is of the same order as inserting a tree-level lepton-number-violating dimension-7 operator to realize a dimension-5 Weinberg operator at one loop via RGE mixing, with $d_\chi=5=d_c$. Therefore, to obtain the operator mixing contribution to the lepton-number-violating dimension-5 Weinberg operator, we solve Eqs.~\eqref{eq:rge-structure} with only the tree-level Wilson coefficients of the lepton-number-violating dimension-7 operator and take the tree-level contribution for the initial $C^{(5) \alpha \beta}_{W}$ on the right-hand side to be vanishing. Clearly, inserting the one-loop matching contribution on the right-hand side would correspond to a higher-loop-order contribution. On the other hand, the one-loop matching contribution for $C^{(5) \alpha \beta}_{W}$ itself must be added to the REG mixing contribution for $C^{(5) \alpha \beta}_{W}$ to obtain a one-loop-order consistent result. In the current UV model case under discussion, for the chosen flavor scenarios below, we will have a vanishing matching contribution; however, the above discussion is valid in general.
\begin{table}[b]
    \centering
    \renewcommand{\arraystretch}{1.15}
    \begin{tabular}{@{}c|ccc|cc@{}}
        \toprule
        &
        $g_1$
        &
        $g_2$
        &
        $g_3$
        &
        $M_{\widetilde R_2}$
        &
        $M_{S_1}$
        \\
        \midrule
       MS-I
        &
        $\mathrm{diag}(1,0,0)$
        &
        $\mathrm{diag}(0,1,0)$
        &
        $\mathrm{diag}(0,0,1)$
        &
        $\Lambda_{\mathrm{NP}}$
        &
        $\Lambda_{\mathrm{NP}}$
        \\
        MS-II
        &
        $\mathrm{diag}(1,0,0)$
        &
        $\mathrm{diag}(0,0,1)$
        &
        $\mathrm{diag}(0,1,0)$
        &
        $\Lambda_{\mathrm{NP}}$
        &
        $\Lambda_{\mathrm{NP}}$
        \\
        MS-III
        &
        $\mathrm{diag}(0,1,0)$
        &
        $\mathrm{diag}(1,0,0)$
        &
        $\mathrm{diag}(0,0,1)$
        &
        $\Lambda_{\mathrm{NP}}$
        &
        $\Lambda_{\mathrm{NP}}$
        \\
        MS-IV
        &
        $\mathrm{diag}(0,1,0)$
        &
        $\mathrm{diag}(0,0,1)$
        &
        $\mathrm{diag}(1,0,0)$
        &
        $\Lambda_{\mathrm{NP}}$
        &
        $\Lambda_{\mathrm{NP}}$
        \\
        MS-V
        &
        $\mathrm{diag}(0,0,1)$
        &
        $\mathrm{diag}(1,0,0)$
        &
        $\mathrm{diag}(0,1,0)$
        &
        $\Lambda_{\mathrm{NP}}$
        &
        $\Lambda_{\mathrm{NP}}$
        \\
        MS-VI
        &
        $\mathrm{diag}(0,0,1)$
        &
        $\mathrm{diag}(0,1,0)$
        &
        $\mathrm{diag}(1,0,0)$
        &
        $\Lambda_{\mathrm{NP}}$
        &
        $\Lambda_{\mathrm{NP}}$
        \\
        \bottomrule
    \end{tabular}
    \caption{Flavor structure ans\"{a}tze for the six benchmark leptoquark coupling flavor-structure settings considered for the UV model example.}
    \label{tab:LQ_benchmarks}
\end{table}

%

%
\begin{figure}[h!]
    \centering
    \includegraphics[width=0.82\linewidth]{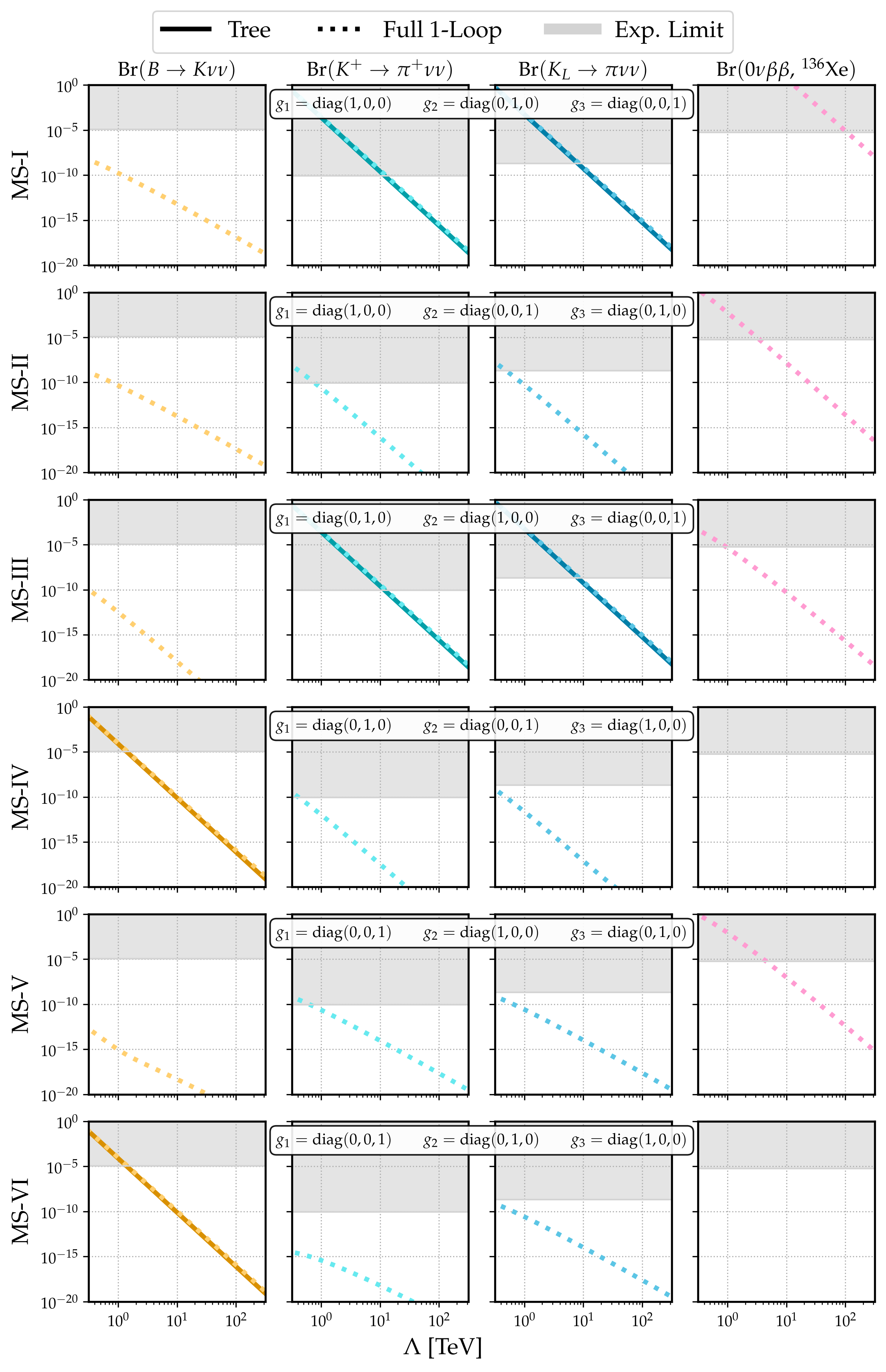}
    \caption{Branching ratios for different observables as a function of the new physics scale $\Lambda$ in the different coupling flavor scenarios in Table~\ref{tab:LQ_benchmarks} for our UV model example. Tree-level branching ratios are displayed as solid lines, while the full one-loop results are shown as dashed lines. Current experimental results already exclude the shaded regions.}
    \label{fig:LQModels}
\end{figure}

\begin{figure}[t!]
    \centering
    \includegraphics[width=\linewidth]{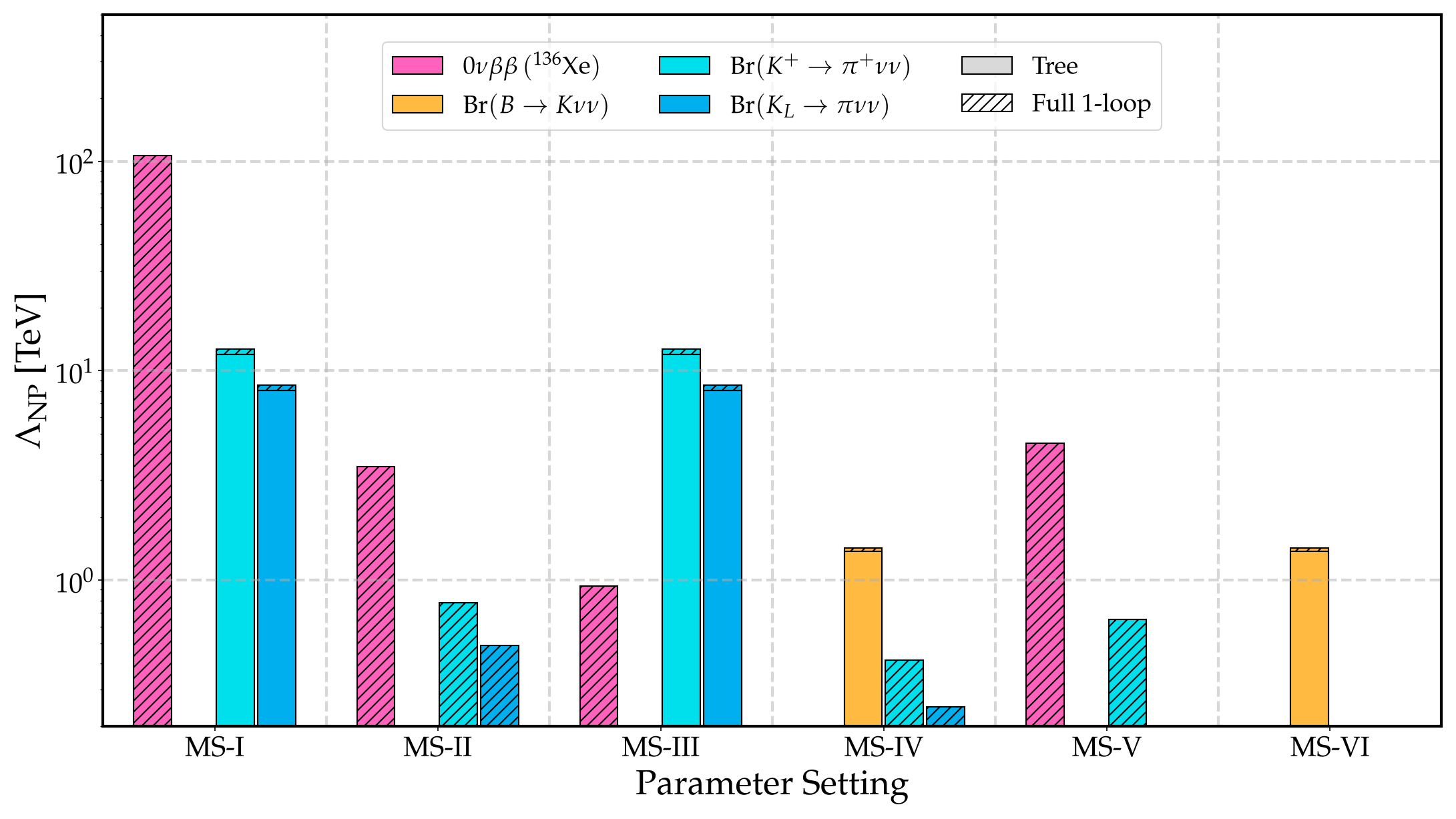}
    \caption{Constraints on the new physics scale $\Lambda$ in the different coupling flavor scenarios in Table~\ref{tab:LQ_benchmarks} for our UV model example (see also Fig.~\ref{fig:LQModels}) from $0\nu\beta\beta$ , $B\rightarrow K\nu\nu$, $K^+\rightarrow\pi^+\nu\nu$, and $K_L\rightarrow\pi\nu\nu$. The constraints derived at tree-level (if present) are displayed as solid bars, and the corresponding one-loop-level results (if tree-level is absent) or corrections are shown as hatched bars.}
    \label{fig:top-down-barplot}
\end{figure}

In what follows, we will explore the phenomenology of the ans\"{a}tze where the new physics couplings $g_{1,2,3}$ couple to only a single fermion generation at a time and, for further simplification, we will consider a diagonal structure for these coupling matrices. This gives six model flavor scenarios, as shown and labeled MS-I to MS-VI in Table~\ref{tab:LQ_benchmarks}. These choices ensure vanishing tree-level contributions to dimension-6 SMEFT operators which can induce lepton-number-conserving meson decay modes. Furthermore, we assume the heavy new-physics mass scales are of the same order by setting $\mu=M_{\tilde R_2} = M_{S_2}=\Lambda_{NP}$. For a relevant discussion of the impact of a hierarchy between these mass scales, see, e.g., Ref.~\cite{Fridell:2024pmw}. 

In Figure~\ref{fig:LQModels}, we show the branching ratios for meson decays $B\rightarrow K\nu\nu$, $K\rightarrow\pi\nu\nu$ as well as \0 as a function of the heavy new physics mass scale for the six different flavor ans\"{a}tze given in Table~\ref{tab:LQ_benchmarks}. For ease of comparison and convenience, we have used a \0 branching ratio defined as
\begin{align}
    \mathrm{Br}(0\nu\beta\beta) = \frac{T_{1/2}^{2\nu}}{T_{1/2}^{0\nu}},
\end{align}
where we assume the recommended experimental \textsuperscript{136}Xe half-life of $T_{2\nu}^{1/2}=2.18\times10^{21}\,\mathrm{yr}$~\cite{EXO-200:2013xfn, KamLAND-Zen:2016pfg, Barabash:2020nck}. The gray shaded regions show the excluded parts from the current best experimental limits, c.f. Table~\ref{tab:observables}. The solid curves (if present) show the tree-level contribution to the corresponding observable. The dashed curves, on the other hand, represent the predictions at the one-loop level; therefore, in cases where there are only dashed curves and no solid curve, the corresponding contribution is generated purely at the one-loop level via SMEFT operator mixing governed by the RGEs. Finally, in Fig.~\ref{fig:top-down-barplot} we show a comparison of the constraints on the heavy new physics scale obtained from the different observables for the six flavor ans\"{a}tze in Table~\ref{tab:LQ_benchmarks}. The solid bars (if present) show tree-level constraints from the corresponding observables. The hatched bars show the full one-loop-level constraints.

We note that in the flavor scenarios where $g_1$ couples to the first fermion generation (corresponding to rows 1 and 2 of Table~\ref{tab:observables}), the kaon decays $K\rightarrow\pi\nu\nu$ provide limits in the range of $\mathcal{O}(1-10)\,\mathrm{TeV}$. However the one-loop induced \0 provides stronger constraints in the $\mathcal{O}(10-100)\,\mathrm{TeV}$ range. For the flavor scenario where $g_1$ and $g_2$ couple to the second and first generation, respectively (corresponding to row 3 of Table~\ref{tab:observables}), $K\rightarrow\pi\nu\nu$ is induced at tree level and provides the strongest constraint at $\mathcal{O}(10)\,\mathrm{TeV}$.  In the cases where $g_1$ and $g_2$ couple to the second and third generations, respectively, or vice versa (corresponding to rows 4 and 6 of Table~\ref{tab:observables}), $B\rightarrow K\nu\nu$  is induced at tree level and provides the strongest constraints of $\mathcal{O}(1)\,\mathrm{TeV}$. Finally, in the other remaining flavor setting with no tree-level contribution to either meson decays or \0 (corresponding to row 5 of Table~\ref{tab:observables}), the one-loop induced \0 limits provide the strongest constraints at $\sim\mathcal{O}(10)\,\mathrm{TeV}$. 


\section{Conclusions}
\label{sec:conclusions}
In this study, we have investigated the interplay between \0 and rare meson decays as complementary probes of lepton-number-violating new physics underlying the $\Delta L=2$ dimension-7 SMEFT operator. A central aim of our analysis is to consistently implement one-loop mixing and matching effects above the electroweak scale across different relevant operators and flavors in the bottom-up and top-down EFT approaches. In particular, SMEFT operator mixing governed by RGEs induces non-trivial mixing among operators and flavor structures, establishing connections between different lepton-number-violating operators and their flavor combinations underlying \0, $K\to\pi\nu\nu$, and $B\to K^{(*)}\nu\nu$ that are absent in a tree-level treatment. As a consequence, observables involving very different external-state flavors cannot always be taken to probe independent lepton-number-violating new physics.

We first presented this complementarity in the bottom-up EFT approach by analyzing the full set of $\Delta L=2$ dimension-7 SMEFT operators under the assumption of single-operator dominance in two limits of the underlying flavor structure. For a lepton flavor universal and diagonal case, we find that \0 provides the strongest constraint for most quark-flavor assignments once one-loop mixing is included. Remarkably, this also applies to several operators involving second- and third-generation quarks, for which \0 is absent or considerably weaker at tree level. The effect can be particularly pronounced for third-generation flavor, where large SM Yukawa couplings enhance the mixing into operators strongly constrained by \0. On the other hand, rare kaon decays remain the leading probe for selected flavor structures. The picture changes substantially when general lepton- and quark-flavor structures are totally independent. In such a case, there are a sizable number of scenarios for which $K\to\pi\nu\nu$, and in some cases $B\to K^{(*)}\nu\nu$, provide the dominant sensitivity, while \0 continues to provide the dominant sensitivities in the remaining cases. These results demonstrate that the relative importance of \0 and rare meson decays is intrinsically tied to the flavor structure of the underlying lepton-number-violating new physics interactions.

We also illustrated these effects in a top-down EFT approach using a UV-complete model with two scalar leptoquarks, $\widetilde R_2$ and $S_1$. By exploring representative flavor textures that suppress tree-level lepton-number-conserving meson-decay contributions, we demonstrated how the relative sensitivities among different lepton-number-violating observables emerge when we consistently account for one-loop matching and mixing effects. We showed that, depending on the underlying flavor structure Ans\"atze, at the one-loop level the leading constraint can originate either from \0 or rare meson decays. 

Our findings therefore emphasize that a consistent interpretation of lepton-number-violating searches requires treating flavor and scale evolution simultaneously to capture the full picture. Lepton-number-violating operator mixing above the electroweak symmetry breaking scale can turn observables that appear unrelated at tree level into powerful complementary probes of the same high-scale heavy new physics. The continued improvement of experimental search sensitivity for \0, $K\to\pi\nu\nu$, and $B\to K^{(*)}\nu\nu$ searches will therefore provide complementary information not only on the scale of possible lepton-number-violating new physics, but also on its underlying flavor structure. Incorporating their combined information within a bottom-up or top-down EFT framework will be essential for identifying the origin of any future experimental signal of lepton-number-violating new physics beyond the SM.

\section*{Acknowledgements}
The authors thank Jordy de Vries for some valuable comments. L.~G. is supported by Charles University through the project PRIMUS/24/SCI/013 and by the Dutch Research Council (NWO) through the project number VI.Veni.222.318. C.~H. is funded by the Generalitat Valenciana under Plan Gen-T via CDEIGENT grant No. CIDEIG/2022/16. A.~M. is funded via grant No. CIDEXG/2022/20 by Generalitat Valenciana. C.~H. and A.~M.~ also acknowledge support in part from the Spanish grants PID2023-147306NB-I00 and CEX2023-001292-S (MCIU/AEI/10.13039/501100011033). O. S. acknowledges support from the Alexander von Humboldt Foundation under the Feodor Lynen Research Fellowship program and from the National Science Foundation under cooperative agreement 2020275.
\appendix

\section{Illustrative derivation of RGE-induced operator mixing}\label{Ap:RGE}
 Here we will briefly summarize some of the key steps and their interpretation in the derivation of RGEs and the relevant anomalous dimension matrix from the counterterms and wave-function renormalization constants. The UV divergences arising in loop calculations are encoded in the counterterms and, in dimensional regularization, appear as poles in $1/\epsilon$, with $d=4-2\epsilon$. The residues of these poles determine the anomalous dimensions and hence the renormalization-group evolution and mixing of the Wilson coefficients. After the UV divergences are absorbed into the corresponding counterterms, the renormalized quantities are finite and the limit $\epsilon\to 0$ can be taken to recover the four-dimensional theory.

 The scale dependence of the couplings $h_i$ in $d=4-2\epsilon$ dimensions contains a tree-level contribution,
\begin{equation}
    \mu \frac{d h_i}{d\mu}
    =
    -\epsilon\, n_i' h_i
    + \mathcal{O}(\hbar),
\end{equation}
where $h_i$ runs over all couplings, including the Wilson coefficients, and $n_i'$ denotes the corresponding tree-level scaling coefficient. Requiring the bare Wilson coefficients to be independent of the renormalization scale then gives, at one-loop order,
\begin{equation}
    \mu \frac{d C_r}{d \mu}
    =
    \epsilon
    \left(
        \sum_i n_i' h_i
        \frac{\partial Z_r}{\partial h_i}
    \right) C_r .
\end{equation}
The quantity on the right-hand side determines the one-loop beta function of $C_r$. For contributions that are linear in the Wilson coefficients, these beta functions can be organized in terms of an anomalous-dimension matrix, while more general terms can also describe nonlinear mixing among Wilson coefficients of different operator dimensions.

The renormalization constant associated with a Wilson coefficient $C_r$ can be written as
\begin{equation}
    Z_r
    =
    1
    -\sum_\varphi \frac{1}{2} n_\varphi \delta Z_\varphi
    +\frac{\delta C_r}{C_r},
\end{equation}
where $\delta Z_\varphi$ denotes the wave-function renormalization of the field $\varphi$, $n_\varphi$ is the number of fields of type $\varphi$ appearing in the corresponding operator, and $\delta C_r$ denotes the corresponding Wilson coefficient counterterm.

To illustrate how operator mixing arises from this relation, we consider a representative contribution to the RGE of the dimension-seven coefficient $C_{LH}^{(7),\alpha\beta}$. In the notation used throughout this work, the relevant term is~\cite{Zhang:2023ndw}
\begin{equation}
    16\pi^2 \mu
    \frac{d C_{LH}^{(7),\alpha\beta}}{d\mu}
    \supset
    -2\,C_{LH}^{(5)*,\alpha\beta}
    \frac{3}{4}
    \left(g_1^2-g_2^2+4\lambda\right)
    C_{HD}^{(6)} .
\end{equation}
This contribution is generated at one loop through the combined insertion of the dimension-five coefficient $C_{LH}^{(5)}$ and the dimension-six coefficient $C_{HD}^{(6)}$, together with the SM couplings $g_1$, $g_2$, and $\lambda$.

\begin{figure}[t]
    \centering
\includegraphics[width=0.9\linewidth]{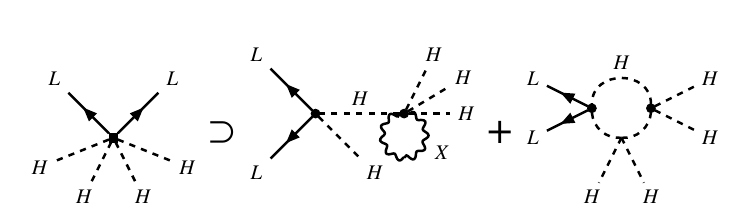}
    \caption{Mixing contribution for $C^{(7)}_{LH}$ operator, where $X=B,W$ denotes one of the gauge bosons. Note that not all the possible topologies are illustrated in this figure.}
    \label{fig:mixing}
\end{figure}

The corresponding contribution can be obtained from the UV divergence of the relevant one-loop Green functions in dimensional regularization. After reduction to the physical operator basis, the relevant part of the counterterm is
\begin{equation}
    \delta C_{LH}^{(7),\alpha\beta}
    \supset
    -\frac{1}{16\pi^2\epsilon}\,
    C_{LH}^{(5)*,\alpha\beta}
    \left[
        \frac{3}{4}
        \left(g_1^2-g_2^2+4\lambda\right)
        C_{HD}^{(6)}
    \right].
\end{equation}
Representative one-loop topologies contributing to this mixing are shown in
Figure~\ref{fig:mixing}.

The tree-level scaling coefficients entering the $d=4-2\epsilon$ dimensional running are
\begin{equation}
    n'_{C_{LH}^{(5)}}=2,
    \qquad
    n'_{C_{HD}^{(6)}}=2,
    \qquad
    n'_{g_{1,2}}=1,
    \qquad
    n'_{\lambda}=2,
    \qquad
    n'_{C_{LH}^{(7)}}=4.
\end{equation}
Consequently, each term appearing in the numerator of
$\delta C_{LH}^{(7)}$ has total tree-level scaling weight six,
\begin{equation}
    n'_{C_{LH}^{(5)}}+2n'_{g_{1,2}}
    +n'_{C_{HD}^{(6)}}
    =
    n'_{C_{LH}^{(5)}}+n'_{\lambda}
    +n'_{C_{HD}^{(6)}}
    =6,
\end{equation}
whereas the coefficient $C_{LH}^{(7)}$ itself has scaling weight four.
For the particular mixing contribution considered here, the wave-function
renormalization terms do not generate the structure
$C_{LH}^{(5)*} C_{HD}^{(6)}$. Therefore, using
\begin{equation}
    Z_{LH}^{(7)}
    \supset
    \frac{\delta C_{LH}^{(7)}}{C_{LH}^{(7)}},
\end{equation}
the weighted derivative entering the RGE gives
\begin{equation}
    \left(
        \sum_i n'_i h_i \frac{\partial}{\partial h_i}
    \right)
    \frac{\delta C_{LH}^{(7)}}{C_{LH}^{(7)}}
    =
    (6-4)
    \frac{\delta C_{LH}^{(7)}}{C_{LH}^{(7)}}
    =
    2
    \frac{\delta C_{LH}^{(7)}}{C_{LH}^{(7)}}.
\end{equation}
It then follows from the general RGE relation that
\begin{equation}
    \mu
    \frac{d C_{LH}^{(7),\alpha\beta}}{d\mu}
    \supset
    2\epsilon\,
    \delta C_{LH}^{(7),\alpha\beta}.
\end{equation}
Substituting the $1/\epsilon$ pole of the counterterm, we obtain
\begin{equation}
    16\pi^2 \mu
    \frac{d C_{LH}^{(7),\alpha\beta}}{d\mu}
    \supset
    -2\,C_{LH}^{(5)*,\alpha\beta}
    \left[
        \frac{3}{4}
        \left(g_1^2-g_2^2+4\lambda\right)
        C_{HD}^{(6)}
    \right],
\end{equation}
which reproduces the desired contribution to the RGE.

\section{Bottom-up EFT constraints for individual observables}\label{Ap:EFT_Limits}
This appendix provides the observable-wise constraints for the bottom-up EFT
analysis of Section~\ref{sec:eftlimits}. While the results presented in the main text retain
only the most constraining observable for each Wilson-coefficient flavor assignment, here
we show separately the limits obtained from rare meson decays and from
$0\nu\beta\beta$. This representation makes explicit the flavor structures accessible to
each observable and, in particular, the additional sensitivity generated by SMEFT operator mixing. We use the same conventions and
flavor scenarios as in Section~\ref{sec:eftlimits}: FS-I corresponds to diagonal lepton
flavor universality, while in FS-II a single independent flavor combination is switched on
at the high scale. As in the main analysis, only limits with
$\Lambda_{\rm NP}>0.5~{\rm TeV}$ are displayed. The black frames identify flavor
assignments for which the corresponding observable receives a tree-level contribution;
colored entries without a black frame therefore arise through loop-induced operator
mixing.

\begin{figure}[p]
    \centering   \includegraphics[width=0.99\linewidth]{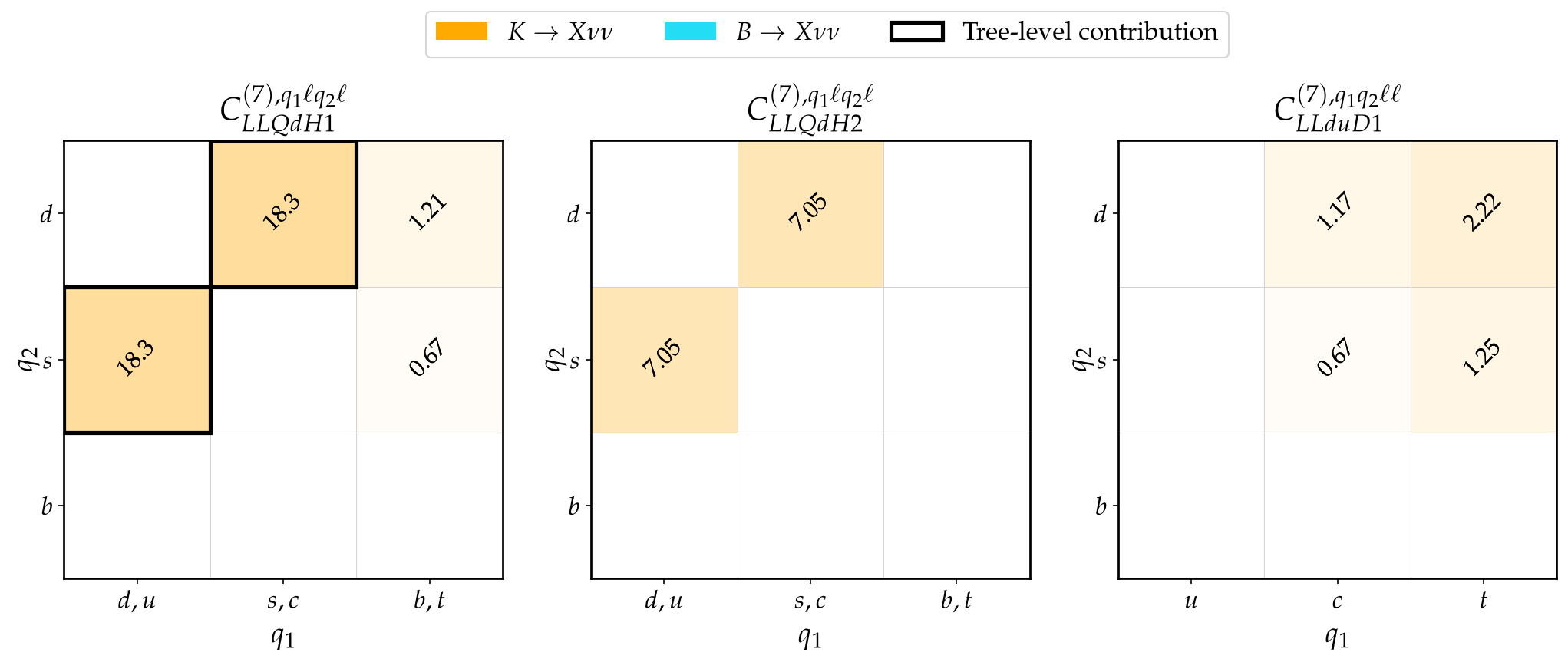}\\
\includegraphics[width=0.99\linewidth]{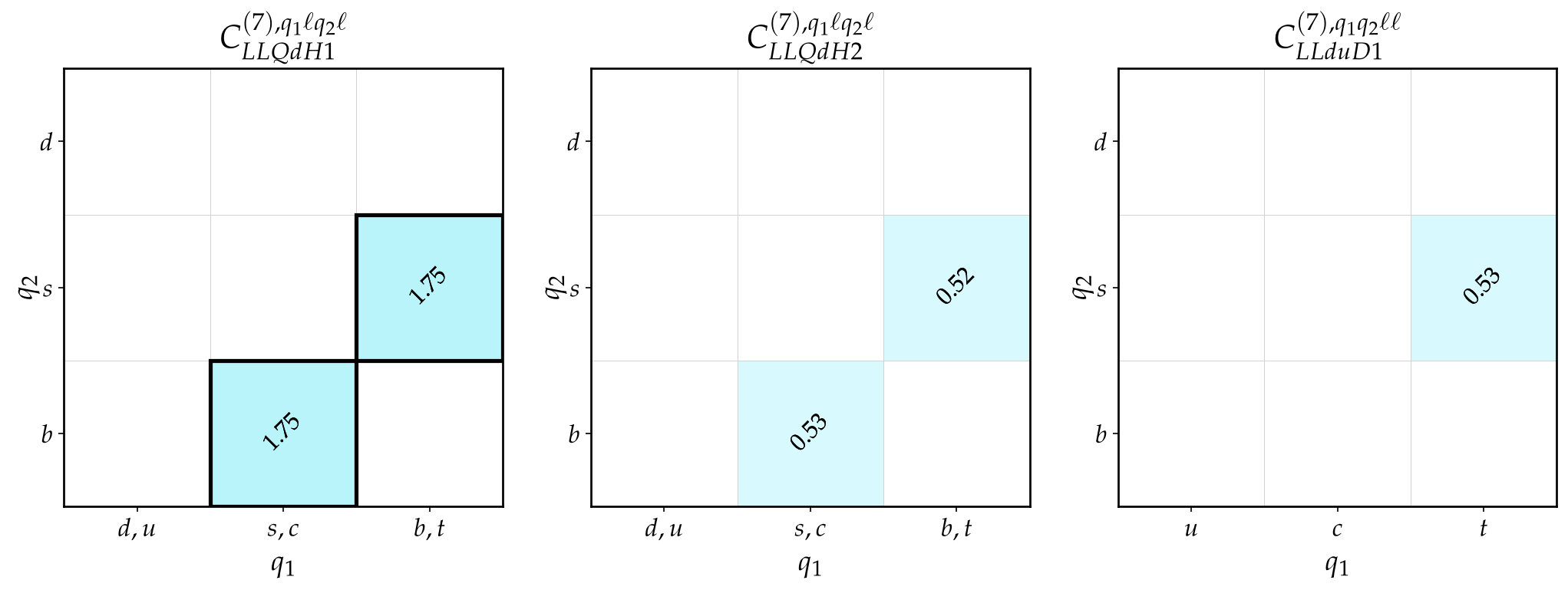}
    \caption{Relevant limits derived on the new physics scale $\Lambda_{\text{NP}}$ in TeV from $B$ and $K$ decays for the $\Delta L=2$ dimension-7 SMEFT operators for scenario FS-I lepton (flavor universal and diagonal case). The black frame denotes which flavor combination of quarks contributes at the tree-level.}
    \label{fig:LFU_K_and_B}
\end{figure}

\begin{figure}[p]
    \centering   
\includegraphics[width=0.80\linewidth]{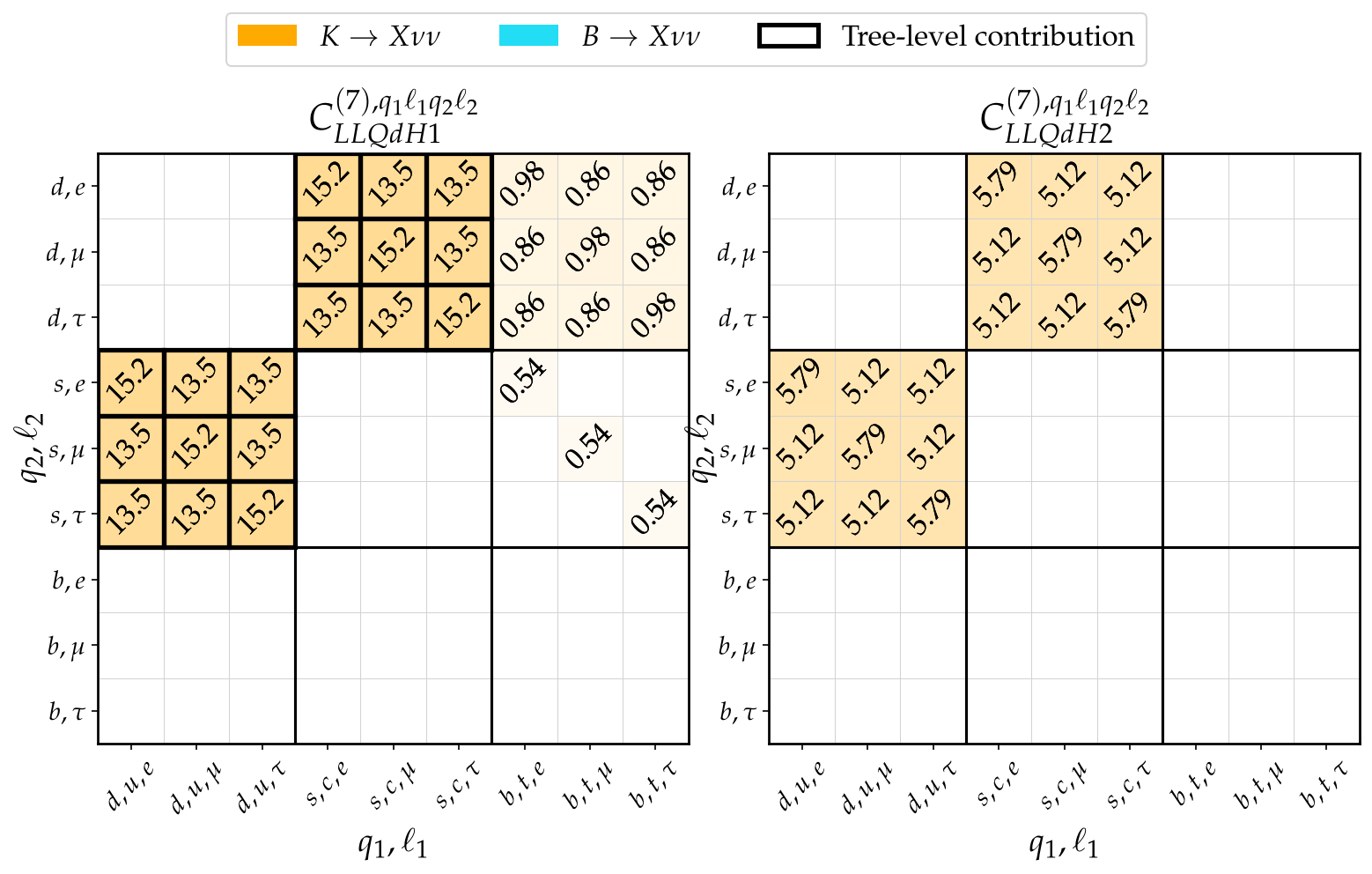}\
\includegraphics[width=0.40\linewidth]{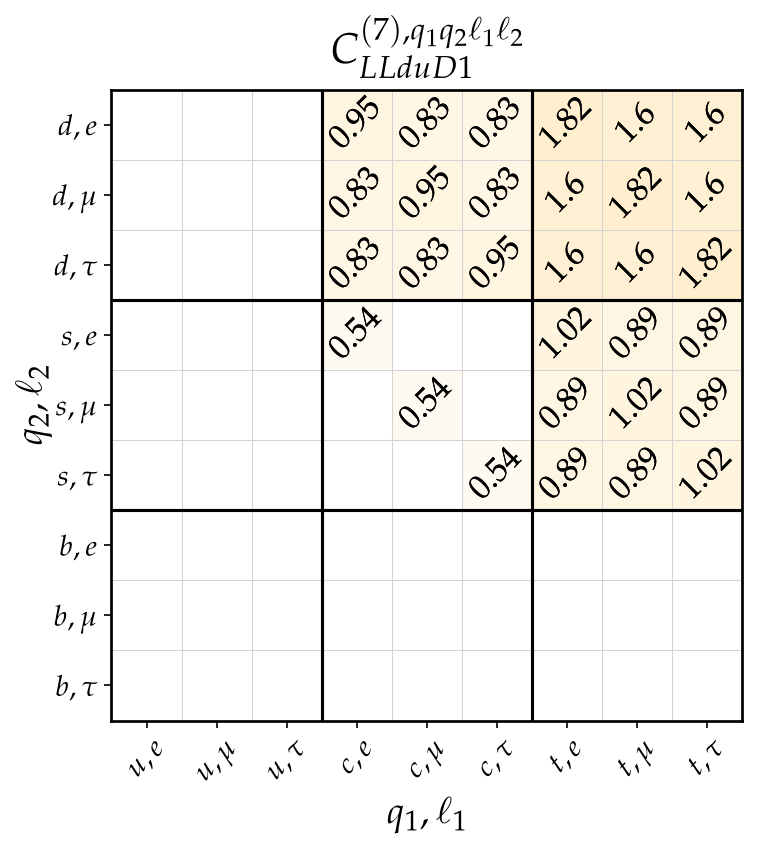}
\includegraphics[width=0.40\linewidth]{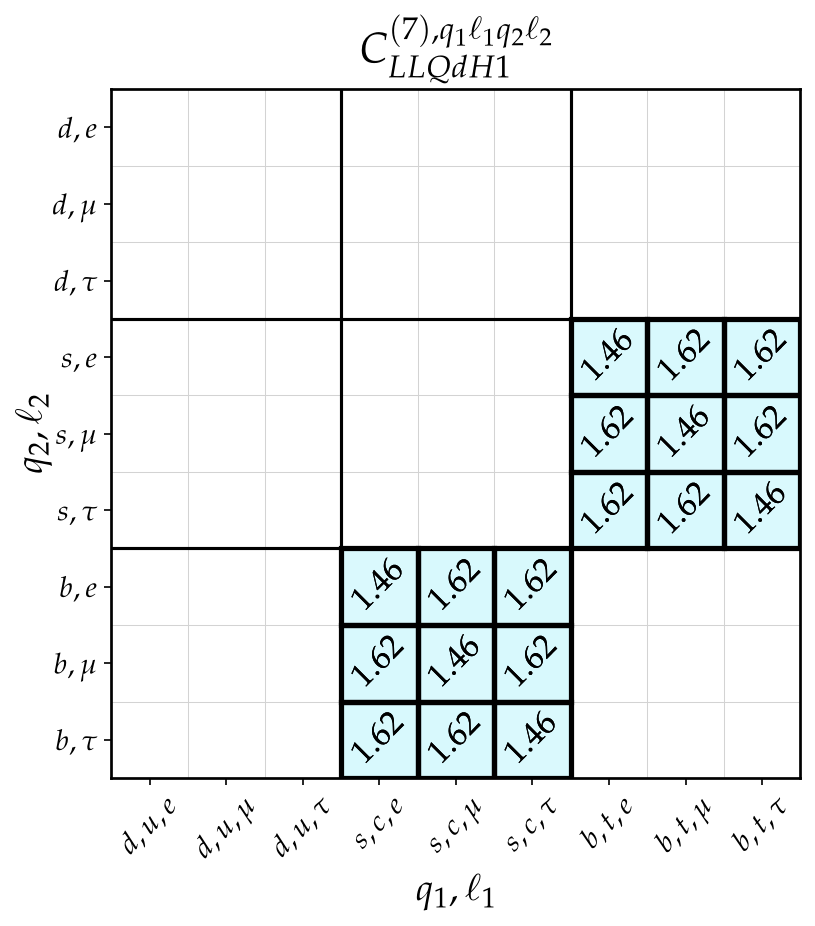}
    \caption{Relevant limits derived on the new physics scale $\Lambda_{\text{NP}}$ in TeV from $B$ and $K$ decays for the $\Delta L=2$ dimension-7 SMEFT operators for the scenario FS-II (one single active lepton or flavor combination at the time). The black frame denotes which flavor combination of quarks contributes at the tree level.}
    \label{fig:full_B_and_K}
\end{figure}

Figures~\ref{fig:LFU_K_and_B} and ~\ref{fig:full_B_and_K} show the constraints from
$K\to\pi\nu\nu$ and $B\to K^{(*)}\nu\nu$ in FS-II and FS-I, respectively. The
tree-level flavor structure directly reflects the underlying flavor-changing transitions.
For kaon decays, $\mathcal{O}^{(7)}_{LLQdH1}$ generates the relevant $s\to d$
transition already at tree level. The operator mixing substantially extends this sensitivity:
$\mathcal{O}^{(7)}_{LLQdH2}$ and $\mathcal{O}^{(7)}_{LLduD1}$, which do not contribute
directly to these modes at tree level, become accessible through mixing, and additional
high-scale quark-flavor assignments can also be probed. The resulting kaon limits therefore
illustrate both an enlargement of the operator space and a broadening of the flavor
structures accessible at low energies.

For $B$ decays, the pattern is more restricted. In the down-quark-diagonal basis used
here, the tree-level contribution from $\mathcal{O}^{(7)}_{LLQdH1}$ requires the
appropriate $b$--$s$ flavor structure, without an additional CKM rotation in the
SMEFT--LEFT matching. 
In the LFU scenario of Figure~\ref{fig:LFU_K_and_B}, the enhanced combined
sensitivity brings selected loop-induced contributions from
$\mathcal{O}^{(7)}_{LLQdH2}$ and $\mathcal{O}^{(7)}_{LLduD1}$ above this threshold.
In the single-flavor scenario FS-II of
Figure~\ref{fig:full_B_and_K}, only the corresponding
$\mathcal{O}^{(7)}_{LLQdH1}$ limits remain above our $0.5~{\rm TeV}$ threshold.
Thus, for $B$ decays, operator mixing primarily enlarges the set of SMEFT operators
that can be tested, while the accessible quark-flavor structure remains comparatively
closely tied to the underlying $b\to s$ transition.

\begin{figure}[p]
    \centering
    \includegraphics[width=0.99\linewidth]{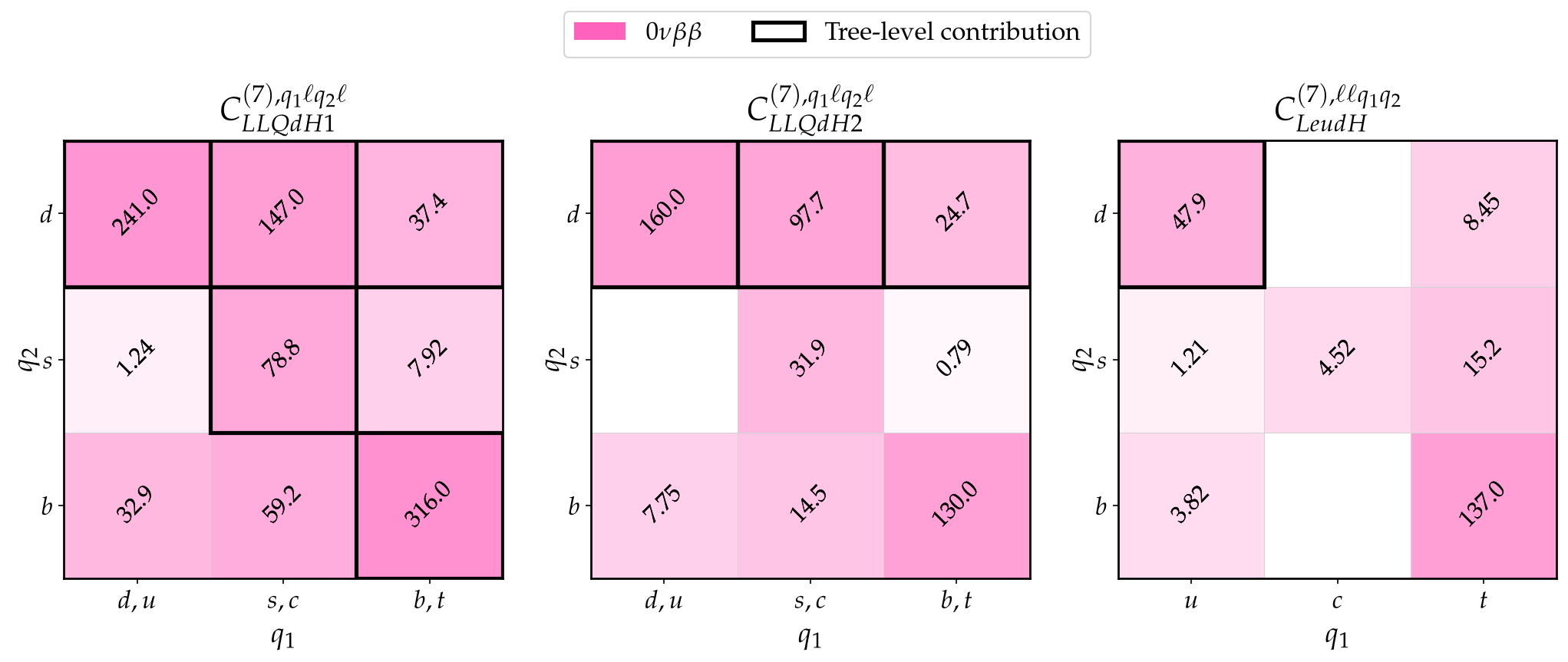}\\

    \includegraphics[width=0.66\linewidth]{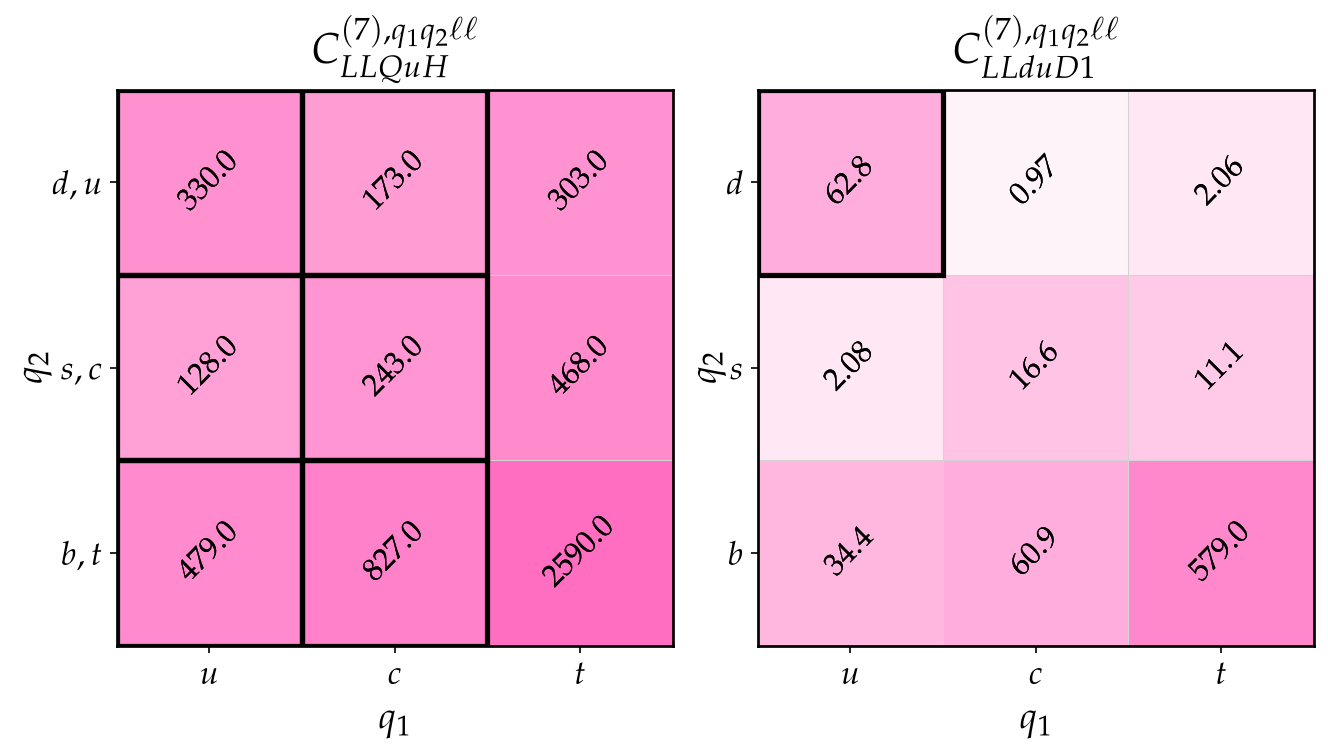}\\
    \includegraphics[width=0.80\linewidth]{plots/LFU_combined_3.png}\\
    
    \caption{Relevant limits derived on the new physics scale $\Lambda_{\text{NP}}$ in TeV from $0\nu\beta\beta$ for the $\Delta L=2$ dimension-7 SMEFT operators for scenario FS-I (flavor universal and diagonal case). The black frame denotes which flavor combination of quarks contributes at the tree level.}
    \label{fig:LFU_0vbb}
\end{figure}

\begin{figure}[p]
    \centering    \includegraphics[width=0.99\linewidth]{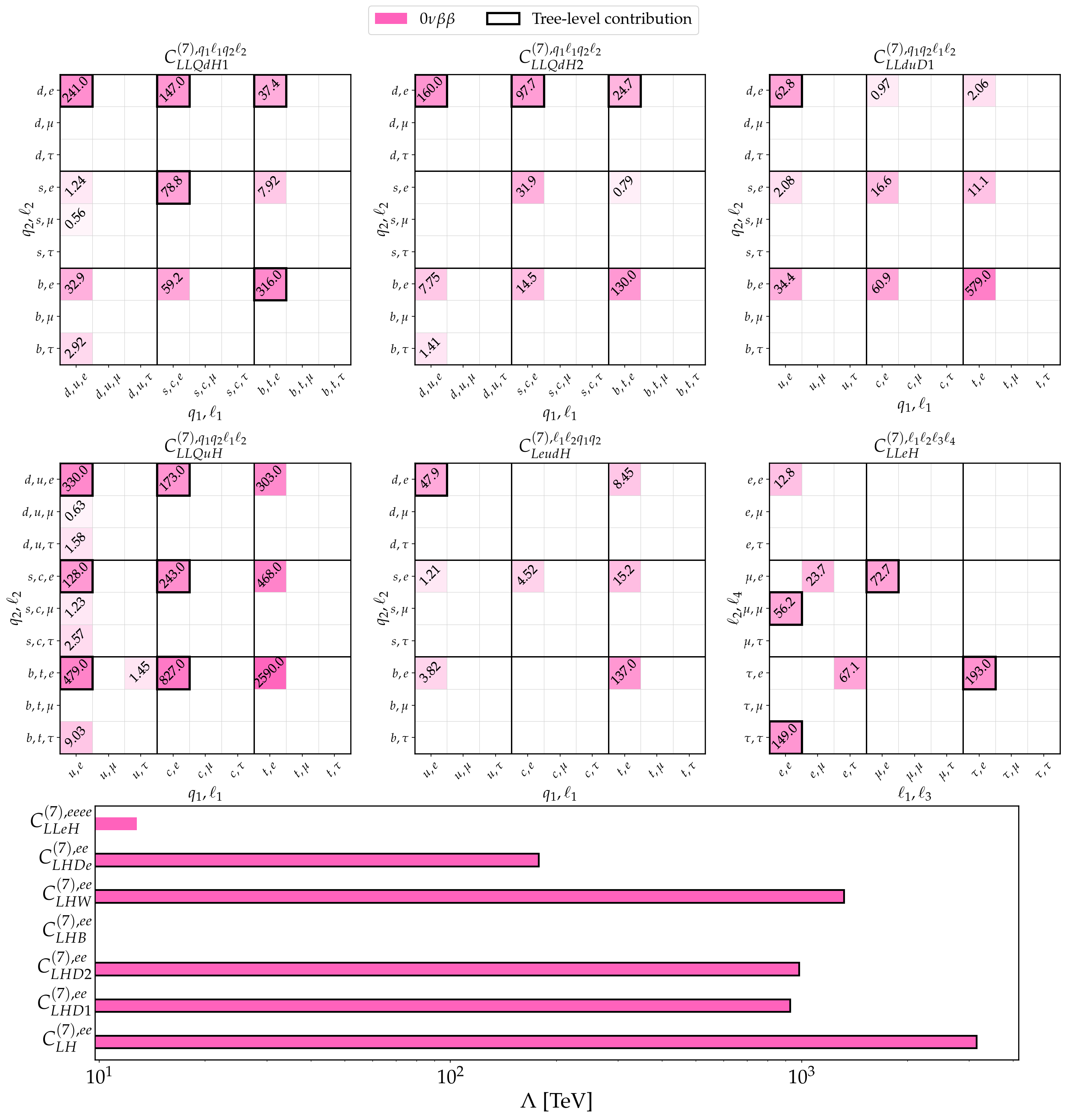}
    \caption{Relevant limits derived on the new physics scale $\Lambda_{\text{NP}}$ in TeV from $0\nu\beta\beta$ for the $\Delta L=2$ dimension-7 SMEFT operators for scenario FS-II (one single active lepton or flavor combination at the time). The black frame denotes which flavor combination of quarks contributes at the tree level.}
    \label{fig:full_0vbb}
\end{figure}

The corresponding observable-wise constraints from $0\nu\beta\beta$ are collected in
Figures~\ref{fig:LFU_0vbb} and~\ref{fig:full_0vbb}. Figure~\ref{fig:LFU_0vbb} shows the analogous $0\nu\beta\beta$ constraints in the
LFU scenario. Once the lepton-flavor structure is fixed, the impact of the quark-flavor
dependence and its RG evolution becomes particularly transparent. For several operators,
most notably $\mathcal{O}^{(7)}_{LLQuH}$ and
$\mathcal{O}^{(7)}_{LLduD1}$, coefficients involving third-generation quarks are subject
to particularly strong loop-induced constraints. This behavior originates from the
Yukawa-enhanced mixing associated with the heavier SM fermions and is the same mechanism
illustrated explicitly for $\mathcal{O}^{(7)}_{LLQuH}$ in
Figure~\ref{fig:barplot_LLQuH}. The pattern is nevertheless operator dependent:
third-generation flavor assignments are not uniformly the most constrained for every
Wilson coefficient. Figure~\ref{fig:LFU_0vbb} therefore provides the complete
operator-by-operator picture behind the more compact comparison presented in the main
text.

Figure~\ref{fig:full_0vbb}
shows the FS-II results and therefore retains the dependence on the individual quark and
lepton flavor indices. The black-framed entries indicate the flavor assignments that already
generate a contribution to $0\nu\beta\beta$ at tree level. Notice that these need not
correspond exclusively to first-generation quark indices at the SMEFT scale: for operators
involving up-type quarks, CKM rotations entering the SMEFT--LEFT matching can already
connect heavier-generation quark coefficients to the first-generation low-energy
interactions relevant for $0\nu\beta\beta$. Beyond these direct contributions, the many
unframed entries illustrate the considerably broader sensitivity generated by SMEFT
running. This is particularly apparent for
$\mathcal{O}^{(7)}_{LLQdH1}$,
$\mathcal{O}^{(7)}_{LLQdH2}$,
$\mathcal{O}^{(7)}_{LLQuH}$,
$\mathcal{O}^{(7)}_{LLduD1}$, and
$\mathcal{O}^{(7)}_{LeudH}$, for which flavor structures involving heavier-generation
quarks can feed into interactions probed by $0\nu\beta\beta$. The four-lepton operator
$\mathcal{O}^{(7)}_{LLeH}$ exhibits a correspondingly richer dependence on the lepton
flavor indices, illustrating that the strongest constraint need not always originate from
the simplest $ee$ flavor assignment. The lower panel collects the constraints for operators
without quark flavor indices.

Taken together, Figures~\ref{fig:full_B_and_K}--\ref{fig:LFU_0vbb} show that the
effect of SMEFT running is not limited to changing the numerical strength of constraints
that are already present at tree level. Operator mixing can also make additional SMEFT
operators and high-scale flavor structures accessible to a given low-energy observable.
The observable-wise presentation given here complements the summary plots of
Section~\ref{sec:eftlimits}, where this information is reduced to the strongest constraint
for each flavor assignment.

\clearpage

\clearpage

\clearpage


\bibliography{references}

@article{Belle-II:2023esi,
    author = "Adachi, I. and others",
    collaboration = "Belle-II",
    title = "{Evidence for B+{\textrightarrow}K+{\ensuremath{\nu}}{\ensuremath{\nu}}{\textasciimacron} decays}",
    eprint = "2311.14647",
    archivePrefix = "arXiv",
    primaryClass = "hep-ex",
    reportNumber = "Belle II Preprint 2023-017, KEK Preprint 2023-35",
    doi = "10.1103/PhysRevD.109.112006",
    journal = "Phys. Rev. D",
    volume = "109",
    number = "11",
    pages = "112006",
    year = "2024"
}

@article{NA62:2024pjp,
    author = "Cortina Gil, Eduardo and others",
    collaboration = "NA62",
    title = "{Observation of the $ {K}^{+}\to {\pi}^{+}\nu \overline{\nu} $ decay and measurement of its branching ratio}",
    eprint = "2412.12015",
    archivePrefix = "arXiv",
    primaryClass = "hep-ex",
    reportNumber = "CERN-EP-2024-343",
    doi = "10.1007/JHEP02(2025)191",
    journal = "JHEP",
    volume = "02",
    pages = "191",
    year = "2025"
}

@article{KOTO:2024zbl,
    author = "Ahn, J. K. and others",
    collaboration = "KOTO",
    title = "{Search for the KL{\textrightarrow}{\ensuremath{\pi}}0{\ensuremath{\nu}}{\ensuremath{\nu}}{\textasciimacron} Decay at the J-PARC KOTO Experiment}",
    eprint = "2411.11237",
    archivePrefix = "arXiv",
    primaryClass = "hep-ex",
    doi = "10.1103/PhysRevLett.134.081802",
    journal = "Phys. Rev. Lett.",
    volume = "134",
    number = "8",
    pages = "081802",
    year = "2025"
}

@article{Buras:2026hbe,
    author = "Buras, Andrzej J.",
    title = "{Hunting New Animalcula with Flavour Changing Processes}",
    eprint = "2605.27525",
    archivePrefix = "arXiv",
    primaryClass = "hep-ph",
    doi = "10.5506/APhysPolB.57.6-A15",
    journal = "Acta Phys. Polon. B",
    volume = "57",
    number = "6",
    pages = "6--A15",
    year = "2026"
}

@article{Fuentes-Martin:2022jrf,
    author = {Fuentes-Mart\'\i{}n, Javier and K\"onig, Matthias and Pag\`es, Julie and Thomsen, Anders Eller and Wilsch, Felix},
    title = "{A proof of concept for matchete: an automated tool for matching effective theories}",
    eprint = "2212.04510",
    archivePrefix = "arXiv",
    primaryClass = "hep-ph",
    reportNumber = "MITP-22-105, TUM-HEP-1443/22, ZU-TH-58/22",
    doi = "10.1140/epjc/s10052-023-11726-1",
    journal = "Eur. Phys. J. C",
    volume = "83",
    number = "7",
    pages = "662",
    year = "2023"
}

@article{Carmona:2021xtq,
    author = "Carmona, Adrian and Lazopoulos, Achilleas and Olgoso, Pablo and Santiago, Jose",
    title = "{Matchmakereft: automated tree-level and one-loop matching}",
    eprint = "2112.10787",
    archivePrefix = "arXiv",
    primaryClass = "hep-ph",
    doi = "10.21468/SciPostPhys.12.6.198",
    journal = "SciPost Phys.",
    volume = "12",
    number = "6",
    pages = "198",
    year = "2022"
}

@article{Jenkins:2013zja,
    author = "Jenkins, Elizabeth E. and Manohar, Aneesh V. and Trott, Michael",
    title = "{Renormalization Group Evolution of the Standard Model Dimension Six Operators I: Formalism and lambda Dependence}",
    eprint = "1308.2627",
    archivePrefix = "arXiv",
    primaryClass = "hep-ph",
    doi = "10.1007/JHEP10(2013)087",
    journal = "JHEP",
    volume = "10",
    pages = "087",
    year = "2013"
}

@article{Jenkins:2013wua,
    author = "Jenkins, Elizabeth E. and Manohar, Aneesh V. and Trott, Michael",
    title = "{Renormalization Group Evolution of the Standard Model Dimension Six Operators II: Yukawa Dependence}",
    eprint = "1310.4838",
    archivePrefix = "arXiv",
    primaryClass = "hep-ph",
    reportNumber = "CERN-PH-TH/2015-247",
    doi = "10.1007/JHEP01(2014)035",
    journal = "JHEP",
    volume = "01",
    pages = "035",
    year = "2014"
}

@article{Alonso:2013hga,
    author = "Alonso, Rodrigo and Jenkins, Elizabeth E. and Manohar, Aneesh V. and Trott, Michael",
    title = "{Renormalization Group Evolution of the Standard Model Dimension Six Operators III: Gauge Coupling Dependence and Phenomenology}",
    eprint = "1312.2014",
    archivePrefix = "arXiv",
    primaryClass = "hep-ph",
    reportNumber = "CERN-PH-TH-2013-305, CERN-PH-TH/2013-305",
    doi = "10.1007/JHEP04(2014)159",
    journal = "JHEP",
    volume = "04",
    pages = "159",
    year = "2014"
}

@article{Fridell:2024pmw,
    author = "Fridell, K{\r{a}}re and Gr{\'a}f, Luk{\'a}{\v{s}} and Harz, Julia and Hati, Chandan",
    title = "{Radiative neutrino masses from dim-7 SMEFT: a simplified multi-scale approach}",
    eprint = "2412.14268",
    archivePrefix = "arXiv",
    primaryClass = "hep-ph",
    reportNumber = "KEK-TH-2673, MITP-24-090",
    doi = "10.1007/JHEP09(2025)050",
    journal = "JHEP",
    volume = "09",
    pages = "050",
    year = "2025"
}

@article{Graf:2022lhj,
    author = "Gr\'af, Luk\'a\v{s} and Lindner, Manfred and Scholer, Oliver",
    title = "{Unraveling the 0\ensuremath{\nu}\ensuremath{\beta}\ensuremath{\beta} decay mechanisms}",
    eprint = "2204.10845",
    archivePrefix = "arXiv",
    primaryClass = "hep-ph",
    doi = "10.1103/PhysRevD.106.035022",
    journal = "Phys. Rev. D",
    volume = "106",
    number = "3",
    pages = "035022",
    year = "2022"
}

@article{Graf:2023dzf,
    author = "Gr{\'a}f, Luk{\'a}{\v{s}} and Jana, Sudip and Scholer, Oliver and Volmer, Nele",
    title = "{Neutrinoless double beta decay without vacuum Majorana neutrino mass}",
    eprint = "2312.15016",
    archivePrefix = "arXiv",
    primaryClass = "hep-ph",
    reportNumber = "N3AS-23-033",
    doi = "10.1016/j.physletb.2024.139111",
    journal = "Phys. Lett. B",
    volume = "859",
    pages = "139111",
    year = "2024"
}

@article{Scholer:2023bnn,
    author = "Scholer, Oliver and de Vries, Jordy and Gr\'af, Luk\'a\v{s}",
    title = "{\ensuremath{\nu}DoBe \textemdash{} A Python tool for neutrinoless double beta decay}",
    eprint = "2304.05415",
    archivePrefix = "arXiv",
    primaryClass = "hep-ph",
    doi = "10.1007/JHEP08(2023)043",
    journal = "JHEP",
    volume = "08",
    pages = "043",
    year = "2023"
}

@article{Zhang:2023ndw,
    author = "Zhang, Di",
    title = "{Revisiting renormalization group equations of the SMEFT dimension-seven operators}",
    eprint = "2310.11055",
    archivePrefix = "arXiv",
    primaryClass = "hep-ph",
    reportNumber = "TUM-HEP 1475/23",
    doi = "10.1007/JHEP02(2024)133",
    journal = "JHEP",
    volume = "02",
    pages = "133",
    year = "2024"
}

@article{Zhang:2023kvw,
    author = "Zhang, Di",
    title = "{Renormalization group equations for the SMEFT operators up to dimension seven}",
    eprint = "2306.03008",
    archivePrefix = "arXiv",
    primaryClass = "hep-ph",
    reportNumber = "TUM-HEP 1460/23",
    doi = "10.1007/JHEP10(2023)148",
    journal = "JHEP",
    volume = "10",
    pages = "148",
    year = "2023"
}

@article{Cirigliano:2022oqy,
    author = "Cirigliano, Vincenzo and others",
    title = "{Neutrinoless Double-Beta Decay: A Roadmap for Matching Theory to Experiment}",
    eprint = "2203.12169",
    archivePrefix = "arXiv",
    primaryClass = "hep-ph",
    reportNumber = "LA-UR-22-22587",
    month = "3",
    year = "2022"
}

@article{GERDA:2020xhi,
    author = "Agostini, M. and others",
    collaboration = "GERDA",
    title = "{Final Results of GERDA on the Search for Neutrinoless Double-$\beta$ Decay}",
    eprint = "2009.06079",
    archivePrefix = "arXiv",
    primaryClass = "nucl-ex",
    doi = "10.1103/PhysRevLett.125.252502",
    journal = "Phys. Rev. Lett.",
    volume = "125",
    number = "25",
    pages = "252502",
    year = "2020"
}

@article{Agostini:2022zub,
    author = "Agostini, Matteo and Benato, Giovanni and Detwiler, Jason A. and Men{\'e}ndez, Javier and Vissani, Francesco",
    title = "{Toward the discovery of matter creation with neutrinoless {\ensuremath{\beta}}{\ensuremath{\beta}} decay}",
    eprint = "2202.01787",
    archivePrefix = "arXiv",
    primaryClass = "hep-ex",
    doi = "10.1103/RevModPhys.95.025002",
    journal = "Rev. Mod. Phys.",
    volume = "95",
    number = "2",
    pages = "025002",
    year = "2023"
}

@article{Augier:2022znx,
    author = "Augier, C. and others",
    title = "{Final results on the $0\nu \beta \beta $ decay half-life limit of $^{100}$Mo from the CUPID-Mo experiment}",
    eprint = "2202.08716",
    archivePrefix = "arXiv",
    primaryClass = "nucl-ex",
    doi = "10.1140/epjc/s10052-022-10942-5",
    journal = "Eur. Phys. J. C",
    volume = "82",
    number = "11",
    pages = "1033",
    year = "2022"
}

@article{CUORE:2024ikf,
    author = "Adams, D. Q. and others",
    collaboration = "CUORE",
    title = "{Constraints on lepton number violation with the 2 tonne {\textperiodcentered} year CUORE dataset}",
    eprint = "2404.04453",
    archivePrefix = "arXiv",
    primaryClass = "nucl-ex",
    doi = "10.1126/science.adp6474",
    journal = "Science",
    volume = "390",
    number = "6777",
    pages = "1029--1032",
    year = "2025"
}

@article{LEGEND:2021bnm,
    author = "Abgrall, N. and others",
    collaboration = "LEGEND",
    title = "{The Large Enriched Germanium Experiment for Neutrinoless $\beta\beta$ Decay}: {LEGEND-1000 Preconceptual Design Report}",
    eprint = "2107.11462",
    archivePrefix = "arXiv",
    primaryClass = "physics.ins-det",
    month = "7",
    year = "2021"
}

@article{CUPID:2022wpt,
    author = "Armatol, A. and others",
    collaboration = "CUPID",
    title = "{Toward CUPID-1T}",
    eprint = "2203.08386",
    archivePrefix = "arXiv",
    primaryClass = "nucl-ex",
    month = "3",
    year = "2022"
}

@article{SNO:2021xpa,
    author = "Albanese, V. and others",
    collaboration = "SNO+",
    title = "{The SNO+ experiment}",
    eprint = "2104.11687",
    archivePrefix = "arXiv",
    primaryClass = "physics.ins-det",
    doi = "10.1088/1748-0221/16/08/P08059",
    journal = "JINST",
    volume = "16",
    number = "08",
    pages = "P08059",
    year = "2021"
}

@article{NuDoubt:2024jax,
    author = {B{\"o}hles, Manuel and others},
    collaboration = "NuDoubt++",
    title = "{Combining hybrid and opaque scintillator techniques in the search for double beta plus decays}",
    eprint = "2407.05999",
    archivePrefix = "arXiv",
    primaryClass = "physics.ins-det",
    doi = "10.1140/epjc/s10052-025-13847-1",
    journal = "Eur. Phys. J. C",
    volume = "85",
    number = "2",
    pages = "121",
    year = "2025",
    note = "[Erratum: Eur.Phys.J.C 85, 755 (2025)]"
}

@article{Schechter:1981bd,
    author = "Schechter, J. and Valle, J. W. F.",
    title = "{Neutrinoless Double beta Decay in SU(2) x U(1) Theories}",
    reportNumber = "SU-4217-213, COO-3533-213",
    doi = "10.1103/PhysRevD.25.2951",
    journal = "Phys. Rev. D",
    volume = "25",
    pages = "2951",
    year = "1982"
}

@article{Takasugi:1984xr,
    author = "Takasugi, Eiichi",
    title = "{Can the Neutrinoless Double Beta Decay Take Place in the Case of Dirac Neutrinos?}",
    reportNumber = "OS-GE 84-08",
    doi = "10.1016/0370-2693(84)90426-X",
    journal = "Phys. Lett. B",
    volume = "149",
    pages = "372--376",
    year = "1984"
}

@article{Duerr:2011zd,
    author = "Duerr, Michael and Lindner, Manfred and Merle, Alexander",
    title = "{On the Quantitative Impact of the Schechter-Valle Theorem}",
    eprint = "1105.0901",
    archivePrefix = "arXiv",
    primaryClass = "hep-ph",
    doi = "10.1007/JHEP06(2011)091",
    journal = "JHEP",
    volume = "06",
    pages = "091",
    year = "2011"
}

@article{Cirigliano:2018yza,
    author = "Cirigliano, V. and Dekens, W. and de Vries, J. and Graesser, M. L. and Mereghetti, E.",
    title = "{A neutrinoless double beta decay master formula from effective field theory}",
    eprint = "1806.02780",
    archivePrefix = "arXiv",
    primaryClass = "hep-ph",
    reportNumber = "LA-UR-18-24895, Nikhef 2018-023, NIKHEF-2018-023, DESY-18-072",
    doi = "10.1007/JHEP12(2018)097",
    journal = "JHEP",
    volume = "12",
    pages = "097",
    year = "2018"
}

@article{Weinberg:1979sa,
	author = "Weinberg, Steven",
	title = "{Baryon and Lepton Nonconserving Processes}",
	reportNumber = "HUTP-79-A050",
	doi = "10.1103/PhysRevLett.43.1566",
	journal = "Phys. Rev. Lett.",
	volume = "43",
	pages = "1566--1570",
	year = "1979"
}

@article{Babu:2001ex,
	author = "Babu, K. S. and Leung, Chung Ngoc",
	title = "{Classification of effective neutrino mass operators}",
	eprint = "hep-ph/0106054",
	archivePrefix = "arXiv",
	reportNumber = "OSU-HEP-01-02, UDHEP-02-01",
	doi = "10.1016/S0550-3213(01)00504-1",
	journal = "Nucl. Phys. B",
	volume = "619",
	pages = "667--689",
	year = "2001"
}

@article{Lehman:2014jma,
	author = "Lehman, Landon",
	title = "{Extending the Standard Model Effective Field Theory with the Complete Set of Dimension-7 Operators}",
	eprint = "1410.4193",
	archivePrefix = "arXiv",
	primaryClass = "hep-ph",
	doi = "10.1103/PhysRevD.90.125023",
	journal = "Phys. Rev. D",
	volume = "90",
	number = "12",
	pages = "125023",
	year = "2014"
}

@article{Liao:2016hru,
	author = "Liao, Yi and Ma, Xiao-Dong",
	title = "{Renormalization Group Evolution of Dimension-seven Baryon- and Lepton-number-violating Operators}",
	eprint = "1607.07309",
	archivePrefix = "arXiv",
	primaryClass = "hep-ph",
	doi = "10.1007/JHEP11(2016)043",
	journal = "JHEP",
	volume = "11",
	pages = "043",
	year = "2016"
}

@article{Deppisch:2012nb,
	author = "Deppisch, Frank F. and Hirsch, Martin and Pas, Heinrich",
	title = "{Neutrinoless Double Beta Decay and Physics Beyond the Standard Model}",
	eprint = "1208.0727",
	archivePrefix = "arXiv",
	primaryClass = "hep-ph",
	reportNumber = "IFIC-12-56",
	doi = "10.1088/0954-3899/39/12/124007",
	journal = "J. Phys. G",
	volume = "39",
	pages = "124007",
	year = "2012"
}

@article{Cirigliano:2017djv,
	author = "Cirigliano, V. and Dekens, W. and de Vries, J. and Graesser, M. L. and Mereghetti, E.",
	title = "{Neutrinoless double beta decay in chiral effective field theory: lepton number violation at dimension seven}",
	eprint = "1708.09390",
	archivePrefix = "arXiv",
	primaryClass = "hep-ph",
	reportNumber = "LA-UR-17-27799, NIKHEF-2017-039",
	doi = "10.1007/JHEP12(2017)082",
	journal = "JHEP",
	volume = "12",
	pages = "082",
	year = "2017"
}

@article{Graf:2018ozy,
	author = "Graf, Lukas and Deppisch, Frank F. and Iachello, Francesco and Kotila, Jenni",
	title = "{Short-Range Neutrinoless Double Beta Decay Mechanisms}",
	eprint = "1806.06058",
	archivePrefix = "arXiv",
	primaryClass = "hep-ph",
	doi = "10.1103/PhysRevD.98.095023",
	journal = "Phys. Rev. D",
	volume = "98",
	number = "9",
	pages = "095023",
	year = "2018"
}

@article{Deppisch:2020ztt,
	author = "Deppisch, Frank F. and Graf, Lukas and Iachello, Francesco and Kotila, Jenni",
	title = "{Analysis of light neutrino exchange and short-range mechanisms in $0\nu\beta\beta$ decay}",
	eprint = "2009.10119",
	archivePrefix = "arXiv",
	primaryClass = "hep-ph",
	doi = "10.1103/PhysRevD.102.095016",
	journal = "Phys. Rev. D",
	volume = "102",
	number = "9",
	pages = "095016",
	year = "2020"
}

@article{Li:2019fhz,
	author = "Li, Tong and Ma, Xiao-Dong and Schmidt, Michael A.",
	title = "{Implication of $K\to \pi \nu \bar{\nu}$ for generic neutrino interactions in effective field theories}",
	eprint = "1912.10433",
	archivePrefix = "arXiv",
	primaryClass = "hep-ph",
	doi = "10.1103/PhysRevD.101.055019",
	journal = "Phys. Rev. D",
	volume = "101",
	number = "5",
	pages = "055019",
	year = "2020"
}

@article{Jenkins:2017jig,
	author = "Jenkins, Elizabeth E. and Manohar, Aneesh V. and Stoffer, Peter",
	title = "{Low-Energy Effective Field Theory below the Electroweak Scale: Operators and Matching}",
	eprint = "1709.04486",
	archivePrefix = "arXiv",
	primaryClass = "hep-ph",
	doi = "10.1007/JHEP03(2018)016",
	journal = "JHEP",
	volume = "03",
	pages = "016",
	year = "2018"
}

@article{Brod:2010hi,
	author = "Brod, Joachim and Gorbahn, Martin and Stamou, Emmanuel",
	title = "{Two-Loop Electroweak Corrections for the $K \to \pi \nu \bar{\nu}$ Decays}",
	eprint = "1009.0947",
	archivePrefix = "arXiv",
	primaryClass = "hep-ph",
	doi = "10.1103/PhysRevD.83.034030",
	journal = "Phys. Rev. D",
	volume = "83",
	pages = "034030",
	year = "2011"
}

@article{Gubernari:2018wyi,
	author = "Gubernari, Nico and Kokulu, Ahmet and van Dyk, Danny",
	title = "{$B\to P$ and $B\to V$ Form Factors from $B$-Meson Light-Cone Sum Rules beyond Leading Twist}",
	eprint = "1811.00983",
	archivePrefix = "arXiv",
	primaryClass = "hep-ph",
	reportNumber = "EOS-2018-02, TUM-HEP-1172/18",
	doi = "10.1007/JHEP01(2019)150",
	journal = "JHEP",
	volume = "01",
	pages = "150",
	year = "2019"
}

@article{Bharucha:2015bzk,
	author = "Bharucha, Aoife and Straub, David M. and Zwicky, Roman",
	title = "{$B\to V\ell^+\ell^-$ in the Standard Model from light-cone sum rules}",
	eprint = "1503.05534",
	archivePrefix = "arXiv",
	primaryClass = "hep-ph",
	reportNumber = "TUM-HEP-957-14, CP3-Origins-2015-010, DIAS-2015-10",
	doi = "10.1007/JHEP08(2016)098",
	journal = "JHEP",
	volume = "08",
	pages = "098",
	year = "2016"
}

@article{Belle:2013tnz,
	author = "Lutz, O. and others",
	collaboration = "Belle",
	title = "{Search for $B \to h^{(*)} \nu \bar{\nu}$ with the full Belle $\Upsilon(4S)$ data sample}",
	eprint = "1303.3719",
	archivePrefix = "arXiv",
	primaryClass = "hep-ex",
	reportNumber = "BELLE-PREPRINT-2013-1, KEK-PREPRINT-2012-37",
	doi = "10.1103/PhysRevD.87.111103",
	journal = "Phys. Rev. D",
	volume = "87",
	number = "11",
	pages = "111103",
	year = "2013"
}

@article{deGouvea:2007qla,
	author = "de Gouvea, Andre and Jenkins, James",
	title = "{A Survey of Lepton Number Violation Via Effective Operators}",
	eprint = "0708.1344",
	archivePrefix = "arXiv",
	primaryClass = "hep-ph",
	reportNumber = "NUHEP-TH-07-10",
	doi = "10.1103/PhysRevD.77.013008",
	journal = "Phys. Rev. D",
	volume = "77",
	pages = "013008",
	year = "2008"
}

@article{Deppisch:2020oyx,
	author = "Deppisch, Frank F. and Fridell, K\r{a}re and Harz, Julia",
	title = "{Constraining lepton number violating interactions in rare kaon decays}",
	eprint = "2009.04494",
	archivePrefix = "arXiv",
	primaryClass = "hep-ph",
	reportNumber = "TUM-HEP-1274/20",
	doi = "10.1007/JHEP12(2020)186",
	journal = "JHEP",
	volume = "12",
	pages = "186",
	year = "2020"
}

@article{Bolton:2021pey,
	author = "Bolton, Patrick D. and Deppisch, Frank F. and Fridell, K\r{a}re and Harz, Julia and Hati, Chandan and Kulkarni, Suchita",
	title = "{Probing active-sterile neutrino transition magnetic moments with photon emission from CE$\nu$NS}",
	eprint = "2110.02233",
	archivePrefix = "arXiv",
	primaryClass = "hep-ph",
	reportNumber = "TUM-HEP-1357/21",
	doi = "10.1103/PhysRevD.106.035036",
	journal = "Phys. Rev. D",
	volume = "106",
	number = "3",
	pages = "035036",
	year = "2022"
}

@article{Buras:2015qea,
	author = "Buras, Andrzej J. and Buttazzo, Dario and Girrbach-Noe, Jennifer and Knegjens, Robert",
	title = "{$ {K}^{+}\to {\pi}^{+}\nu \overline{\nu} $ and $ {K}_L\to {\pi}^0\nu \overline{\nu} $ in the Standard Model: status and perspectives}",
	eprint = "1503.02693",
	archivePrefix = "arXiv",
	primaryClass = "hep-ph",
	reportNumber = "FLAVOUR(267104)-ERC-88",
	doi = "10.1007/JHEP11(2015)033",
	journal = "JHEP",
	volume = "11",
	pages = "033",
	year = "2015"
}

@article{Giunti:2014ixa,
	author = "Giunti, Carlo and Studenikin, Alexander",
	title = "{Neutrino electromagnetic interactions: a window to new physics}",
	eprint = "1403.6344",
	archivePrefix = "arXiv",
	primaryClass = "hep-ph",
	doi = "10.1103/RevModPhys.87.531",
	journal = "Rev. Mod. Phys.",
	volume = "87",
	pages = "531",
	year = "2015"
}

@article{Miranda:2021kre,
	author = "Miranda, O. G. and Papoulias, D. K. and Sanders, O. and T\'ortola, M. and Valle, J. W. F.",
	title = "{Low-energy probes of sterile neutrino transition magnetic moments}",
	eprint = "2109.09545",
	archivePrefix = "arXiv",
	primaryClass = "hep-ph",
	doi = "10.1007/JHEP12(2021)191",
	journal = "JHEP",
	volume = "12",
	pages = "191",
	year = "2021"
}

@article{Liao:2019tep,
	author = "Liao, Yi and Ma, Xiao-Dong",
	title = "{Renormalization Group Evolution of Dimension-seven Operators in Standard Model Effective Field Theory and Relevant Phenomenology}",
	eprint = "1901.10302",
	archivePrefix = "arXiv",
	primaryClass = "hep-ph",
	doi = "10.1007/JHEP03(2019)179",
	journal = "JHEP",
	volume = "03",
	pages = "179",
	year = "2019"
}

@article{Grzadkowski:2010es,
	author = "Grzadkowski, B. and Iskrzynski, M. and Misiak, M. and Rosiek, J.",
	title = "{Dimension-Six Terms in the Standard Model Lagrangian}",
	eprint = "1008.4884",
	archivePrefix = "arXiv",
	primaryClass = "hep-ph",
	reportNumber = "IFT-9-2010, TTP10-35",
	doi = "10.1007/JHEP10(2010)085",
	journal = "JHEP",
	volume = "10",
	pages = "085",
	year = "2010"
}

@article{Henning:2015alf,
	author = "Henning, Brian and Lu, Xiaochuan and Melia, Tom and Murayama, Hitoshi",
	title = "{2, 84, 30, 993, 560, 15456, 11962, 261485, ...: Higher dimension operators in the SM EFT}",
	eprint = "1512.03433",
	archivePrefix = "arXiv",
	primaryClass = "hep-ph",
	reportNumber = "UCB-PTH-15-14, IPMU15-0207",
	doi = "10.1007/JHEP08(2017)016",
	journal = "JHEP",
	volume = "08",
	pages = "016",
	year = "2017",
	note = "[Erratum: JHEP 09, 019 (2019)]"
}

@article{Felkl:2021uxi,
	author = "Felkl, Tobias and Li, Sze Lok and Schmidt, Michael A.",
	title = "{A tale of invisibility: constraints on new physics in b \textrightarrow{} s\ensuremath{\nu}\ensuremath{\nu}}",
	eprint = "2111.04327",
	archivePrefix = "arXiv",
	primaryClass = "hep-ph",
	reportNumber = "CPPC-2021-13",
	doi = "10.1007/JHEP12(2021)118",
	journal = "JHEP",
	volume = "12",
	pages = "118",
	year = "2021"
}

@article{Liao:2020zyx,
	author = "Liao, Yi and Ma, Xiao-Dong and Wang, Quan-Yu",
	title = "{Extending low energy effective field theory with a complete set of dimension-7 operators}",
	eprint = "2005.08013",
	archivePrefix = "arXiv",
	primaryClass = "hep-ph",
	doi = "10.1007/JHEP08(2020)162",
	journal = "JHEP",
	volume = "08",
	pages = "162",
	year = "2020"
}

@article{Fridell:2023rtr,
    author = "Fridell, K\r{a}re and Gr\'af, Luk\'a\v{s} and Harz, Julia and Hati, Chandan",
    title = "{Probing lepton number violation: a comprehensive survey of dimension-7 SMEFT}",
    eprint = "2306.08709",
    archivePrefix = "arXiv",
    primaryClass = "hep-ph",
    reportNumber = "KEK-TH-2486, N3AS-23-014, MITP-23-027, ULB-TH/23-06",
    doi = "10.1007/JHEP05(2024)154",
    journal = "JHEP",
    volume = "05",
    pages = "154",
    year = "2024"
}

@article{Kobach:2016ami,
    author = "Kobach, Andrew",
    title = "{Baryon Number, Lepton Number, and Operator Dimension in the Standard Model}",
    eprint = "1604.05726",
    archivePrefix = "arXiv",
    primaryClass = "hep-ph",
    reportNumber = "PHYS.LETT.-B758-(2016)-455-457",
    doi = "10.1016/j.physletb.2016.05.050",
    journal = "Phys. Lett. B",
    volume = "758",
    pages = "455--457",
    year = "2016"
}

@article{nEXO:2021ujk,
    author = "Adhikari, G. and others",
    collaboration = "nEXO",
    title = "{nEXO: neutrinoless double beta decay search beyond 10$^{28}$ year half-life sensitivity}",
    eprint = "2106.16243",
    archivePrefix = "arXiv",
    primaryClass = "nucl-ex",
    doi = "10.1088/1361-6471/ac3631",
    journal = "J. Phys. G",
    volume = "49",
    number = "1",
    pages = "015104",
    year = "2022"
}

@article{Graf:2025cfk,
    author = "Gr{\'a}f, Luk{\'a}{\v{s}} and Hati, Chandan and Mart{\'\i}n-Gal{\'a}n, Ana and Scholer, Oliver",
    title = "{Importance of Loop Effects in Probing Lepton Number Violation}",
    eprint = "2504.00081",
    archivePrefix = "arXiv",
    primaryClass = "hep-ph",
    month = "3",
    year = "2025"
}

@article{Hyvarinen:2015bda,
    author = {Hyv{\"a}rinen, Juhani and Suhonen, Jouni},
    title = "{Nuclear matrix elements for $0\nu\beta\beta$ decays with light or heavy Majorana-neutrino exchange}",
    doi = "10.1103/PhysRevC.91.024613",
    journal = "Phys. Rev. C",
    volume = "91",
    number = "2",
    pages = "024613",
    year = "2015"
}

@article{Menendez:2017fdf,
    author = "Men{\'e}ndez, J.",
    title = "{Neutrinoless $\beta\beta$ decay mediated by the exchange of light and heavy neutrinos: The role of nuclear structure correlations}",
    eprint = "1804.02105",
    archivePrefix = "arXiv",
    primaryClass = "nucl-th",
    doi = "10.1088/1361-6471/aa9bd4",
    journal = "J. Phys. G",
    volume = "45",
    number = "1",
    pages = "014003",
    year = "2018"
}

@article{KamLAND-Zen:2024eml,
    author = "Abe, S. and others",
    collaboration = "KamLAND-Zen",
    title = "{Search for Majorana Neutrinos with the Complete KamLAND-Zen Dataset}",
    eprint = "2406.11438",
    archivePrefix = "arXiv",
    primaryClass = "hep-ex",
    doi = "10.1103/jkf6-48j8",
    journal = "Phys. Rev. Lett.",
    volume = "135",
    number = "26",
    pages = "262501",
    year = "2025"
}

@article{Belle:2017oht,
    author = "Grygier, J. and others",
    collaboration = "Belle",
    title = "{Search for $\boldsymbol{B\to h\nu\bar{\nu}}$ decays with semileptonic tagging at Belle}",
    eprint = "1702.03224",
    archivePrefix = "arXiv",
    primaryClass = "hep-ex",
    doi = "10.1103/PhysRevD.96.091101",
    journal = "Phys. Rev. D",
    volume = "96",
    number = "9",
    pages = "091101",
    year = "2017",
    note = "[Addendum: Phys.Rev.D 97, 099902 (2018)]"
}

@article{Kitahara:2019lws,
    author = "Kitahara, Teppei and Okui, Takemichi and Perez, Gilad and Soreq, Yotam and Tobioka, Kohsaku",
    title = "{New physics implications of recent search for $K_L \to \pi^0 \nu\bar{\nu}$ at KOTO}",
    eprint = "1909.11111",
    archivePrefix = "arXiv",
    primaryClass = "hep-ph",
    reportNumber = "KEK-TH-2157, CERN-TH-2019-151",
    doi = "10.1103/PhysRevLett.124.071801",
    journal = "Phys. Rev. Lett.",
    volume = "124",
    number = "7",
    pages = "071801",
    year = "2020"
}

@article{NA62:2020fhy,
    author = "Cortina Gil, Eduardo and others",
    collaboration = "NA62",
    title = "{An investigation of the very rare $ {K}^{+}\to {\pi}^{+}\nu \overline{\nu} $ decay}",
    eprint = "2007.08218",
    archivePrefix = "arXiv",
    primaryClass = "hep-ex",
    reportNumber = "CERN-EP-2020-132",
    doi = "10.1007/JHEP11(2020)042",
    journal = "JHEP",
    volume = "11",
    pages = "042",
    year = "2020"
}

@article{Buras:2006gb,
    author = "Buras, Andrzej J. and Gorbahn, Martin and Haisch, Ulrich and Nierste, Ulrich",
    title = "{Charm quark contribution to K+ ---{\ensuremath{>}} pi+ nu anti-nu at next-to-next-to-leading order}",
    eprint = "hep-ph/0603079",
    archivePrefix = "arXiv",
    reportNumber = "TUM-HEP-600-05, IPPP-05-74, DCPT-05-148, TTP06-06, ZU-TH-23-05, FERMILAB-PUB-05-512-T",
    doi = "10.1007/JHEP11(2012)167",
    journal = "JHEP",
    volume = "11",
    pages = "002",
    year = "2006",
    note = "[Erratum: JHEP 11, 167 (2012)]"
}

@article{deVries:2026ujz,
    author = "de Vries, J. and Fajfer, S. and Leal, L. P. S. and Sumensari, O. and Funchal, R. Zukanovich",
    title = "{Neutrinoless Double-Beta Decays from Operator Mixing}",
    eprint = "2608.07657",
    archivePrefix = "arXiv",
    primaryClass = "hep-ph",
    month = "8",
    year = "2026"
}

@article{Liao:2025lxg,
    author = "Liao, Yi and Ma, Xiao-Dong and Wang, Hao-Lin and Zhao, Xiang",
    title = "{RGE solver for the complete dim-7 SMEFT interactions and its application to 0{\ensuremath{\nu}}{\ensuremath{\beta}}{\ensuremath{\beta}} decay}",
    eprint = "2505.06499",
    archivePrefix = "arXiv",
    primaryClass = "hep-ph",
    doi = "10.1007/JHEP08(2025)138",
    journal = "JHEP",
    volume = "08",
    pages = "138",
    year = "2025"
}

@article{Jenkins:2017dyc,
    author = "Jenkins, Elizabeth E. and Manohar, Aneesh V. and Stoffer, Peter",
    title = "{Low-Energy Effective Field Theory below the Electroweak Scale: Anomalous Dimensions}",
    eprint = "1711.05270",
    archivePrefix = "arXiv",
    primaryClass = "hep-ph",
    doi = "10.1007/JHEP01(2018)084",
    journal = "JHEP",
    volume = "01",
    pages = "084",
    year = "2018",
    note = "[Erratum: JHEP 12, 042 (2023)]"
}

@article{Endo:2026qof,
    author = "Endo, Motoi and Fridell, K{\r{a}}re and Iwamoto, Sho and Mura, Yushi and Yamamoto, Kei",
    title = "{Feasibility study of lepton number violation in rare B and K meson decays}",
    eprint = "2601.16422",
    archivePrefix = "arXiv",
    primaryClass = "hep-ph",
    reportNumber = "KEK-TH-2791",
    doi = "10.1007/JHEP06(2026)132",
    journal = "JHEP",
    volume = "06",
    pages = "132",
    year = "2026"
}

@article{Gonzalez:2023him,
    author = "Gonz{\'a}lez, Marcela and Neill, Nicol{\'a}s A.",
    title = "{QCD running in lepton number violating meson and tau decays}",
    eprint = "2309.14445",
    archivePrefix = "arXiv",
    primaryClass = "hep-ph",
    doi = "10.1103/PhysRevD.111.015041",
    journal = "Phys. Rev. D",
    volume = "111",
    number = "1",
    pages = "015041",
    year = "2025"
}

@article{Buras:2024ewl,
    author = "Buras, Andrzej J. and Harz, Julia and Mojahed, Martin A.",
    title = "{Disentangling new physics in $ K\to \pi \nu \overline{\nu} $ and $ B\to K\left({K}^{\ast}\right)\nu \overline{\nu} $ observables}",
    eprint = "2405.06742",
    archivePrefix = "arXiv",
    primaryClass = "hep-ph",
    reportNumber = "MITP-24-049, AJB-24-1",
    doi = "10.1007/JHEP10(2024)087",
    journal = "JHEP",
    volume = "10",
    pages = "087",
    year = "2024"
}

@article{Bell:2006wi,
    author = "Bell, Nicole F. and Gorchtein, Mikhail and Ramsey-Musolf, Michael J. and Vogel, Petr and Wang, Peng",
    title = "{Model independent bounds on magnetic moments of Majorana neutrinos}",
    eprint = "hep-ph/0606248",
    archivePrefix = "arXiv",
    reportNumber = "CALT-68-2603, KRL-MAP-321",
    doi = "10.1016/j.physletb.2006.09.055",
    journal = "Phys. Lett. B",
    volume = "642",
    pages = "377--383",
    year = "2006"
}

@article{Canas:2015yoa,
    author = "Canas, B. C. and Miranda, O. G. and Parada, A. and Tortola, M. and Valle, Jose W. F.",
    title = "{Updating neutrino magnetic moment constraints}",
    eprint = "1510.01684",
    archivePrefix = "arXiv",
    primaryClass = "hep-ph",
    reportNumber = "IFIC-15-XX",
    doi = "10.1016/j.physletb.2015.12.011",
    journal = "Phys. Lett. B",
    volume = "753",
    pages = "191--198",
    year = "2016",
    note = "[Addendum: Phys.Lett.B 757, 568--568 (2016)]"
}

@article{Raffelt:1998xu,
    author = "Raffelt, Georg G.",
    title = "{Comment on neutrino radiative decay limits from the infrared background}",
    eprint = "astro-ph/9808299",
    archivePrefix = "arXiv",
    doi = "10.1103/PhysRevLett.81.4020",
    journal = "Phys. Rev. Lett.",
    volume = "81",
    pages = "4020",
    year = "1998"
}

@article{Tandean:1999mg,
    author = "Tandean, Jusak",
    title = "{New physics and short distance s ---{\ensuremath{>}} d gamma transition in Omega- ---{\ensuremath{>}} Xi-gamma decay}",
    eprint = "hep-ph/9912497",
    archivePrefix = "arXiv",
    reportNumber = "ISU-HET-99-15",
    doi = "10.1103/PhysRevD.61.114022",
    journal = "Phys. Rev. D",
    volume = "61",
    pages = "114022",
    year = "2000"
}

@article{Finauri:2023kte,
    author = "Finauri, Gael and Gambino, Paolo",
    title = "{The q$^{2}$ moments in inclusive semileptonic B decays}",
    eprint = "2310.20324",
    archivePrefix = "arXiv",
    primaryClass = "hep-ph",
    reportNumber = "TUM-HEP 1477/23",
    doi = "10.1007/JHEP02(2024)206",
    journal = "JHEP",
    volume = "02",
    pages = "206",
    year = "2024"
}

@article{Jacob:1959at,
    author = "Jacob, M. and Wick, G. C.",
    title = "{On the General Theory of Collisions for Particles with Spin}",
    doi = "10.1006/aphy.2000.6022",
    journal = "Annals Phys.",
    volume = "7",
    pages = "404--428",
    year = "1959"
}

@article{Gratrex:2015hna,
    author = "Gratrex, James and Hopfer, Markus and Zwicky, Roman",
    title = "{Generalised helicity formalism, higher moments and the $B \to K_{J_K}(\to K \pi) \bar{\ell}_1 \ell_2$ angular distributions}",
    eprint = "1506.03970",
    archivePrefix = "arXiv",
    primaryClass = "hep-ph",
    reportNumber = "CP3-ORIGINS-2015-017, DIAS-2015-17, CP3-Origins-2015-017 DNRF90, DIAS-2015-17",
    doi = "10.1103/PhysRevD.93.054008",
    journal = "Phys. Rev. D",
    volume = "93",
    number = "5",
    pages = "054008",
    year = "2016"
}

@article{Das:2017ebx,
    author = "Das, Diganta and Hiller, Gudrun and Nisandzic, Ivan",
    title = "{Revisiting $B\to K^\ast( \to K\pi) \nu\bar{\nu}$ decays}",
    eprint = "1702.07599",
    archivePrefix = "arXiv",
    primaryClass = "hep-ph",
    reportNumber = "DO-TH-16-32",
    doi = "10.1103/PhysRevD.95.073001",
    journal = "Phys. Rev. D",
    volume = "95",
    number = "7",
    pages = "073001",
    year = "2017"
}

@article{Weinberg:1978kz,
    author = "Weinberg, Steven",
    editor = "Deser, S.",
    title = "{Phenomenological Lagrangians}",
    reportNumber = "HUTP-78-A051A",
    doi = "10.1016/0378-4371(79)90223-1",
    journal = "Physica A",
    volume = "96",
    number = "1-2",
    pages = "327--340",
    year = "1979"
}

@article{Manohar:1983md,
    author = "Manohar, Aneesh and Georgi, Howard",
    title = "{Chiral Quarks and the Nonrelativistic Quark Model}",
    reportNumber = "HUTP-83/A042a",
    doi = "10.1016/0550-3213(84)90231-1",
    journal = "Nucl. Phys. B",
    volume = "234",
    pages = "189--212",
    year = "1984"
}

@article{Jenkins:2013sda,
    author = "Jenkins, Elizabeth E. and Manohar, Aneesh V. and Trott, Michael",
    title = "{Naive Dimensional Analysis Counting of Gauge Theory Amplitudes and Anomalous Dimensions}",
    eprint = "1309.0819",
    archivePrefix = "arXiv",
    primaryClass = "hep-ph",
    reportNumber = "CERN-PH-TH-2013-213",
    doi = "10.1016/j.physletb.2013.09.020",
    journal = "Phys. Lett. B",
    volume = "726",
    pages = "697--702",
    year = "2013"
}

@article{Buchalla:2013eza,
    author = "Buchalla, Gerhard and Cat{\'a}, Oscar and Krause, Claudius",
    title = "{On the Power Counting in Effective Field Theories}",
    eprint = "1312.5624",
    archivePrefix = "arXiv",
    primaryClass = "hep-ph",
    reportNumber = "LMU-ASC-81-13",
    doi = "10.1016/j.physletb.2014.02.015",
    journal = "Phys. Lett. B",
    volume = "731",
    pages = "80--86",
    year = "2014"
}

@article{Gavela:2016bzc,
    author = "Gavela, B. M. and Jenkins, E. E. and Manohar, A. V. and Merlo, L.",
    title = "{Analysis of General Power Counting Rules in Effective Field Theory}",
    eprint = "1601.07551",
    archivePrefix = "arXiv",
    primaryClass = "hep-ph",
    reportNumber = "CERN-TH-2016-015, FTUAM-16-2, IFT-UAM-CSIC-16-006",
    doi = "10.1140/epjc/s10052-016-4332-1",
    journal = "Eur. Phys. J. C",
    volume = "76",
    number = "9",
    pages = "485",
    year = "2016"
}

@article{Buchalla:2022vjp,
    author = {Buchalla, Gerhard and Heinrich, Gudrun and M{\"u}ller-Salditt, Ch. and Pandler, Florian},
    title = "{Loop counting matters in SMEFT}",
    eprint = "2204.11808",
    archivePrefix = "arXiv",
    primaryClass = "hep-ph",
    reportNumber = "LMU-ASC{\textasciitilde}16/22, KA-TP-10-2022, P3H-22-041",
    doi = "10.21468/SciPostPhys.15.3.088",
    journal = "SciPost Phys.",
    volume = "15",
    number = "3",
    pages = "088",
    year = "2023"
}

@article{FlavourLatticeAveragingGroupFLAG:2021npn,
    author = "Aoki, Y. and others",
    collaboration = "Flavour Lattice Averaging Group (FLAG)",
    title = "{FLAG Review 2021}",
    eprint = "2111.09849",
    archivePrefix = "arXiv",
    primaryClass = "hep-lat",
    reportNumber = "CERN-TH-2021-191, JLAB-THY-21-3528, FERMILAB-PUB-21-620-SCD-T",
    doi = "10.1140/epjc/s10052-022-10536-1",
    journal = "Eur. Phys. J. C",
    volume = "82",
    number = "10",
    pages = "869",
    year = "2022"
}

@article{Descotes-Genon:2019bud,
    author = "Descotes-Genon, S{\'e}bastien and Khodjamirian, Alexander and Virto, Javier",
    title = "{Light-cone sum rules for $B\to K\pi$ form factors and applications to rare decays}",
    eprint = "1908.02267",
    archivePrefix = "arXiv",
    primaryClass = "hep-ph",
    reportNumber = "LPT-ORSAY/19-31, SI-HEP-2019-11, TUM-HEP-1161/18, MIT-CTP/5058,
  NIOBE-2019-01",
    doi = "10.1007/JHEP12(2019)083",
    journal = "JHEP",
    volume = "12",
    pages = "083",
    year = "2019"
}

@article{Barabash:2020nck,
    author = "Barabash, Alexander",
    title = "{Precise Half-Life Values for Two-Neutrino Double-{\ensuremath{\beta}} Decay: 2020 Review}",
    eprint = "2009.14451",
    archivePrefix = "arXiv",
    primaryClass = "nucl-ex",
    doi = "10.3390/universe6100159",
    journal = "Universe",
    volume = "6",
    number = "10",
    pages = "159",
    year = "2020"
}

@article{KamLAND-Zen:2016pfg,
    author = "Gando, A. and others",
    collaboration = "KamLAND-Zen",
    title = "{Search for Majorana Neutrinos near the Inverted Mass Hierarchy Region with KamLAND-Zen}",
    eprint = "1605.02889",
    archivePrefix = "arXiv",
    primaryClass = "hep-ex",
    doi = "10.1103/PhysRevLett.117.082503",
    journal = "Phys. Rev. Lett.",
    volume = "117",
    number = "8",
    pages = "082503",
    year = "2016",
    note = "[Addendum: Phys.Rev.Lett. 117, 109903 (2016)]"
}

@article{EXO-200:2013xfn,
    author = "Albert, J. B. and others",
    collaboration = "EXO-200",
    title = "{Improved measurement of the $2\nu\beta\beta$ half-life of $^{136}$Xe with the EXO-200 detector}",
    eprint = "1306.6106",
    archivePrefix = "arXiv",
    primaryClass = "nucl-ex",
    doi = "10.1103/PhysRevC.89.015502",
    journal = "Phys. Rev. C",
    volume = "89",
    number = "1",
    pages = "015502",
    year = "2014"
}

@article{Graf:2026rie,
    author = "Gr{\'a}f, Luk{\'a}{\v{s}} and Kotila, Jenni and Scholer, Oliver",
    title = "{Positron-Emitting and Electron-Capturing Double-Beta Processes in the Standard Model and Beyond}",
    eprint = "2606.26097",
    archivePrefix = "arXiv",
    primaryClass = "hep-ph",
    month = "6",
    year = "2026"
}

@article{Esser:2026ehc,
    author = "Esser, Fabian and Gr{\'a}f, Luk{\'a}{\v{s}} and Hati, Chandan",
    title = "{Cartography of LNV dim-9 SMEFT: implications for radiative neutrino masses and 0{\ensuremath{\nu}}{\ensuremath{\beta}}{\ensuremath{\beta}}}",
    eprint = "2602.18395",
    archivePrefix = "arXiv",
    primaryClass = "hep-ph",
    doi = "10.1007/JHEP07(2026)206",
    journal = "JHEP",
    volume = "07",
    pages = "206",
    year = "2026"
}

@article{deVries:2025hqa,
    author = "de Vries, Jordy and Gr{\'a}f, Luk{\'a}{\v{s}} and Plakkot, Vaisakh and Star{\'y}, Dominik",
    title = "{Scalarful double beta decay}",
    eprint = "2511.19173",
    archivePrefix = "arXiv",
    primaryClass = "hep-ph",
    doi = "10.1007/JHEP03(2026)102",
    journal = "JHEP",
    volume = "03",
    pages = "102",
    year = "2026"
}

@article{Brune:2018sab,
    author = {Brune, Tim and P{\"a}s, Heinrich},
    title = "{Massive Majorons and constraints on the Majoron-neutrino coupling}",
    eprint = "1808.08158",
    archivePrefix = "arXiv",
    primaryClass = "hep-ph",
    reportNumber = "DO-TH 18/23",
    doi = "10.1103/PhysRevD.99.096005",
    journal = "Phys. Rev. D",
    volume = "99",
    number = "9",
    pages = "096005",
    year = "2019"
}

@article{Doi:1987rx,
    author = "Doi, M. and Kotani, T. and Takasugi, E.",
    title = "{The Neutrinoless Double Beta Decay With Majoron Emission}",
    reportNumber = "OS-GE-87-07-REV, OS-GE-87-07",
    doi = "10.1103/PhysRevD.37.2575",
    journal = "Phys. Rev. D",
    volume = "37",
    pages = "2575",
    year = "1988"
}

@article{Cepedello:2018zvr,
    author = "Cepedello, Ricardo and Deppisch, Frank F. and Gonz{\'a}lez, Lorena and Hati, Chandan and Hirsch, Martin",
    title = "{Neutrinoless Double-$\beta$ Decay with Nonstandard Majoron Emission}",
    eprint = "1811.00031",
    archivePrefix = "arXiv",
    primaryClass = "hep-ph",
    reportNumber = "IFIC/18-38",
    doi = "10.1103/PhysRevLett.122.181801",
    journal = "Phys. Rev. Lett.",
    volume = "122",
    number = "18",
    pages = "181801",
    year = "2019"
}

@article{Graf:2020cbf,
    author = "Graf, Lukas and Jana, Sudip and Lindner, Manfred and Rodejohann, Werner and Xu, Xun-Jie",
    title = "{Flavored neutrinoless double beta decay}",
    eprint = "2010.15109",
    archivePrefix = "arXiv",
    primaryClass = "hep-ph",
    doi = "10.1103/PhysRevD.103.055007",
    journal = "Phys. Rev. D",
    volume = "103",
    number = "5",
    pages = "055007",
    year = "2021"
}

@article{Bolton:2026gku,
    author = "Bolton, Patrick D. and Boudjema, Noor-In{\`e}s and Deppisch, Frank F. and Hati, Chandan and Majumdar, Chayan",
    title = "{Displaced Signals from Long-lived Particles in Neutrinoless Double Beta Decay}",
    eprint = "2609.05352",
    archivePrefix = "arXiv",
    primaryClass = "hep-ph",
    month = "9",
    year = "2026"
}
\bibliographystyle{JHEP}

\end{document}